\documentclass[a4paper,11pt]{article}
\usepackage{jheppub}
\usepackage[T1]{fontenc}
\usepackage{graphicx}
\usepackage{amsmath}
\usepackage{amsmath}
\usepackage{xspace}
\usepackage{subfig}
\usepackage{color}
\usepackage[utf8]{inputenc}
\usepackage{commath}
\usepackage{slashed}
\usepackage{fancyvrb}

\usepackage[utf8]{inputenc}
\usepackage[english]{babel}
\usepackage{amsfonts}
\usepackage{amssymb}
\usepackage{color}
\usepackage{cancel}
\newcommand{\be}{\begin{equation}}
\newcommand{\ee}{\end{equation}}
\newcommand{\bea}{\begin{eqnarray}}
\newcommand{\eea}{\end{eqnarray}}

\newcommand{\nn}{\nonumber}

\catcode`@=12

\newcommand{\MMr}{\textcolor{magenta}}

\preprint{IP/BBSR/2022-05}

\title{\boldmath WIMP and FIMP Dark Matter in Singlet-Triplet Fermionic Model  }

\author[a]{Genevi\`{e}ve B\'{e}langer,}
\author[b]{Sandhya Choubey,}
\author[c]{Rohini M. Godbole,}
\author[d]{Sarif Khan,}
\author[e,f]{Manimala Mitra,}
\author[e,f]{Abhishek Roy,}

\newcommand{\AddrHBNI}{
	Homi Bhabha National Institute, BARC Training School Complex, Anushakti Nagar, Mumbai 400094, India }
\affiliation[a]{LAPTh, CNRS, USMB, 9 Chemin de Bellevue, 74940 Annecy, France}
\affiliation[b]{Department of Physics, School of Engineering Sciences, KTH Royal Institute of Technology, AlbaNova
University Center, 106 91 Stockholm, Sweden}
\affiliation[c]{Centre for High Energy Physics, Indian Institute of Science, Bengaluru - 560012, India}
\affiliation[d]{Institut f\"{u}r Theoretische Physik, Georg-August-Universit\"{a}t G\"{o}ttingen,
Friedrich-Hund-Platz 1, 37077 G\"{o}ttingen, Germany}
\affiliation[e]{Institute of Physics, Sachivalaya Marg, Bhubaneswar, Odisha 751005, India}
\affiliation[f]{\AddrHBNI}

\emailAdd{belanger@lapth.cnrs.fr}
\emailAdd{choubey@kth.se}
\emailAdd{rohini@iisc.ac.in}
\emailAdd{sarif.khan@uni-goettingen.de}
\emailAdd{manimala@iopb.res.in}
\emailAdd{abhishek.r@iopb.res.in}

\abstract{We present  an extension of the SM involving three triplet fermions, one triplet scalar and one singlet fermion, which can explain both  neutrino masses and dark matter. One triplet of fermions and the singlet are odd under a $Z_2$ symmetry, thus the  model features two possible dark matter candidates. The two remaining  $Z_2$-even triplet fermions can reproduce  the  neutrino masses and oscillation parameters consistent with observations. We consider the case where the singlet has feeble couplings  while the triplet is weakly interacting and investigate the different possibilities for reproducing the observed dark matter relic density. This includes  production of the  triplet WIMP from freeze-out and from decay of the singlet as well as freeze-in production of the singlet from decay of particles  that belong to the  thermal bath or are thermally decoupled. While freeze-in production is usually dominated by decay processes, we also show cases where the annihilation of bath particles give substantial contribution to the final relic density. This occurs when the new scalars are below the TeV scale, thus in the reach of the LHC.  The next-to-lightest odd particle can be long-lived and can alter the successful BBN predictions for the abundance of light elements, these constraints are relevant in both the scenarios where the singlet or the triplet are  the long-lived particle.In the case where the triplet is the DM, the model is subject to constraints from ongoing direct, indirect and collider experiments.   When the singlet is the DM, the triplet which is the next-to-lightest odd particle can be long-lived and can be probed at the proposed MATHUSLA detector. Finally we also address the detection prospects of  triplet fermions and scalars at the LHC. 
}

\begin{document} 

\maketitle
\flushbottom

\section{Introduction}
Numerous cosmological and astrophysical observations  indicate that $23 \%$ of the energy budget of the universe is in the form of dark matter (DM). However, we are still in the dark about the nature and the origin of the dark matter. The absence of a DM candidate within the Standard Model (SM) gives  compelling evidence for physics beyond the Standard Model (BSM). Weakly Interacting Massive Particles (WIMP) have been the leading candidate for DM for decades.  WIMPs have a mass around the electroweak scale and  interact with constituents of the plasma in the early Universe via electroweak interactions. WIMPs decouple from the thermal bath through the  freeze-out mechanism and can naturally explain the DM relic density extracted from precise measurements of the Cosmic Microwave Background 
\cite{Ade:2015xua}. 
However  the  null results from  various direct and indirect searches for WIMP DM motivate the exploration of alternative DM production mechanisms. 
The production of DM through the thermal  freeze-in mechanism \cite{Hall:2009bx} is one of the most popular alternatives. 
This mechanism involves a Feebly Interacting Massive Particle (FIMP)  which interacts so  feebly   that it  never attains chemical equilibrium with the thermal bath.
The DM is produced through decay and/or annihilations of SM and BSM particles which are in equilibrium with the thermal plasma. 
Another possibility is to have DM produced at a late epoch of the Universe through the out-of-equilibrium decay of particles that have themselves frozen-in or frozen-out. The latter is known as  the superWIMP mechanism~\cite{Feng:2003xh}.

In addition to  the DM problem,  the SM fails to address  the issue of neutrino masses and mixings for which their exist strong 
experimental evidence.
 One of the most promising model to explain the neutrino masses relies on an extension of the SM with fermion triplets that generate the active neutrino masses through the Type-III seesaw mechanism \cite{Foot:1988aq}. 
 In order to incorporate a DM candidate, this model can further be extended with a dark sector comprising  a fermion triplet $\rho$ having gauge interaction with SM states and a gauge singlet fermionic state $N$. The dark sector is odd under a $Z_{2}$ symmetry which ensures that either of the neutral components $\rho$ or $N$ is stable depending on their masses 
and can thus be a good dark matter candidate.
 In addition, this model also contains a scalar triplet ($\Delta$) having zero hypercharge. Electroweak symmetry breaking (EWSB) induces a vev for the neutral component of $\Delta$, this vev then generates a mixing between $\rho$ and N. The neutrino singlet-triplet fermionic model ($
\nu$STFM) can thus explain neutrino masses while providing a DM candidate that can either be the singlet or the triplet. 

The properties of WIMP DM in the singlet-triplet fermionic model  were studied in \cite{Choubey:2017yyn, Chardonnet:1993wd}. It was shown that the presence of the additional state in the dark sector impacts significantly the phenomenology as compared with  scenarios  with only a singlet or a triplet of fermions in the dark sector.  In this article we explore further avenues for DM formation in the $\nu$STFM model and entertain the possibility that the singlet state in the dark sector is feebly interacting.  This requires that the Yukawa coupling between the singlet and triplet dark states and the triplet scalar be feeble.  
Under these conditions, DM can be either a FIMP, when the singlet is lightest,  or a WIMP, when the triplet is lightest. In this model, several mechanisms can contribute to DM formation.
The WIMP can be produced through the freeze-out mechanism and through the decay of the heavier FIMP in the dark sector, we refer to the latter as non-thermal freeze-in production. When the singlet FIMP is the lightest state, it  can be produced through the decay or scattering of any SM or BSM particle in the thermal bath, this is  thermal freeze-in production,  or can also  receive a contribution from the out-of-equilibrium decay of the heavier WIMP produced through freeze-out, also called  non-thermal freeze-in production. 


To explore all possibilities for DM formation we consider two scenarios.
In the first scenario, $\rho$ is the DM candidate and N is the next-to-lightest odd particle (NLOP). In the minimal DM model with only a fermion triplet, it was shown that  the triplet is required to be rather heavy, $M_\rho> 2.4 \  {\rm TeV}$~\cite{Cirelli:2005uq} to have a large enough relic abundance. Indeed the thermally averaged cross section $\langle\sigma v\rangle$ increases as the mass of the $\rho$ decreases, thus for lower values of $M_\rho$ the relic density of  $\rho$  is below the observed value. The presence of the singlet NLOP can change this conclusion. The NLOP which never attains thermal and chemical equilibrium with bath particles can be produced through the freeze-in mechanism and at late times decay into $\rho$, thus increasing the relic density of $\rho$.

In the second  scenario,  N is the lightest dark sector particle and the DM candidate while $\rho$ is the NLOP.  The production of $N$ through the freeze-in mechanism is directly proportional to the dark sector Yukawa coupling $Y_{\rho\Delta}$, for large enough coupling this can be sufficient to reproduce the DM relic density. When the Yukawa coupling is small N can also be produced through out-of-equilibrium decay of $\rho$. For this to be efficient the abundance of $\rho$ must be large enough, which requires  $M_\rho>2.4 \ {\rm TeV}$.  
In both scenarios the production of the FIMP can arise from decays or from scattering of particles in the thermal bath, 
an important contribution from annihilation processes however requires that the new scalar fields  be rather light, at the TeV scale or below. In this context,  the  discovery prospects of scalar triplets   ($\Delta$) at the LHC are improved. 

The late decay of  the heavier state in the dark sector for the two scenarios considered can disrupt the successful predictions of big bang nucleosynthesis (BBN). In particular,  constraints from hadronic energy injection become
important when the lifetime of the late decaying particle exceeds 100 sec. The impact of these constraints on the model parameters is elaborately discussed. Moreover when the triplet is long-lived, the charged component can be searched for at the LHC through its disappearing track signature.

The manuscript is organised as follows. The model is described in   Section \ref{model}. In Section \ref{boundscal} and \ref{boundDM}, we discuss theoretical constraints on the scalar sector as well as constraints on  DM from astrophysical, cosmological and collider observables.  Section \ref{DM1} contains most of our results,  dark matter production is explored  in detail for the two scenarios considered and a subsection is dedicated to the case where the BSM scalar sector is light. Moreover the impact of  BBN constraints arising from the late production of DM through out-of-equilibrium decay is investigated. 
The prospects for searches for  triplet fermions and triplet  scalars at the LHC are presented  in Section \ref{collider1}. We present our conclusions in Section \ref{conclusion}.  Necessary calculational details are provided in the Appendix \ref{appendix}.

\section{The Model}\label{model}
In addition to the SM fields, the  
$\nu$STFM model contains $SU(2)$ triplet fermions $\rho_{i} (i=1,2,3)$, a SM gauge singlet fermion $N^{\prime}$ and a $SU(2)$ triplet scalar field $\Delta$.  We show the particle content of the model in Table.~\ref{tab:tab2}. The Lagrangian possesses a discrete $\mathbb{Z}_2$ symmetry, in addition to the SM gauge symmetries. The latter is required to stabilise the DM fields. The SM fields, the scalar triplet and two of the fermionic triplets are even under this symmetry while one of the fermionic triplet are odd under the $\mathbb{Z}_2$ symmetry and form the dark sector.
The Lagrangian reads, 
\begin{eqnarray}
&\mathcal{L} &= \mathcal{L}_{SM} + \sum^{3}_{i=1} Tr \left[ \bar{\rho_i}\,i\, \gamma^{\mu} D_{\mu} \rho_i \right]
+ \bar{N^{\prime}}\,i\, \gamma^{\mu} D_{\mu} N^{\prime}
+Tr[(D_{\mu}\Delta)^{\dagger} (D^{\mu} \Delta)] -V(\phi_{h}, \Delta) \nn \\
&&- \sum^{(3,2)}_{(i,j) = (1,1)} \lambda_{ij} \bar{L_i}  \phi_h \rho^c_j 
- Y_{\rho \Delta}\, (Tr[\bar{\rho_3}\, \Delta]\, N^{\prime} + h.c.) 
-\sum^{3}_{i = 1} M_{\rho_i}\, Tr[\bar{\rho^{c}_i} \rho_i]- M_{N^{\prime}}\, \bar{N^{\prime c}} N^{\prime},  \nn \\
\label{mix-lag}
\end{eqnarray}

\begin{table}[h!]
	\begin{tabular}{|c|c|c|c|}
		\hline
		\begin{tabular}{c}
			Symmetry\\
			Group\\ 
			\hline
			$SU(3)_{c}$\\ 
			\hline
			$SU(2)_{L}$\\ 
			\hline
			$U(1)_{Y}$\\ 
			\hline
			$\mathbb{Z}_{2}$\\     
		\end{tabular}
		&
		
		\begin{tabular}{c|c|c}
			\multicolumn{3}{c}{Baryon Fields}\\ 
			\hline
			$Q_{L}^{i}$&$u_{R}^{i}$&$d_{R}^{i}$\\
			\hline
			$3$&$3$&$3$\\  
			\hline
			$2$&$1$&$1$\\ 
			\hline
			$1/6$&$2/3$&$-1/3$\\
			\hline
			$+$&$+$&$+$\\ 
			
		\end{tabular}
		&
		\begin{tabular}{c|c|c|c|c|c}
			\multicolumn{6}{c}{Fermion Fields}\\
			\hline
			$L_{L}^{i}$ & ~$e_{R}^{i}$~ & ~$N^{\prime}$~ & ~$\rho_1$~& ~$\rho_2$~ & ~$\rho_3$~  \\
			\hline
			$1$&$1$&$1$&$1$&$1$&$1$\\
			\hline
			$2$&$1$&$1$&$3$&$3$&$3$\\
			\hline
			$-1/2$&$-1$&$0$&$0$&$0$&$0$\\
			\hline
			$+$&$+$&$-$&$+$&$+$&$-$\\
			
		\end{tabular}
		&
		\begin{tabular}{c|c}
			\multicolumn{2}{c}{Scalar Fields}\\
			\hline
			$\phi_{h}$&$\Delta$\\
			\hline
			~~$1$~~&$1$\\
			\hline
			~~$2$~~&$3$\\
			\hline
			~~$1/2$~~&$0$\\
			\hline
			~~$+$~~&$+$\\
			
		\end{tabular}\\
		\hline
	\end{tabular}
	\caption{Particle content and their corresponding
		charges under various symmetry groups.} 
	\label{tab:tab2}
\end{table}
where the triplet fermion takes the following form,
\begin{eqnarray}
\rho_i=
\begin{pmatrix}
\frac{\rho^0_{i}}{2} & \frac{\rho^{+}_i}{\sqrt{2}} \\
\frac{\rho^{-}_i}{\sqrt{2}} & -\frac{\rho^0_{i}}{2}
\end{pmatrix}\,,\,i = 1, 2, 3\,.
\label{rho}
\end{eqnarray}
The triplet scalar field $\Delta$ is represented as 
\begin{eqnarray}
\Delta=\begin{pmatrix}
\frac{\Delta^0}{2} & \frac{\Delta^{+}}{\sqrt{2}} \\
\frac{\Delta^{-}}{\sqrt{2}} & -\frac{\Delta^0}{2}
\end{pmatrix}. 
\label{delta}
\end{eqnarray}
where $\Delta^0$ is a single real field and $\Delta^+$ and $\Delta^-$ are charge conjugate to each other. 
Note that $\mathcal{L}_{SM}$ does not include the potential for the standard Higgs field $\phi_h$.
With $\phi_h$ and $\Delta$, the scalar potential of the model has the following form,
\begin{eqnarray}
V(\phi_{h}, \Delta) &=& -\mu_{h}^{2} \phi_{h}^{\dagger} \phi_{h}
+ \frac{\lambda_{h}}{4} (\phi_{h}^{\dagger} \phi_{h})^{2}
+\mu_{\Delta}^{2} Tr[\Delta^{\dagger} \Delta] + \lambda_{\Delta} (\Delta^{\dagger} \Delta)^{2}
+ \lambda_{1}\, (\phi_{h}^{\dagger} \phi_{h})\, {\rm Tr}\,[\Delta^{\dagger} \Delta] \nonumber \\
&&+ \lambda_{2}\,\left(Tr[\Delta^{\dagger} \Delta]\right)^{2} 
+ \lambda_{3}\,Tr[(\Delta^{\dagger} \Delta)^{2}]
+ \lambda_{4}\, \phi_{h}^{\dagger} \Delta \Delta^{\dagger} \phi_{h}
+ ( \mu \phi_{h}^{\dagger} \Delta \phi_{h} + h.c.)\,.
\end{eqnarray}
In general, a $\phi_{h}^{\dagger} \Delta^{\dagger} \Delta \phi_{h}$   term is also allowed by the gauge symmetry, however, this term  
can be decomposed into two terms  which are similar to the quartic coupling associated with $\lambda_1$
and $\lambda_4$. Hence, we do not include this term in the Lagrangian. 
The quadratic and quartic couplings associated with the potential  obey the following conditions,
\begin{eqnarray}
\mu_{h}^{2} > 0, \,\,\,\, \mu_{\Delta}^{2} > 0, \,\,\,\, \lambda_{h} > 0\,\,\,\, {\rm and}\,\,\,\, \lambda_{\Delta} > 0\,.
\end{eqnarray}

The neutral real component of $\phi_h$ acquires a vacuum expectation value, $v$, and breaks the electroweak symmetry. The  field $\Delta^0$ acquires an induced vev $v_{\Delta}$, 
 which obeys the following  relation, 
\begin{eqnarray}
\langle \Delta^0 \rangle = v_{\Delta} = 
\frac{\mu v^{2}}{2 \left(\mu_{\Delta}^{2} + (\lambda_{4} + 2\lambda_{1})
\frac{v^{2}}{4} + (\lambda_3 + 2 \lambda_2) \frac{v_{\Delta}^2}{2}\right)}
\end{eqnarray} 
Note that, for $\mu=0$ or the electroweak vev $v=0$, the vev of $\Delta$ will also vanish. Hence, the vev $v_{\Delta}$ depends heavily on the electroweak vev.   We expand the fields $\phi_h$ and $\Delta$ as follows, 
\begin{eqnarray}
\phi_{h}=
\begin{pmatrix}
\phi^{+} \\
\dfrac{v+H +  i\, \xi}{\sqrt{2}}
\end{pmatrix}
\,\,\,\,\,\,\,\,\,\,\,\,
\Delta=
\begin{pmatrix}
\frac{\Delta^{0} + v_{\Delta}}{2}  & \frac{\Delta^{+}}{\sqrt{2}} \\
\frac{\Delta^{-}}{\sqrt{2}} & -\frac{\Delta^{0} + v_{\Delta}}{2} 
\end{pmatrix}\,\,.
\label{phih}
\end{eqnarray}

\subsection{  Neutral and charged scalar masses and mixings} After symmetry breaking, the  $2 \times 2$ mass matrix for the CP even neutral Higgs
in the basis $H$ and $\Delta^{0}$ has the following form,
\begin{eqnarray}
M^2_{s}
= \frac{1}{2}
\begin{pmatrix}
\lambda_h\, v^{2} & v\, v_{\Delta} (2 \lambda_{1}  + \lambda_{4}) - 2\, \mu\, v \\
v\, v_{\Delta} (2 \lambda_{1}  + \lambda_{4}) - 2\,\mu\, v &
2\, v_{\Delta}^{2} (\lambda_{3} + 2 \lambda_{2}) + \frac{\mu\, v^{2}}
{v_{\Delta}}
\end{pmatrix}
\end{eqnarray}

The above mass matrix can be diagonalised by a unitary matrix $\mathcal{U}$ and the mass basis fields $H_{1}$, $H_2$ 
are related to the fields  $H$, $\Delta^{0}$ in the following way,
\begin{eqnarray}
H_{1} &=& \cos \alpha\, H + \sin \alpha\, \Delta^{0} \nn \\ 
H_{2} &=& -\sin \alpha\, H + \cos \alpha\, \Delta^{0}
\label{scalardiag}
\end{eqnarray}
where $\alpha$ is the mixing angle between  the neutral scalar fields.  We denote the  masses  of $H_{1,2}$ by $M_{H_{1,2}}$, respectively,

\begin{eqnarray}
M_{H_1}^{2}&=& 2 \lambda_{h} v^{2} + \tan \alpha (v v_{\Delta} (\lambda_{4} + 2\,\lambda_{1}) - v\mu )\, , \nn \\
M_{H_2}^{2}&=& 2 \lambda_{h} v^{2} - \cot \alpha (v v_{\Delta} (\lambda_{4} + 2\,\lambda_{1}) - v\mu )\, , \nn \\
\label{scalar_mass}
\end{eqnarray}

The fields  $\xi$ of the SM Higgs doublet  is ``eaten'' by the SM gauge boson $Z$ and the gauge boson acquires mass. 
The charged scalars $\phi^+$ and $\Delta^+$ also mix 
and one  becomes the Goldstone boson ``eaten'' by 
$W^{\pm}$.   The
physical charged scalar fields are related to the fields $\phi^{\pm}$ and $\Delta^{\pm}$ 
in the following way,
\begin{eqnarray}
G^{\pm} &=& \cos \delta \, \phi^{\pm} + \sin \delta \, \Delta^{\pm} \nn \\ 
H^{\pm} &=& -\sin \delta \, \phi^{\pm} + \cos \delta\, \Delta^{\pm}
\end{eqnarray}
where $\delta$ is the mixing angle and depends on the
ratio of the vevs of  the doublet and
triplet, 
\begin{eqnarray}
\tan \delta = \frac{2\,v_{\Delta}}{v}
\,.
\label{eq:vdelta}
\end{eqnarray}
In the above $G^{\pm}$ is the charged Goldstone and $H^{\pm}$ is the physical charged scalar field with mass $M_{H^{\pm}}$,

\begin{eqnarray}
M_{H^{\pm}}^{2}&=& \frac{\mu v}{\sin \delta \cos \delta}.
\label{chargedscalar_mass}
\end{eqnarray}
 
In the following we will use the masses, mixing angles and vevs as fundamental parameters, the 
 quadratic and quartic couplings can be expressed in terms of these as,  
\begin{eqnarray}
\mu &=& \frac{M_{H^{\pm}}^{2}
	\sin \delta \cos \delta}{v}\,,\nn \\
\lambda_{3} + 2\,\lambda_{2} &=& \dfrac{ M_{H_{1}}^{2} + M_{H_{2}}^{2} +
	(M_{H_{2}}^{2} - M_{H_{1}}^{2})\cos 2 \alpha - 2\,M_{H^{\pm}}^{2} \cos^{2}\delta}{2\,v_{\Delta}^{2}} \,,\nn \\
\lambda_{h} &=& \dfrac{M_{H_{1}}^{2} + M_{H_{2}}^{2} +
	(M_{H_{1}}^{2} - M_{H_{2}}^{2})\cos 2 \alpha}{v^{2}},\nn\\
\lambda_{4} + 2\,\lambda_{1} &=& \dfrac{(M_{H_{1}}^{2}-M_{H_{2}}^{2})
	\sin 2 \alpha + M^{2}_{H^{\pm}} \sin 2\, \delta
}{v\, v_{\Delta}}\,,\nn\\
\mu_{h}^{2} &=& \lambda_{h}\, \frac{v^{2}}{4} +(\lambda_{4} + 2 \lambda_{1})\,\frac{v_{\Delta}^{2}}{4}
- \mu\, v_{\Delta}\,.
\label{constraints-eq}
\end{eqnarray}

Once the SM Higgs doublet and the triplet Higgs acquired vevs, then the 
W-boson mass and Z-boson mass take the following form,
\begin{eqnarray}
M^2_{W} &=& \frac{g^{2} v^{2}}{4} \left( 1 + \frac{4  v^2_{\Delta}}{v^{2}}\right) \nonumber \\
M^2_{Z} &=& \frac{\left(g^{2} + g^{\prime\,2}\right)\,v^{2}}{4} 
\end{eqnarray}
The extra vev shifts the electroweak precision parameter as follows,
\begin{eqnarray}
\rho_{prec} = 1 + \frac{4 v^2_{\Delta}}{v^{2}}
\end{eqnarray}
Note that the vev of the Higgs triplet is constrained by precision electroweak data. Recent results  from a global fit to the SM lead to 
$\rho_{prec} = 1.00038 \pm 0.00020$ \cite{ParticleDataGroup:2020ssz}. From this, a $3\sigma$ upper bound $v_{\Delta} < 3$ GeV can be derived. 
The improvement over the previous bound \cite{Erler:2004nh, Erler:2008ek} on $\rho_{prec}$ is due partly to a more precise determination of $m_t$.
However, if we rescale  $\rho_{prec}$ using  the W-boson mass
 announced by  the CDF-II collaboration \cite{CDF:2022hxs} then the bound on  the vev  is relaxed to  $v_{\Delta} < 6.5$ GeV. In our scans we will take even a wider range to allow for potential BSM effects in the electroweak precision fit, namely we consider   $v_{\Delta} < 12$ GeV, which is obtained using $\rho_{prec}$ of \cite{Erler:2004nh, Erler:2008ek} while neglecting the scalar loop contributions\cite{Chen:2008jg}.
 Note however that the DM analysis that we perform is mostly independent of the choice of $v_{\Delta}$.


\subsection{ Dark Matter Mass } 
The two neutral fermionic states $\rho^0_3$ and $N^{\prime}$ mix  and the mixing term is proportional to $v_{\Delta}$. 
The  mass matrix for the neutral fermions has the  following form,

\begin{eqnarray}
M_{F}
= 
\begin{pmatrix}
M_{\rho_3} & \frac{Y_{\rho\Delta} v_{\Delta}}{2} \\
\frac{Y_{\rho\Delta} v_{\Delta}}{2} & M_{N^{\prime}} 
\end{pmatrix}
\,.
\label{fermion-mass}
\end{eqnarray} 
and  the mass eigenstates and weak eigenstates are related as follows:
\begin{eqnarray}
\rho &=& \cos\beta\, \rho^{0}_{3} + \sin\beta\, N^{\prime c} \nn \\ 
{N} &=& -\sin\beta\, \rho^{0}_{3} + \cos\beta\, N^{\prime c}\,.
\end{eqnarray}
Diagonalising  Eq.\,(\ref{fermion-mass}),  the tree level mass eigenstates can be expressed  as,
\begin{eqnarray}
M_{{N}} &=& \frac{1}{2} \left(M_{\rho_3} + M_{N^{\prime}} - \sqrt{(M_{\rho_3} - M_{N^{\prime}})^{2} +
4\left(\frac{Y_{\rho\Delta} v_{\Delta}}{2}\right)^{2}} \right)\,, \nn \\
M_{{\rho}} &=& \frac{1}{2} \left(M_{\rho_3} + M_{N^{\prime}} + \sqrt{(M_{\rho_3} - M_{N^{\prime}})^{2} +
4\left(\frac{Y_{\rho\Delta} v_{\Delta}}{2}\right)^{2}} \right)\,, \nn \\
\label{eq:fermionmass}
\end{eqnarray}
where the mixing angle is 
\begin{eqnarray}
\tan 2 \beta &=& \frac{Y_{\rho \Delta} v_{\Delta}}{M_{\rho_3} - M_{N^{\prime}}}.
\label{eq:mixing}
\end{eqnarray}
The Yukawa coupling 
$Y_{\rho \Delta}$ can further be expressed  in terms of $M_{\rho}$ and $M_{N}$, 
\begin{eqnarray}
Y_{\rho \Delta} &=& \frac{\Delta M_{\rho N} \sin2\beta}{2 v_{\Delta}}
\label{mix-coup}
\end{eqnarray}
where $\Delta M_{\rho N} = (M_{{\rho}} - M_{{N}})$.
In Eq.\,(\ref{mix-lag}), $Y_{\rho \Delta}$ is the Yukawa term which relates the fermionic triplet
with the fermionic singlet. 
Here we are exploring the  regime where the $Y_{\rho \Delta}$ coupling 
is feeble , $Y_{\rho \Delta} \sim \mathcal{O}(10^{-10})$. Hence, the mixing angle $\beta$ is heavily suppressed, and the masses of $M_{N}$ and $M_{{\rho}}$ simplify to 
 \begin{eqnarray}
M_{{N}} \sim  M_{N^{\prime}}, ~~
M_{{\rho}} \sim M_{\rho_3}  
\label{eq:fermionmasssimple}
\end{eqnarray}
The state $N$ and ${\rho}$  
 therefore primarily consist of  the singlet and triplet  fermions, respectively with a very tiny admixture. 
In this paper, we consider both scenarios, where either $\rho$ or $N$ is the DM, with a basic difference that $\rho$ is a WIMP DM and $N$ is a FIMP DM particle.

\subsection{ Neutrino Mass} 
The state $\rho$ is the DM candidate and does not participate in neutrino mass generation. However, the other two states $\rho^0_{1,2}$ mix with active neutrinos and generate light neutrino masses via the seesaw mechanism.  Since only two triplet fermions participate in neutrino mass generation, hence, the  Dirac mass term has  the following form, 
\begin{eqnarray}
m_{D} =
\begin{pmatrix}
\frac{\lambda_{11} v}{\sqrt{2}} &  \frac{\lambda_{12} v}{\sqrt{2}} \\
\frac{\lambda_{21} v}{\sqrt{2}} &  \frac{\lambda_{22} v}{\sqrt{2}} \\
\frac{\lambda_{31} v}{\sqrt{2}} &  \frac{\lambda_{32} v}{\sqrt{2}}
\end{pmatrix}
\end{eqnarray}
We consider a basis of $\rho^0_{1,2}$ where  the mass matrix of neutral triplet fermion is diagonal, 
\begin{eqnarray}
\tilde{M}_{\rho} = 
\begin{pmatrix}
M_{\rho_1} & 0 \\
0 & M_{\rho_2}
\end{pmatrix}
\end{eqnarray} 
The  light neutrino mass has the following expression, 
\begin{eqnarray}
m_{\nu} \sim  - m_{D} \tilde{M}^{-1}_{\rho} m^{T}_{D}\,,
\end{eqnarray}
where $m_{\nu}$ represents the  $3\times 3$ mass matrix for active
light neutrinos, $\nu_i$. We denote the physical masses of $\rho^0_{1,2}$ by $M^{d}_{\rho}$, which 
has the following form, 
\begin{eqnarray}
M^{d}_{\rho} = \tilde{M}_{\rho}. 
\end{eqnarray}

Note that 
if we consider all the elements of $m_{D}$ and $\tilde{M}_{\rho}$ matrices as real 
then we have eight independent variables. The 
constraints from measurements of the oscillation parameters impose five conditions on the neutrino masses and mixings 
namely three mixing angles $\theta_{12}$, $\theta_{23}$, $\theta_{13}$
and two mass square differences $m^2_{12}$, $|m^2_{13}|$. Moreover, there is a bound
the sum of the light neutrino masses from cosmology,  $\sum_{i} m_{\nu_i} = 0.23$ eV \cite{Ade:2015xua}. Since  the neutrino sector contains eight variables and six constraints, the latter can easily be satisfied.  
The mass of the triplet fermion $\tilde{M}_{\rho}$ is constrained from LHC searches as $M_{\rho} > 900$ GeV \cite{ATLAS:2022yhd}. This LHC search is based on the multi lepton searches from the decay of $\rho_{1,2}$ produced via electroweak gauge bosons. When the collider bound is satisifed it is straightforward  to obtain $\textrm{eV}$ scale neutrino mass, for example  for $M_{\rho} = 1$ TeV one needs $m_{D} \sim 10^{-4}$ GeV. We refrain from further
detailed analysis on the neutrino mass as this does not have any effect on the DM phenomenology that we study in the following.  
The free parameters of the model relevant to the dark sector and in particular for obtaining the DM relic density are taken as,
\begin{eqnarray}
M_\rho\ ,M_N\ , M_{H_2}\ , Y_{\rho\Delta}\ , \sin \alpha \ , \sin \delta
\end{eqnarray}

\section{Constraints on the  scalar sector }\label{boundscal}

In this section, we briefly discuss the  theoretical  constraints which are relevant 
in our analysis of the scalar sector. Specifically, we  consider the bounds which arise  in order to
keep the quartic couplings of the scalar potential  in the perturbative regime. The choice of quartic couplings, scalar masses and mixings heavily influence our  subsequent analysis of DM production, since in this work DM production significantly  depends on the choice of BSM Higgs mass. 

\begin{figure}[]
	\includegraphics[angle=0,height=7.5cm,width=7.5cm]{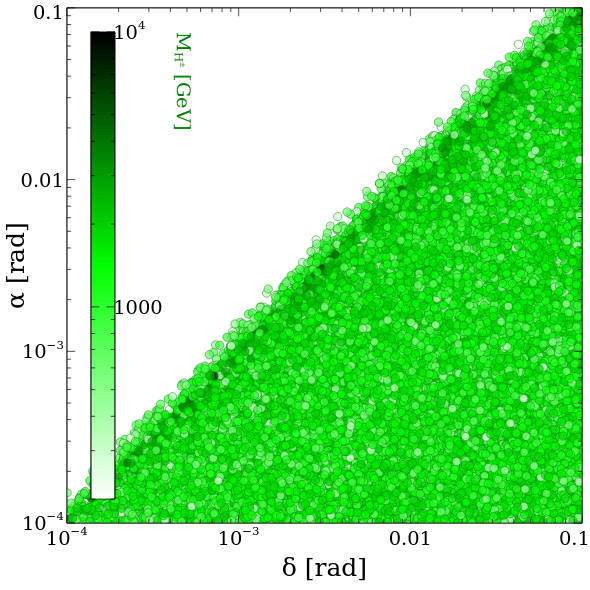}
		\includegraphics[angle=0,height=7.5cm,width=7.5cm]{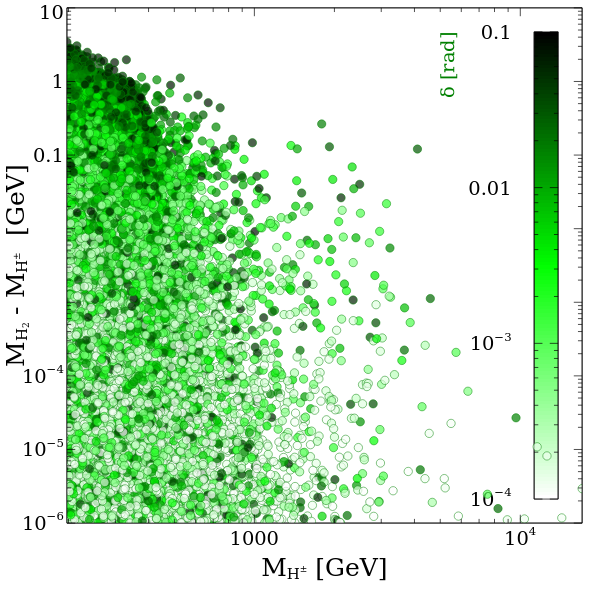}
    \caption{Left: Scatter plot in the $\alpha-\delta$ plane after imposing the
    perturbative limit on the quartic couplings
    $\lambda_i \leq 4 \pi$,  the trivial bound from below, $\lambda_i>0$,   and spontaneous and induced vevs conditions
    $\mu^2_{h}, \mu^2_{\Delta} > 0$, here $M_{H^+}=M_{H_2}$. Right: Scatter plot in the $(M_{H_2}-M_{H^+})$ -- $M_{H^+}$ plane after imposing the same constraints. The color bars indicate the  variation of $M_{H^{\pm}}$ and $\delta$ in the left and right panel, respectively. }
	\label{unitarity-plot}
\end{figure}
As discussed before, $\alpha$ and $\delta$ are the two mixing angles in the scalar sector, $\alpha$ being the neutral BSM Higgs mixing and $\delta$ is the charged Higgs mixing angle.  In Fig.\,(\ref{unitarity-plot}), we 
show scatter plots in  the
$\delta - \alpha$ and $M_{H^{\pm}} - ( M_{H_2} - M_{H^{\pm}})$ planes
after demanding the quartic couplings to be  in the perturbative regime $\lambda_i \leq 4 \pi$  and positive $\lambda_i>0$.
From the left panel(LP), where we consider random values for the masses of the BSM Higgs states {\it i.e.,} $M_{H_2}$, $M_{H^{\pm}}$,  it is evident that  there is a sharp correlation between $\alpha$
and $\delta$ and that they can not be chosen arbitrarily. Throughout our analysis, we consider $\alpha=\delta$ to be consistent with the perturbative constraint.  In the right panel(RP), we show  the scatter plot in  the 
$M_{H^{\pm}} - ( M_{H_2} - M_{H^{\pm}})$ plane where the variation in the 
color bar represents  the charged Higgs mixing angle $\delta$. We
can see that for higher values of $ M_{H^{\pm}}$ a large  
mass gap $M_{H^{\pm}} - ( M_{H_2} - M_{H^{\pm}})$ is disallowed in order
to keep the quartic coupling $\lambda_i$ in the perturbative regime.
In our analysis we have considered these bounds on the masses of $H_2$
and $H^{\pm}$, and throughout the paper we set degenerate masses for the charged and neutral Higgs {\it i.e.,} $M_{H^{\pm}}=M_{H_2}$. 
Note that  it would be possible to relax our assumptions and consider   $\delta > \alpha$  for the  lower mass range of $M_{H_2}, M_{H^{\pm}}$, however $\delta$ is not a crucial parameter for DM observables thus it would not affect our conclusions.

There are a number of experimental constraints which are applicable on the neutral and charged BSM Higgs masses $M_{H_2}$ and $M_{H^{\pm}}$, as well as on  the mixing angle  $\alpha$ between CP even neutral Higgs. We discuss these constraints in detail in Section.~\ref{collider1}. We have specifically considered these following searches
\begin{itemize}
\item
Higgs signal strength measurements  from  $\sqrt{s}=13$ TeV  LHC searches \cite{ATLAS:2020qdt}
\item
Higgs to di-photon $H_1 \to \gamma \gamma$ \cite{ATLAS:2018hxb}
\item
BSM Higgs search via $p p \to H_2 \to Z Z$  \cite{ATLAS:2020tlo}, $pp\to H_2 \to W^{+}W^{-}+ ZZ$ \cite{ATLAS:2020fry} 
\end{itemize} 
The choice of model parameters that we consider for the DM analysis is consistent with these experimental searches. 

\section{DM Constraints}\label{boundDM}

In this work, we consider  scenarios where either the WIMP ($\rho$) or the FIMP ($N$) forms DM. Hence, depending on our choice, different constraints apply on the DM and on the next to lightest odd particle (NLOP).
\subsection{DM relic density}

The DM relic density has been determined  precisely by PLANCK measurements of the CMB \cite{Ade:2015xua}, 

\begin{equation}
\Omega h^{2}= 0.1199\pm0.0027.
\label{PLANCK}
\end{equation}

We will in general impose this $3\sigma$ bound from the relic density on our scenarios, one exception is in section~\ref{sec:lightH2} where we allow for DM to be over-abundant hence consider a larger range.  The predictions for the relic density for the different DM production mechanisms  in the $
\nu$STFM will be discussed in detail in Section~\ref{DM1}.

\subsection{Collider constraints on $\rho$}\label{rho_lhc}
\begin{figure}[]
	\centering
	\includegraphics[angle=0,height=8.5cm,width=10.5cm]{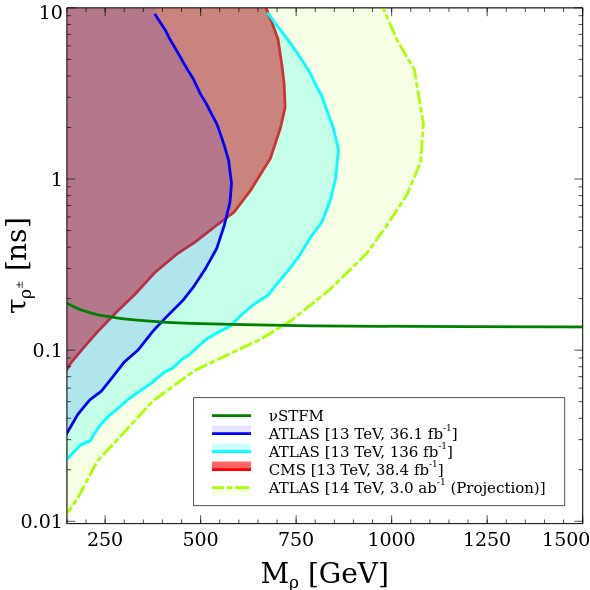}
	\caption{Blue(light) and red(dark) shaded regions show the bound on the 
		DM mass from ATLAS \cite{ATLAS:2017oal,ATLAS:2022rme} 
		and CMS \cite{CMS:2018rea} detectors at LHC
		collider from disappearing charged track, respectively. The light green dot dashed  line represents the projection for HL-LHC \cite{Dainese:2019rgk}.}
	\label{ATLAS-CMS}
\end{figure}

Collider constraints on $\rho$ apply  irrespectively of  our choice of a thermal/non-thermal DM.
The charged component of the triplet, $\rho^\pm$ is nearly degenerate with the neutral component, with a maximum
mass splitting of 167 MeV \cite{Cirelli:2005uq}.  Thus the charged  triplet fermion state is long-lived and constrained from LHC  disappearing track searches.
The charged particle $\rho^{\pm}$ decays to the neutral particle and charged pion 
($\rho^{\pm} \rightarrow \rho\,\pi^{\pm}$) and  the $\pi^{\pm}$ is very difficult  to  reconstruct due to its small 
momentum. 
Therefore, inside the detector, the decay of the charged particle 
 manifest itself as a disappearing track. In Fig.\,\ref{ATLAS-CMS}, we show  the constraints in the plane $M_{\rho^{\pm}}$ - $\tau_{\rho^{\pm}}$ (the $\rho^\pm$ lifetime)  as obtained from  ATLAS \cite{ATLAS:2017oal, ATLAS:2022rme}
and CMS\cite{CMS:2018rea} searches on disappearing tracks. In this figure, the green solid line
represents  the  prediction for the  lifetime  of $\rho^{\pm}$ in the $
\nu$STFM. This figure shows clearly that $M_{\rho^{\pm}} < 580$ GeV is ruled out by ATLAS search with $\mathcal{L}=136$\, $\rm{fb}^{-1}$, which in turn 
means that  the neutral triplet fermion $\rho$ is also ruled out when $M_{\rho} < 579.83 $ GeV.  The light green dotted  line represents the projection for HL-LHC, taken from \cite{Dainese:2019rgk}. 

 \begin{figure}[]
	\includegraphics[angle=0,height=7.5cm,width=7.5cm]{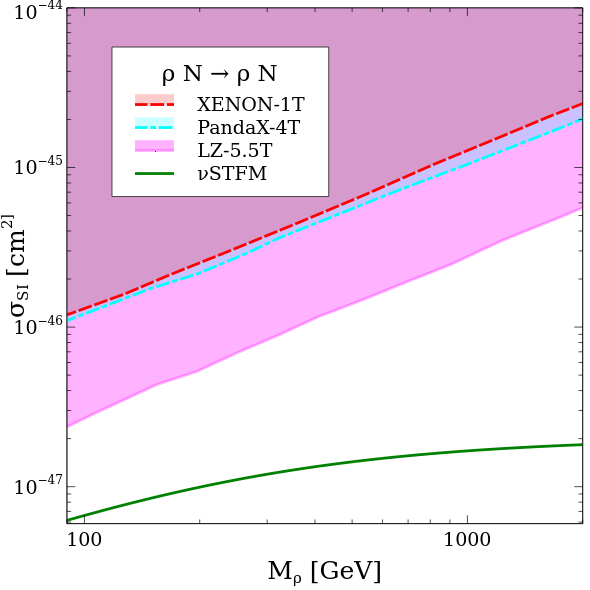}
	\includegraphics[angle=0,height=7.5cm,width=7.5cm]{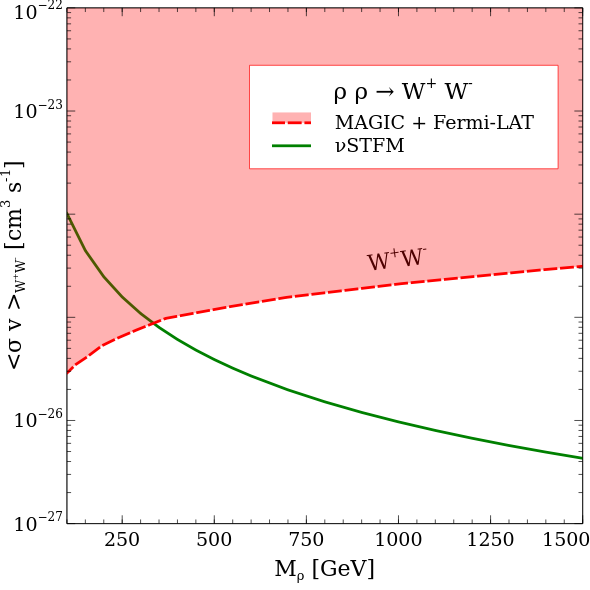}
	\caption{Bounds on the DM
		mass from direct detection search at the Xenon-1T, PandaX-4T and LUX-ZEPLIN (LZ) detectors (left panel)
		\cite{XENON:2018voc,Liu:2022zgu,LUX-ZEPLIN:2022qhg}
		and indirect detection search in $W^{+}W^{-}$ channel 
		by Fermi-LAT (right panel) \cite{MAGIC:2016xys, Reinert:2017aga}. }
	\label{DD-ID-bound}
\end{figure}
\subsection{DM direct and indirect detection}

The direct and indirect detection  constraints are only relevant for WIMP DM, that is when $\rho$ is the DM. For a FIMP DM the suppressed interactions with SM particles  entail that these constraints do not
play a role in our model.\footnote{ Direct detection can constrain FIMPs in the presence of a very light mediator~\cite{Hambye:2018dpi,Belanger:2020npe}, this is not the case here as all mediators are in the range $0.1-10$ TeV.} In Fig.\,\ref{DD-ID-bound}, we show  bounds on the mass of DM, $M_{\rho}$, and its spin independent/annihilation cross-sections that determine the direct detection rate. In the $
\nu$STFM there is no tree-level process for DM elastic scattering on nucleons but the 
process can happen at one loop through the diagram mediated by $W^{\pm}$. The spin-independent cross-section, $\sigma_{SI}$,
for $\rho N \rightarrow \rho N$ (where N is nucleon) direct detection process  is given analytically by \cite{Hisano:2011cs},
\begin{eqnarray}
\sigma_{SI} = \frac{4}{\pi} \mu^2_{r} |f_{N}|^2
\end{eqnarray} 
where $\mu_{r} = \frac{M_{\rho} M_{N}}{M_{\rho} + M_{N}}$ ($M_{N}$ is the
nucleon mass) and $f_{N}$ depends on the DM interaction with the
nucleons and the nuclear form factors which are discussed in detail in \cite{Hisano:2011cs}. In the DD cross section we have also taken into account 
the two loop gluonic which suppress the direct detection cross section as discussed in \cite{Hisano:2011cs}. Note that including only the quark contributions\cite{Cirelli:2005uq}, overestimates the spin independent cross section as the quark and gluon contributions cancel against each other. 
Fig.~\ref{DD-ID-bound} shows that the theoretical prediction for  $\sigma_{SI}$ (green) is lower 
 than the experimental upper limit from  Xenon-1T (red) \cite{XENON:2018voc}, PandaX-4T  (cyan) \cite{Liu:2022zgu} and LUX-ZEPLIN (magenta) 
 \cite{LUX-ZEPLIN:2022qhg}. Therefore,
 in $
\nu$STFM, we do not have any bound on the DM mass from the DD experiments.

 DM can also be detected indirectly by observing gamma-rays originating from DM annihilation in galaxies. In particular, in the $
\nu$STFM 
  DM annihilates  to $W^{+}W^{-}$  via a t-channel process mediated by the charged fermion $\rho^{\pm}$. 
 The predicted cross-section, $\langle \sigma v\rangle_{WW}$ is compared with the upper limit on the same cross-section 
  obtained from analysing data from Dwarf Spheroidal   Galaxies  from  the satellite-based Fermi-LAT and
 earth-based MAGIC collaborations \cite{MAGIC:2016xys, Reinert:2017aga}. In the right panel of In Fig.\,\ref{DD-ID-bound},  we see that this leads to a constraint on the DM mass
 $M_{\rho}>350$ GeV. This is rather weak as compared 
 to the direct detection constraint. 
 In this work to analyse the thermal DM scenario,  we have mostly considered  DM masses in the range of 700 GeV to 1500 GeV which are safe from all kinds of bounds and can be probed by the collider experiment HL-LHC. 
 Note that in the mass range above 1500
 GeV, DM annihilation will get Sommerfeld enhancement \cite{Hisano:2003ec,Hisano:2004ds} and can be 
 in tension with the bound coming from DM indirect detection experiments.

\subsection{BBN constraints}\label{Dm_BBN}
The primordial elements nucleosynthesis occurs approximately between 1 and $10^{3}$ seconds. The long lived particle decaying after 1 sec can inject sufficient energy to the  thermal plasma and perturb the primordial light elements either through hadro-dissociation, $p \leftrightarrow n $ interconversion or photo dissociation. In our analysis, the NLOP can decay to DM after 1 sec depending upon the coupling strength $Y_{\rho\Delta}$. The decay of NLOP also injects hadronic energy to the thermal plasma which disrupts the formation of light elements during BBN. The BBN constraint on the amount of hadronic energy released through late decay of long lived decaying particles were derived in \cite{Kawasaki:2017bqm}. The hadronic energy released through late decay of NLOP is given by,
\begin{eqnarray}\label{hadEvis}
	\zeta_{had}=E_{vis}B_{had}Y_{NLOP},
\end{eqnarray}
where $B_{had}$ is the hadronic branching fraction that is approximately be given by,

\begin{eqnarray}\label{hadBr}
	B_{had}\approx\frac{\sum_{i}\Gamma(NLOP\to DM\  X_{i})Br(X_{i})}{\Gamma_{NLOP}^{total}},\ \ X_{i} \in B/SM
\end{eqnarray}
$ E_{vis}$ is the visible energy released through each NLOP decay,

\begin{eqnarray}
	E_{vis}\approx \frac{M_{NLOP}^{2}-M_{DM}^{2}}{2 M_{NLOP}},
\end{eqnarray}

and $Y_{NLOP}$ is the yield of NLOP before it decays to DM.  The yield is determined through coupled Boltzmann equations  which govern the evolution of DM and NLOP.   

\begin{figure}[]
 	\centering
 	\includegraphics[angle=0,height=7.5cm,width=15.0cm]{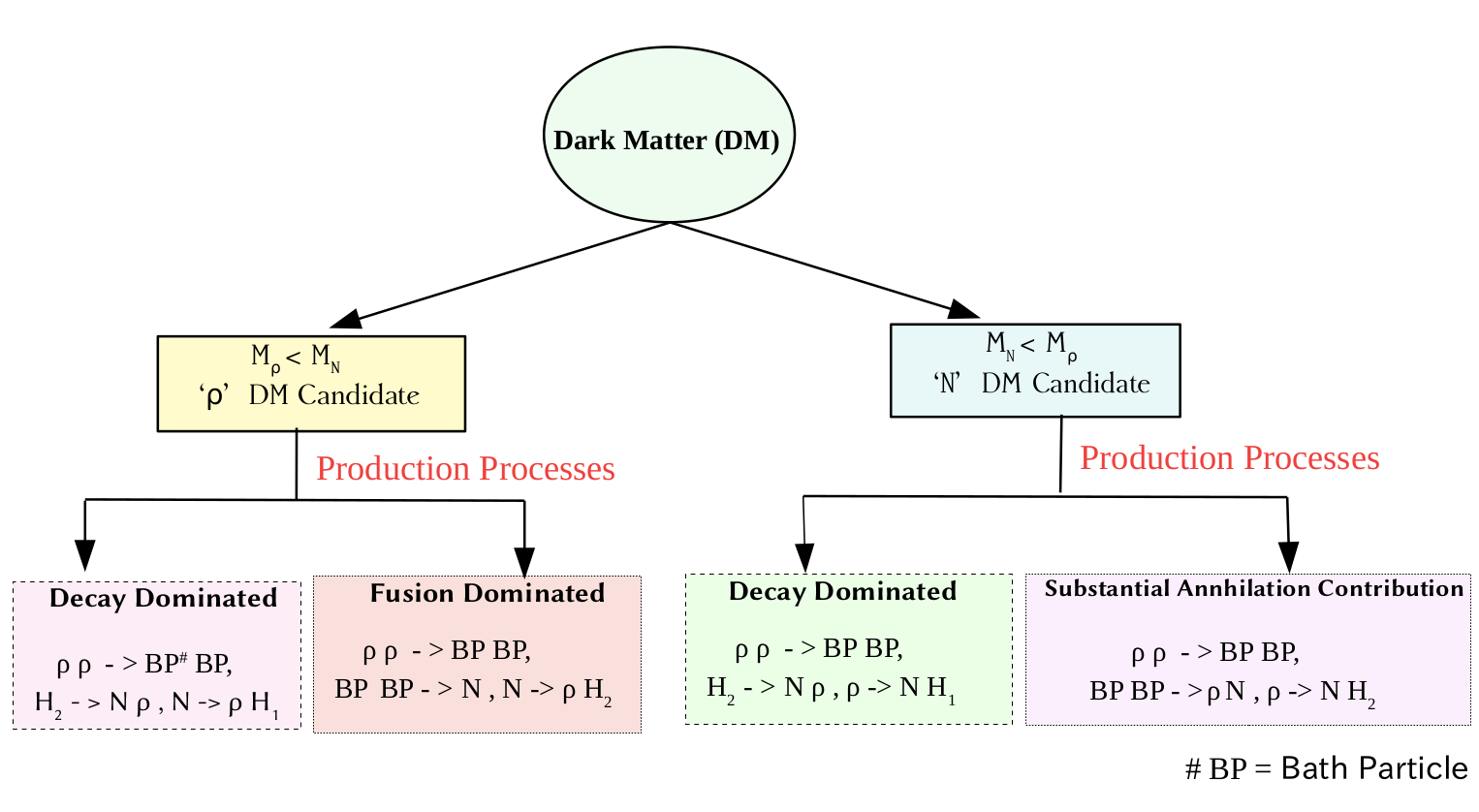}
 	\caption{Schematic diagram representing different scenarios, which we consider in this work.  \label{fig:sche}}
 \end{figure}

\section{DM production}\label{DM1}

The model contains the  thermal fermion triplet $\rho$ and gauge singlet fermion $N$, where either of these can be DM candidate. We  consider the following scenarios: Scenario-I, where $\rho$ is the DM particle and $N$ is the next-to-lightest-odd (NLOP) particle \footnote{Both $\rho$ and $N$ states are odd under $\mathbb{Z}_2$ symmetry.},  Scenario-II  where  $N$ is the DM particle and $\rho$ is the NLOP particle,
 Scenario-III where either $N$ or $\rho$ can be DM but the scalar sector is lighter allowing for  the possibility that  annihilation of bath particles give substantial contributions to DM production. 
 A schematic diagram representing the different DM production possibilities in this model is displayed in Fig.~\ref{fig:sche}. In the first two  scenarios,  DM is primarily produced  from the decay of NLOP states, and Scenario-III corresponds to fusion and annhilation dominated scenarios.
 \begin{figure}[ht!]
	\centering
	\includegraphics[angle=0,height=8.0cm,width=12.5cm]{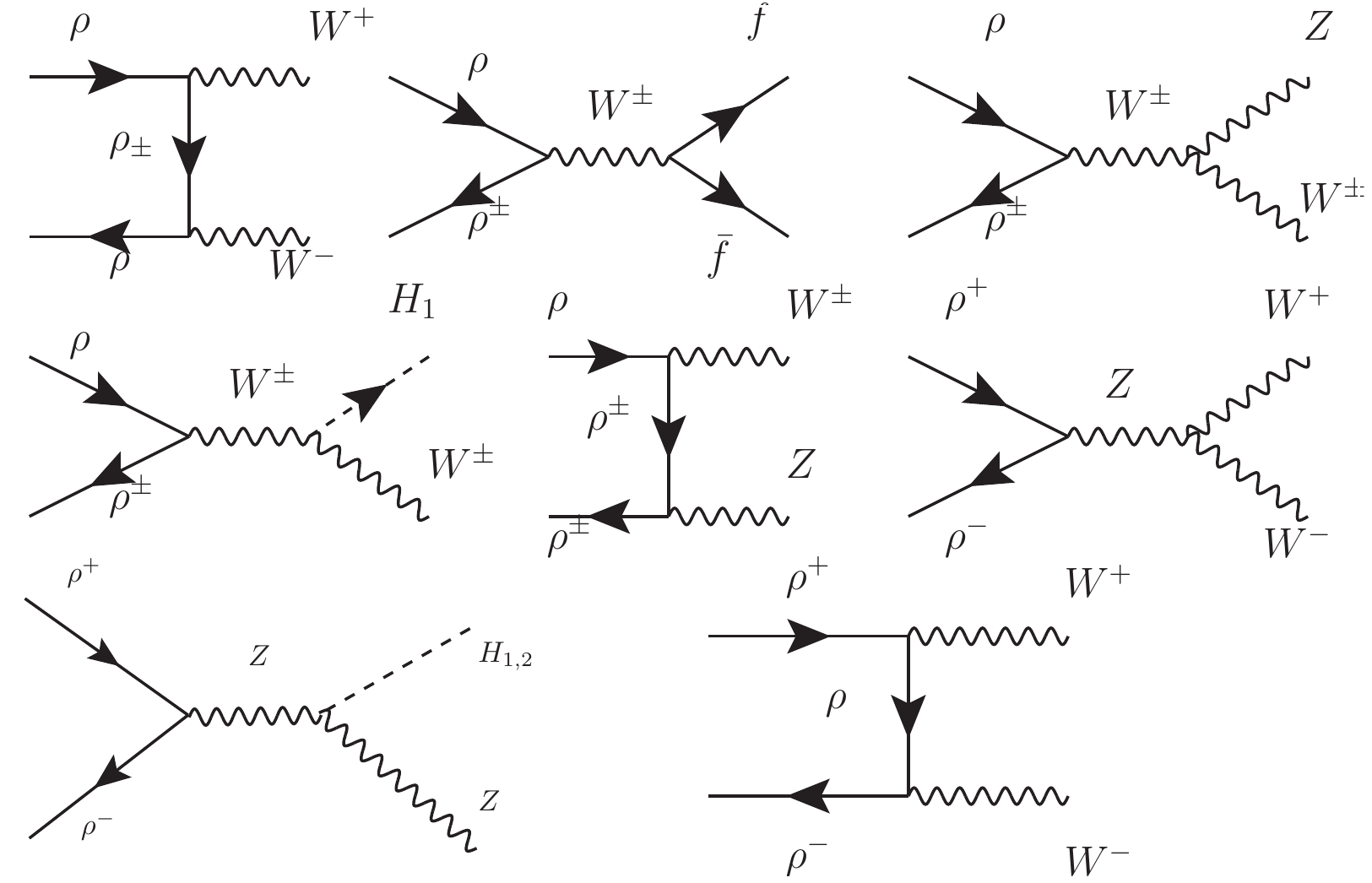}
	\caption{Annhilation and co-annihilation channels of $\rho$ in the early Universe . \label{fdiaTrip}}
\end{figure} 
\subsection{DM production in Scenario I : $M_{N} > M_{\rho}$}\label{S1}

We consider the scenario when  $M_{N} > M_{\rho}$, and $\rho$ is the stable DM. The gauge singlet fermion state $N$ has feeble interactions with other particles
of the model, and hence  $N$ never achieves thermal equilibrium. The DM $\rho$ can be produced through the standard freeze-out mechanism (thermal production) and through the late decay of the FIMP (non-thermal production).
 Due to  abundant  interactions with the gauge bosons, $\rho$ is maintained in thermal equilibrium 
until it freezes-out.  The annihilation and co-annihilation processes which govern the 
thermal production  of $\rho$ include 
$\rho \rho \rightarrow W^{+} W^{-}$,
 $\rho \rho^{\pm} \rightarrow Z W^{\pm}$, $\bar{f} f^{\prime}$,
 $W^{\pm} H_i \,\,(i = 1,\,2)$, 
 $\rho^{+} \rho^{-} \rightarrow Z Z$, $W^{+} W^{-}$, 
 $\bar{f} f$, $W^{+} W^{-}$ and
 $\rho^{\pm} \rho^{\pm} \rightarrow W^{\pm} W^{\pm}$\, see Fig.~\ref{fdiaTrip}.  We provide  
 the expressions for the thermal average cross-section for these processes  in the  Appendix \ref{appendix1}. 
 It has been shown in \cite{Ma:2008cu}, that  through the freeze-out mechanism, the correct DM relic density is satisfied only if $M_{\rho} \sim 2400$ GeV.
As found in \cite{Hisano:2003ec,Hisano:2004ds}, $SU(2)_L$ triplet DM annihilation to W-boson gets zero energy resonance at $M_{\rho} = 2000$ GeV with zero binding energy for the Yukawa type potential. This is known as Sommerfeld enhancement (SE) and happens due to the unsuppressed transition between the two body states of $2 \rho$ and $\rho^{+}\rho^{-}$ for a small mass
difference between charged and neutral components. Therefore,
the $\rho\rho \rightarrow W^{+}W^{-}$ annihilation cross section gets amplified by $\mathcal{O}(10^{3})$  compared to the perturbative estimation of DM annihilation and the enhancement starts from the DM mass $M_{\rho} > 0.5$ TeV. 
 Therefore, such heavy DM  is  ruled out
  from the indirect detection bound since the cross-section benefits from a  
 SE \cite{Hisano:2003ec,Hisano:2004ds}. Moreover, heavy DM of few TeV mass is also beyond 
 the reach of the current collider searches \cite{CMS:2020atg}.  However, when additional non-thermal production for $\rho$ is possible, these restrictions
 can be alleviated. In this case, a relatively light DM  state $\rho$ with mass $M_{\rho} \ll$ 2.4 TeV can be consistent with the DM relic density constraint and direct and indirect measurements.   The thermal contribution for this mass range 
 will only  lead to  under-abundant DM,  however a significant non-thermal contribution will permit to reach  the correct relic density.\ 

The non-thermal  contribution to the DM relic density results   from 
 the decay of the gauge singlet fermion $N$ which is the NLOP. 
The abundance of $N$  is primarily dictated by the decay of the scalar $H_2$, 
 $H_2 \rightarrow \rho\, N$. Note that the process $H_1 \to \rho N$ is kinematically forbidden for our choice of DM and NLOP masses. Since $N$ is not stable it can  eventually decay to
 the DM $\rho$ through the  two body  and three body decay modes $N \rightarrow \rho H_1$,  $N \rightarrow \rho f \bar{f}$, respectively. The latter is sub-dominant  if $N \to \rho H_1$ is open. These contributions are  the late decay contribution of $N$ to the relic density.  
 \subsubsection{Solving for $N$ and $\rho$ abundances}
 Since $N$ itself is a non-thermal particle, in order to compute the late decay contribution of $N$
 to the DM abundance, one first need to compute the distribution function of $N$. 
The general Boltzmann equation is,
\begin{eqnarray}
\hat{L}\,[f_N] = \mathcal{C}\,[f_N]
\end{eqnarray} 
where $\hat{L}$ is the Liouville's operator
\footnote{Liouville operator is,
$\hat{L} = p^{\alpha} \frac{\partial}{\partial x^{\alpha}} -
\Gamma^{\alpha}_{\beta \gamma} p^{\beta} p^{\gamma}
\frac{\partial}{\partial p^{\alpha}}$ where $p^{\alpha}$
is the four momentum and $\Gamma^{\alpha}_{\beta \gamma}$
is Christoffel symbol throung which cosmology enters into the theory} 
which takes the following form,
\begin{eqnarray}
\hat{L} = \frac{\partial}{\partial t} -
H\, p\, \frac{\partial}{\partial p}\,,
\label{liouville}
\end{eqnarray}
$f_N$ is the distribution function of $N$ and implicit function of momentum and temperature and $\mathcal{C}$ 
is the collision function which depends on the interactions among 
the particles. In the Lioville's operator in Eq.\,(\ref{liouville}),
$H$ is the
Hubble parameter and $p$ is magnitude of three momentum. 
By following Ref. \cite{Konig:2016dzg}, we define a new set of 
variables ($\xi_p$,\,$r$) which are related to the old variables 
in the following way,
\begin{eqnarray}
r = \frac{M_{sc}}{T},\,\,\,\, \xi_{p} =
\left(\frac{g_{s} (T_{0})}{g_{s}(T)} \right)^{1/3}
\frac{p}{T}\,,
\end{eqnarray}
where $M_{sc}$ is the mass scale and $g_{s}(T)$ is the entropy d.o.f 
of the Universe at temperature $T$.  In terms of the new variables, the Liouville operator defined in Eq.~(\ref{liouville})
takes the following form, 
\begin{eqnarray}
\hat{L} = r\,H \left( 1 + \frac{T g_{s}^{\prime}}
{3 g_{s}} \right)^{-1}\frac{\partial}{\partial\,r}
\label{eq:L}
\end{eqnarray}
where $\displaystyle g_{s}^{\prime}(T) = \displaystyle \frac{d g_{s}}{d T}$ and we 
have used the time-Temperature relation 
$\displaystyle \frac{dT}{dt}=-H\,T\,\left(1+ \displaystyle \frac{T\,g^{\prime}_s(T)}{3\,g_s(T)}
\right)^{-1}$ in obtaining the above relation.
The Boltzmann equation to determine the distribution function $f_N$ of
$N$ hence can be represented as,
\begin{eqnarray}
\hat{L} f_{N} = 
\mathcal{C}^{H_{2} \rightarrow N \rho}+\mathcal{C}^{AB \to N \rho}
+ \mathcal{C}^{{N} \rightarrow\,\, {all}}\,,
\label{N_prod}
\end{eqnarray}
where the expression of $\hat{L}$ is shown in Eq.\,\ref{eq:L}, $\mathcal{C}^{H_{i} \rightarrow N \rho}$ is the collision term
 for the production of $N$ through the decay of 
$H_1$, $H_2$ and $\mathcal{C}^{{N} \rightarrow\,\, {all}}$
is the collision term for the decay of $N$. We first consider  $H_2$ to be heavy,  $M_{H_2}>M_N >M_{\rho}$, such  that  the production of $N$ 
primarily happens from $H_2 \to \rho N$ decay with subdominant annihilation contributions. In  subsection~\ref{sec:lightH2} we consider a lighter $H_2$ state, where annihilation can also give sizeable contribution in  $N$ production.

\begin{figure}[]
	\centering
	\includegraphics[angle=0,height=3.5cm,width=9.5cm]{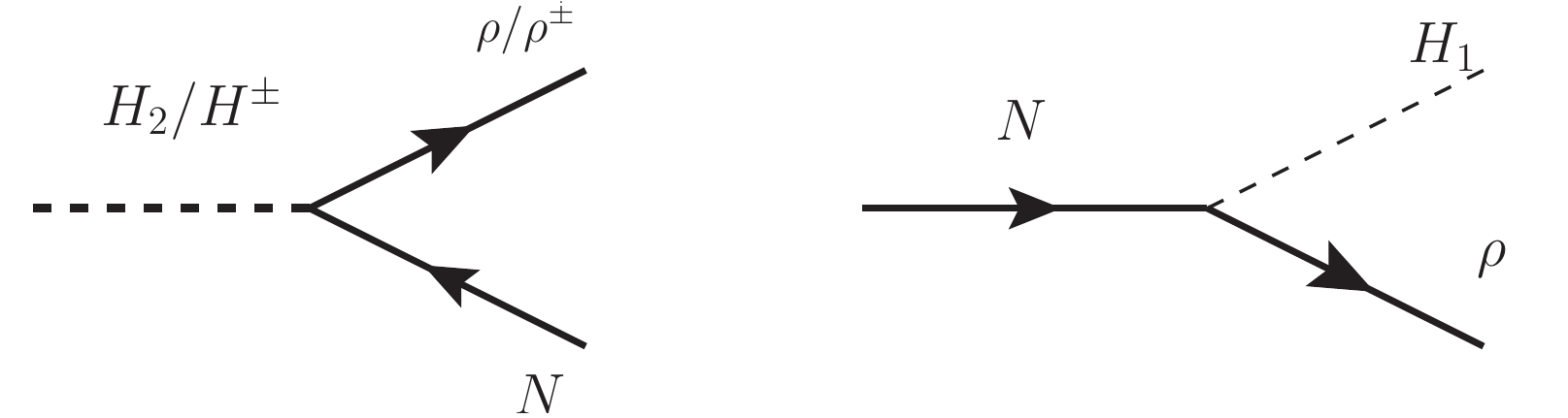}
	\caption{Feynmann Diagram for the dominant production of $N$ as well as its late decay to DM $\rho$. \label{fdiaS1}}
\end{figure}

\subsubsection{DM abundance when $N$ is produced from decays} 
As stated above, for $M_{H_2}>M_N>M_{\rho}$,  $N$ is produced from $H_2 \to N \rho$ decay while $H_1 \to \rho N$ is kinematically forbidden.
The expressions for
$\mathcal{C}^{H_{2} \rightarrow N \rho}$
and $\mathcal{C}^{{N} \rightarrow\,\, {all}}$
are given in Appendix\,\,\,\ref{appendix2}.  
The equation given in Eq.~\ref{N_prod} where we neglect the term $\mathcal{C}^{AB \to N \rho}$ has to be solved numerically.
The number density of $N$ is obtained from the distribution function $f_N$ by the following relation
\begin{eqnarray} 
n_{N}(r) &=& \dfrac{g\, T^3}{2\pi^2} \,
\mathcal{B}(r)^3 \int d\xi_p\,\xi_p^2\, f_{N}(\xi_p,\,r)\,,
\label{eq:masterf}
\end{eqnarray}
where
\begin{eqnarray}
\mathcal{B}(r) = \left(\frac{g_{s}
(T_{0})}{g_{s}(T)} \right)^{1/3} = \left(\frac{g_{s} (M_{sc}/r)}
{g_{s}(M_{sc}/r_0)} \right)^{1/3}\,.
\label{convfact}
\end{eqnarray}
Here $T_0$ is the initial temperature where we assume that the density 
of $N$ is zero and we have chosen $M_{sc} = M_{\rho}$ which is the DM
 mass. Finally, the  co-moving number density  of $N$ can be determined as 
\begin{eqnarray}
Y_{N} = \frac{n_{N}}{s} \,,
\end{eqnarray}
where  $s$ is the entropy density, \cite{Kolb:1990vq},
\begin{eqnarray}
s &=& \dfrac{2\pi^2}{45} \, g_s(T)\,T^3\,.
\label{eq:entropy}
\end{eqnarray}

Finally, to determine the co-moving number density
of DM $\rho$ one  needs to solve the following 
Boltzmann equation,
\begin{eqnarray}
\frac{d Y_{\rho}}{d r}&&= - \sqrt{\frac{\pi}{45 G}}\,\frac{M_{Pl} 
\sqrt{g_{*}(r)}}{r^{2}} \langle \sigma_{eff} |v| 
\rangle \left( Y^2_{\rho} - (Y^{eq}_{\rho})^2 \right)   \nonumber \\
&&+\frac{M_{Pl}\, r\, \sqrt{g_{\star} (r)}}
{1.66\, M_{sc}^{2}\, g_{s}(r)}\left[ \langle \Gamma_{H_2 \rightarrow N\rho}\rangle
(Y_{H_2} - Y_{N} Y_{\rho}) 
+  \langle 
\Gamma_{N \rightarrow \rho A} \rangle_{NTH}\,
(Y_{N} - Y_{\rho} Y_{A}) \right]\,\nonumber\\
\label{be-region-1}
\end{eqnarray}  
where $M_{pl}$ is the Planck mass, $g_{\star}(r) =
\frac{g_s(r)}{\sqrt{g_{\rho}(r)}}\left(1-\frac{1}{3}
\frac{d\,\ln g_s(r)}{d\ln r}\right)$ is
a function of matter ($g_{\rho}(r)$) and entropy ( $g_{s}(r)$) degrees of freedom (d.o.f)  and 
$\langle \sigma_{eff} |v| \rangle$
is the thermally average effective cross section times velocity, whose  
expression is given in Appendix \ref{appendix1}. 
The first term in the above equation represents the annihilation
of $\rho$ through the freeze out mechanism, the second term
is the production of $\rho$ from the decay of  $H_2$
and the third term is the production of $\rho$ from the decay of
$N$ through the two and/or three body decay processes $N \to \rho H_1/\rho f \bar{f}$. In the last term "A" collectively represents 
either a $H_1$ or $f\bar{f}$. Since the scalar field $H_2$ is in thermal equilibrium with the bath,  we 
can use the following relation for the thermal average of the width of 
 $H_2$~\cite{Gondolo:1990dk}, 
\begin{eqnarray}
\langle \Gamma_{H_{2} \rightarrow N\, \rho} \rangle = 
\Gamma_{H_{2} \rightarrow N\, \rho}\,
\frac{K_{1}\left(r\,\frac{M_{H_2}}{M_{sc}}\right)}
{K_{2}\left(r\,\frac{M_{H_2}}{M_{sc}}\right)}\,,
\label{hdk}
\end{eqnarray}
where $K_{1} (x)$, $K_{2} (x)$ are the Modified Bessel functions of 
first and second kind.

Note that, as the other neutral component $N$ has never reached thermal
equilibrium 
we re-iterate that we need to find its distribution function using Eq.~\ref{N_prod}.
The  thermal decay width of $N$ is determined from,
\begin{eqnarray}
\langle \Gamma_{N \rightarrow \rho A}
\rangle_{NTH} = M_{N} 
\Gamma_{{N} \rightarrow \rho A}
\frac{\int \frac{f_{N}(p)}{\sqrt{p^{2}
+M_{N}^{2}}} d^{3} p}
{\int f_{N}(p) d^{3} p}\,.
\label{Ndk}
\end{eqnarray}
The relevant decay width $\Gamma_{N \rightarrow \rho A}$ is  given in
the Appendix \,\ref{appendix3}. 
Finally, after solving the Boltzmann equation  in Eq.\,\ref{N_prod} we find the co-moving 
number density $Y_N$ for $N$.  The
DM relic density can be obtained 
by using the following relation~\cite{Edsjo:1997bg},
\begin{eqnarray} \label{obsDM}
\Omega_{DM}h^2 &=& 2.755\times 10^8
\bigg(\dfrac{M_{\rho}}{\rm GeV}\bigg)
\,Y_{\rho}(T_{\rm Now}). 
\label{relic}
\end{eqnarray}
\begin{figure}[]
	\includegraphics[angle=0,height=6.00cm,width=7.50cm]{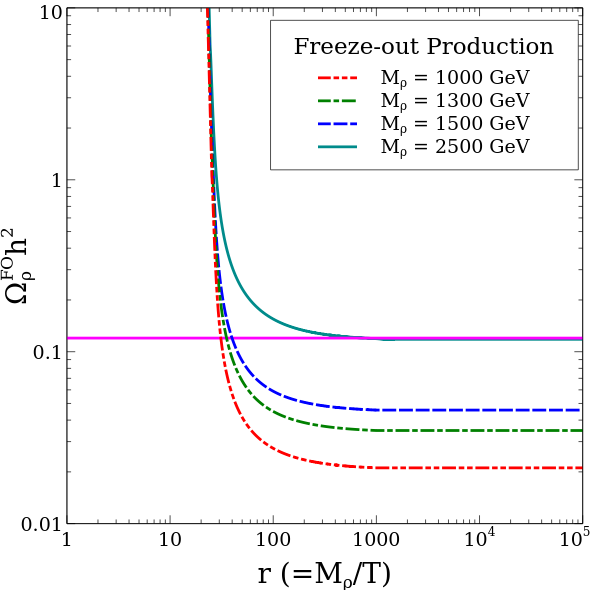}
	\includegraphics[angle=0,height=6.00cm,width=7.50cm]{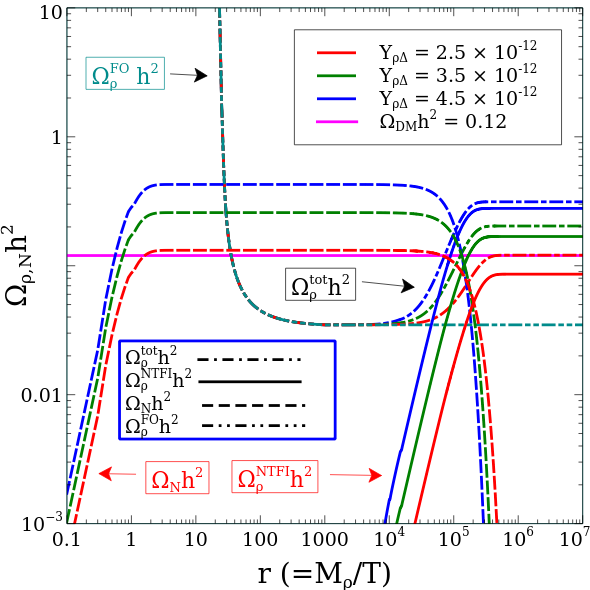}\\
	\center{\includegraphics[angle=0,height=6.10cm,width=7.50cm]{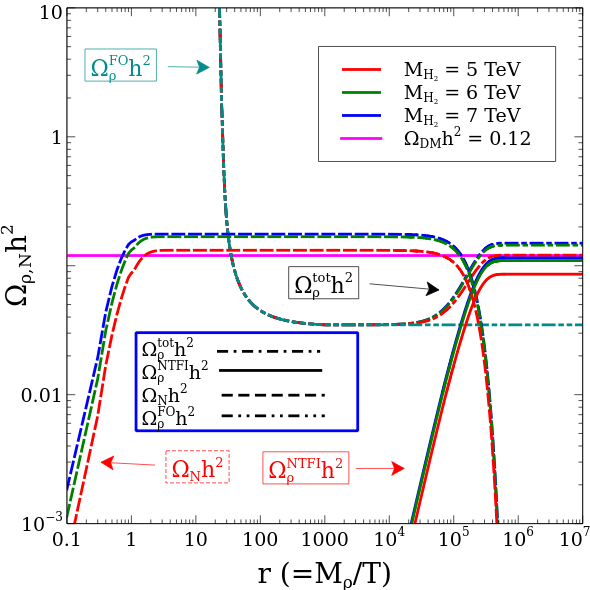}}
	\caption{Top left panel: DM production by only freeze out mechanism for few different values of $M_{\rho}$. Top right panel: DM production from  freeze-out, freeze-in 
		and late decay of $N$ for few different values of 
		$Y_{\rho \Delta}$. Blue dashed dot line is for 
		$Y_{\rho\Delta} = 4.5 \times 10^{-12}$ and the other dashed dot
		lines with different colors are for the descending values of $Y_{\rho\Delta}$. In the same plot, we also show the abundance of NLOP $N$ denoted as $\Omega_N h^2$. Lower panel: the same for few different values of heavy Higgs mass $M_{H_2}$. In generating the plots we have fixed
		the model parameters to the following values unless it is
		varied, $M_{N} = 2000$ GeV, $M_{\rho} = 1300$ GeV, 
		$Y_{\rho \Delta} = 2.5\times10^{-12}$. The  magenta 
		line represents  the reference value of $\Omega_{\rho} h^{2} = 0.12$. In the plots, $\Omega^{tot}_{\rho}h^{2}$ is the total
		contribution represented by the dashed dot line,
		$\Omega^{NTFI}_{\rho}h^{2}$ is the freeze-in 
		contribution when the decaying particle is non-thermal
		represented by solid line, 
		$\Omega_{N}h^{2}$ is the density of $N$ represented by 
		dashed line and $\Omega^{FO}_{\rho}h^{2}$ is the freeze-out
		contribution represented by dashed dot dot line.}   
		\label{diff-production-mode}
\end{figure}
\subsubsection{Results}
We first consider few illustrative benchmark points and analyse the DM and NLOP production. In section~\ref{scenario1:scan} we explore a wide range of parameters.
In this scenario there are two main production mechanisms for DM:  
the thermal production denoted as  $\Omega^{FO}_{\rho} h^{2}$ which takes into account production by the freeze-out mechanism and the  non-thermal freeze-in production referred as $\Omega^{NTFI}_{\rho} h^{2}$ which primarily takes into account the decay contribution from $N \to \rho H_1$.  Here the mother particle $N$ never reaches 
thermal equilibrium\footnote{ The freeze-in production of DM, for instance from $H_2 \to \rho N$ decay,  has no effect on the thermal DM production since it occurs around  $r \sim 1$ when the DM is in thermal equilibrium with the  bath and has a very large number density.}.

Fig.\,\ref{diff-production-mode} illustrates the evolution of the abundances for different parameter choices.
The top left panel (LP)   shows that  DM production 
 via  the freeze-out mechanism increases with the DM mass. Indeed
the annihilation of DM decreases with the increase of its mass. This  implies that 
DM with higher masses will decouple from the thermal 
bath at an early epoch following the condition  $\frac{\langle \sigma v \rangle}{H} < 1$ and an early de-coupling results in a large co-moving number density. 

 In the top right panel (RP), in addition to the thermal production (dashed dot dot line), we also  show DM production via the freeze-in mechanism
which includes  both contributions from $N \to \rho H_1$ and $H_2 \to \rho N$ processes. Although the later process has no effect in determinig the DM relic abundance. For the former process, which is dominant, the mother particle is  
out of equilibrium. The freeze-in production of $N$ denoted as $\Omega^{NTFI}_{N} h^{2}$ strongly depends on the value of the Yukawa coupling 
$Y_{\rho \Delta}$.
As can be seen very clearly, the amount of $N$ produced by the freeze-in mechanism increases with 
$Y_{\rho\Delta}$ since the production rate of $N$ is directly proportional to $Y_{\rho \Delta}$. Subsequently the yield of $\rho$  increases through the late decay of $N$ even though  
 the yield of $\rho$ is independent of $Y_{\rho \Delta}$ because of the feeble coupling. 
 The total production of DM after taking into account both the production processes are shown by the dashed dot lines. 
 In addition to the DM relic density, Fig.\,\ref{diff-production-mode}  also shows the abundance of $N$, referred as $\Omega_{N} h^{2}$. 
The obtained abundance closely follows the analytical expression, relevant for the 
 generic process $A \rightarrow N\, C$,
\begin{eqnarray}
\Omega^{analytical}_{N} h^{2} \simeq \frac{1.09 \times 10^{27}}
{g^{s}_{*}\sqrt{g^{\rho}_{*}}} M_{N} \frac{g_{A} \Gamma_{A}}{M^2_{A}}
\label{eq:omegadecay}
\end{eqnarray}   
where $M_{A}$, $\Gamma_{A}$ and $g_{A}$ are the mass, decay width and $d.o.f$
of $A$. In the present scenario,  the process $H_2 \rightarrow N \rho$
and $H^{\pm} \rightarrow N \rho^{\pm}$ both contribute to the 
 production of $N$. 
We assume that initial abundance of $N$ is equal to zero {\it i.e.} 
$Y^{ini}_{N} = 0$.
Note that as  $N$ completely decays to DM $\rho$  by pre-dominantly two body decay
process,  the produced DM abundance closely follows the following equation, 
$\Omega^{NTFI}_{\rho} h^{2} \simeq \frac{M_{\rho}}{M_{N}} \Omega_{N} h^{2}$.

Note that we chose  masses of $N$ and $\rho$ which satisfy $M_{N} > M_{\rho} + M_{H_1}$ 
and $M_{N} < M_{\rho} + M_{H_2}$. As a result, 
$N$ decays by two body process {\it i.e.} $N \to \rho H_1$, while $N \to \rho H_2$ is forbidden . 
In the case   $M_{N} < M_{\rho} + M_{H_1}$, $N$ will decay through three
body decay processes  with a longer lifetime, if the decay occurs after $T \sim 1$ MeV it might  alter the BBN predictions as will be seen in the next subsection.   
There is also a possibility of the production of  $\rho^{\pm}$ by the three body decay of $N$, this decay is suppressed by the mediator mass, hence $\rho^{\pm}$  will be 
produced in negligible amount. Even though the produced $\rho^{\pm}$ can be long-lived since it decays primarily to DM and charged pion ($\pi^{\pm}$),
 this contribution is negligible and is safe from BBN constraints on the abundance of light elements \cite{Kawasaki:2017bqm,Jedamzik:2006xz}. 
 
Next we discuss  the dependence  of the DM relic density on the remaining model parameters.
In the  lower middle panel (LMP) of Fig.\,\ref{diff-production-mode}, we show the
variation of $\Omega_\rho h^{2}$ and $\Omega_N h^{2}$ with $r\equiv M_\rho/T$ for three different values of the BSM 
Higgs mass  $ M_{H_2} = 5$ TeV, $6$ TeV and $7$ TeV.  $\Omega_N h^{2}$ 
 is inversely proportional to the $H_2$ mass if $M_{H_2}$ is sufficiently large. 
 However,  there is a  phase space suppression in the    $H_2\rightarrow N\rho$  channel and
altogether the relic density is proportional to 
 $\Omega_N h^{2} \propto \frac{1}{ M_{H_2}}\,\left(1 - \frac{(M_{N} + M_{\rho})^{2}}{M^{2}_{H_2}}\right)^{\frac{3}{2}}$.
Thus for $ M_{H_2} = 5$ TeV, which corresponds to the lowest value  among the three benchmarks, the phase space suppression is largest leading to a suppressed production of $N$ and thus of $\Omega^{NTFI}_\rho h^{2}$. We find that the relic density nearly coincides  for $M_{H_2}= 6\  {\rm and}\  7$ TeV, indeed  the larger phase space suppression factor at 6 TeV is compensated by the mass term in the denominator.       

\begin{figure}[]

	\includegraphics[angle=0,height=7.5cm,width=7.5cm]{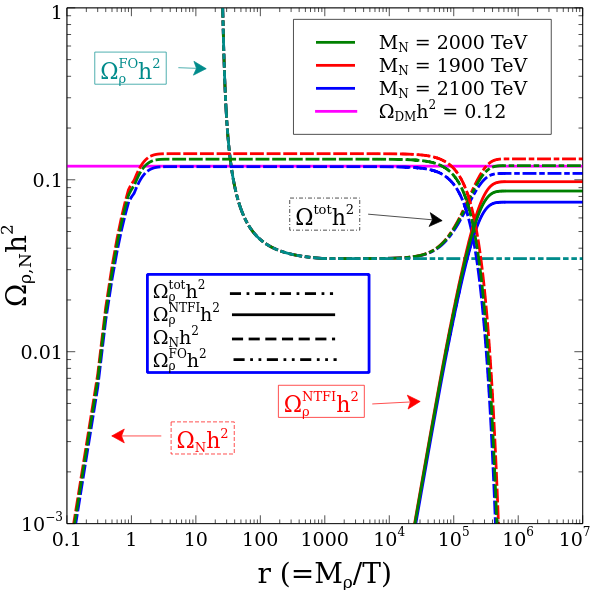}
		\includegraphics[angle=0,height=7.5cm,width=7.5cm]{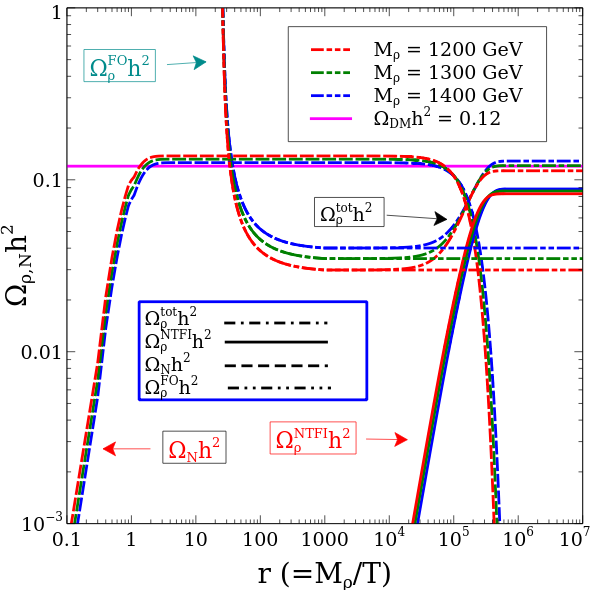}
    \caption{LP shows the variation of DM relic density produced
    by three mechanisms namely freeze-out, thermal freeze-in and non-thermal 
    freeze-in for three
    different values of next to stable particle ($N$) mass, 
    $M_{N}$. In the RP,
    we have shown the variation of DM relic density produced by different
    mechanisms for three different values of DM mass, $M_{\rho}$.
    The other parameters are fixed at the following value unless they are
    varied, $M_{N} = 2000$ GeV, $M_{\rho} = 1300$ GeV, 
    $ M_{H_2} = 5000$ GeV and $Y_{\rho\Delta} = 2.5\times10^{-12}$. Same convention as in Fig. \ref{diff-production-mode}.}
	\label{vary-mN-mrho}
\end{figure}

The abundances of $N$ and $\rho$ also depend on the masses of the NLOP and of DM. 
In Fig.\,\ref{vary-mN-mrho}  we show the dependence of $N$ and $\rho$ abundances  on the value of  the NLOP mass $M_{N}=1.9,2.,2.1\  {\rm TeV}$ (LP) or on the  DM mass $M_{\rho}=1.2,1.3,1.4\  {\rm TeV}$ (RP).   Other
parameters  are fixed, to the value  explicitly mentioned in the figure caption.
In the LP we can see that the production of $N$ is enhanced as $M_N$ decreases. This is  due to an enhancement in the freeze-in production of $N$ from  $H_2$ decay related to the increase in  the phase space factor $\sqrt{1 - \frac{(M_{N} + M_{\rho})^{2}}{M^{2}_{H_2}}}$   as we lower $M_{N}$.
 Moreover, the DM production from the decay of the non-thermal particle 
$N$ is also increased since it is  inversely proportional to  $M_{N}$.
The production of DM by  the freeze-out mechanism is unaffected by the variation of $M_N$, since its effect is suppressed by the feeble coupling $Y_{\rho \Delta}$.
In the RP,  we  show the impact  of the DM mass on the relic density. 
The production of DM in  the freeze-out mechanism decreases with the DM mass as discussed above. On the other hand  
when DM is produced by the non thermal freeze-in mechanism there is no  significant variation in the DM relic density. This happens because $M_\rho< M_N$, thus $M_\rho$  has a smaller effect  on the phase space factor. Nevertheless in Fig.\,\ref{vary-mN-mrho}, the full  lines show that there is small shift
in the DM relic density because of the slight variation in the ratio of DM mass.

\subsubsection{Scan on parameter space}
\label{scenario1:scan}
To investigate further the dependence of the relic density on the model parameters we perform a scan varying the parameters in the following range

\begin{eqnarray}
700\,\,\,{\rm GeV} < &M_{\rho}& < \,\,\,1500\,\,\,{\rm GeV}\,, \nonumber \\
125\,\,\,{\rm GeV} < &M_{N} - M_{\rho}& < \,\,\,3000\,\,\,{\rm GeV}\,, \nonumber \\
1500\,\,\,{\rm GeV} < & M_{H_2}& < \,\,\,20000\,\,\,{\rm GeV}\,, \nonumber \\
10^{-13} < & Y_{\rho\Delta} & < 10^{-10}\,\nonumber \\
10^{-3} < & \alpha & < 0.1\,.
\label{parameter-range}
\end{eqnarray}
We keep only the parameter points for which the DM relic density lies in the $3\sigma$ range, Eq.~\ref{PLANCK}.
When computing the relic abundance we  take into account both the freeze-out and thermal and non-thermal freeze-in contributions, although the  thermal contribution from $H_2 \to \rho N$ does not have any impact in determining the DM relic abundance. We also impose the BBN constraint which we will discuss in detail in the later part of this section.

\begin{figure}[]
	\includegraphics[angle=0,height=7.5cm,width=7.5cm]{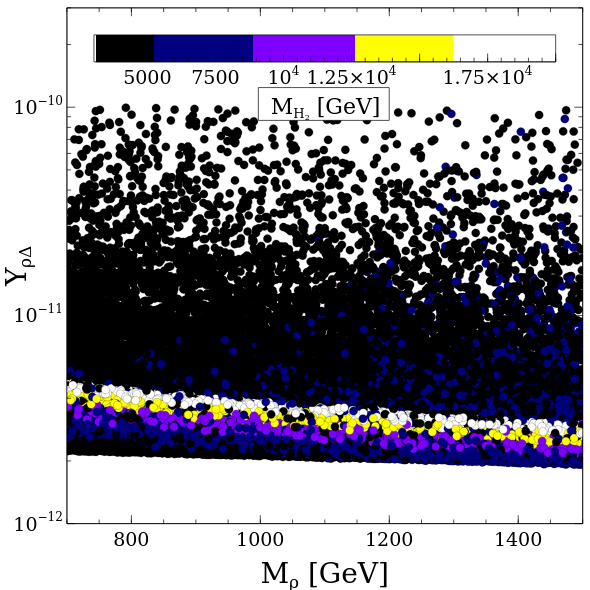}
		\includegraphics[angle=0,height=7.5cm,width=7.5cm]{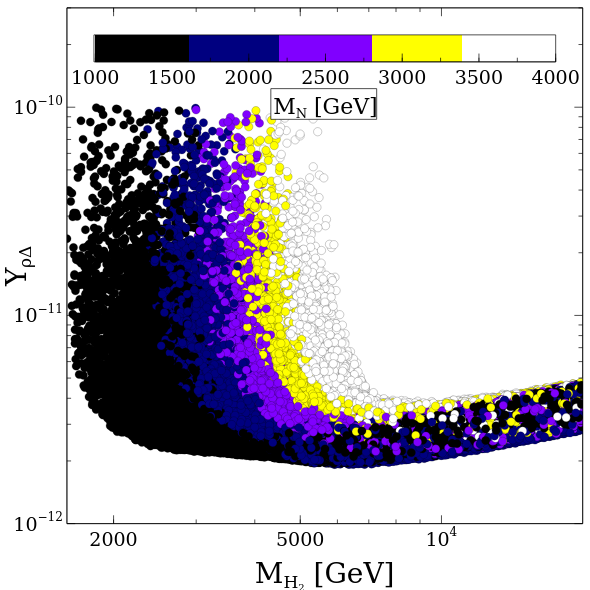}
    \caption{ Allowed parameter space in the  in 
    $Y_{\rho\Delta} - M_{\rho}$ (Left) and $Y_{\rho\Delta} - M_{H_2}$ (Right)
    planes after imposing the relic density as well BBN constraint. DM
    production from both freeze-out and non-thermal freeze-in mechanism is
    included.}
	\label{vary-mrho-mh2-yrhodelta}
\end{figure}

From the electroweak precision data, we have the maximum allowed range of the triplet vev $v_{\Delta} < 12$ GeV. This value puts an upper bound
on the charged Higgses mixing angle {\it i.e.} $\delta < 0.1$. Moreover,
to make the quartic couplings always in the perturbative regime ($\lambda_{i} < 4 \pi$),
 we choose $\delta = \alpha$.
 Therefore,  combining precision data and perturbativity bound 
 we consider an upper limit on the SM-BSM neutral Higgses mixing angle $\alpha = 0.1$ which is consistent with the LHC searches.

In Fig.\,\ref{vary-mrho-mh2-yrhodelta} (LP), we  show  the allowed points in 
the   $Y_{\rho\Delta} - M_{\rho}$ plane. It is evident that  with the increase in  DM mass $M_{\rho}$,  a  lower value of Yukawa coupling $Y_{\rho \Delta}$ is required 
to obtain the DM relic density in the above-mentioned range. This is because the dominant contribution to the DM relic density, which arises from the late decay contribution of $N$ varies as $\Omega^{NTFI}_{\rho} h^{2} \propto Y^{2}_{\rho\Delta} M_{\rho}$.
In the color bar of the same figure,  we show the variation of the BSM Higgs mass 
$ M_{H_2}$. From this we can see  that  lower values of $ M_{H_2}$ correspond to lower as well as higher values $Y_{\rho\Delta}$,  
represented by black points. The higher values of $Y_{\rho \Delta}$ correspond to lower $M_{H_2}$ occurs due to  phase space suppression  in the process $H_2 \to \rho N$. Whereas the lower values of $Y_{\rho\Delta}$ arises due to no phase space supression in the process $H_2 \to \rho N$. Therefore, to satisy the DM relic density in $3\sigma$ range requires smaller $Y_{\rho\Delta}$ for smaller $M_{H_2}$ since  $\Omega^{NTFI}_{\rho} h^{2} \propto \frac{Y^2_{\rho\Delta}}{ M_{H_2}}$. 
In the  RP, we show the allowed points in  
the $Y_{\rho\Delta} - M_{H_2}$ plane, the color bar  represents  the variation of  $M_{N}$.
As has been explained above and also evident from this figure  with the increase in $M_{H_2}$, higher values of $Y_{\rho \Delta}$ are required.
Note that for very  large values of $ M_{H_2}>10^4$ GeV, there is no correlation with $M_N$, as in this region $H_2 \to \rho N$ decay process does not encounter any phase-space suppression.
On the contrary  for $ M_{H_2}<10^4$ GeV, an  increase in  $M_{N}$ results in an increase in the  required  value of $Y_{\rho \Delta}$.  

\begin{figure}[]
	\centering
	\includegraphics[angle=0,height=8.5cm,width=8.5cm]{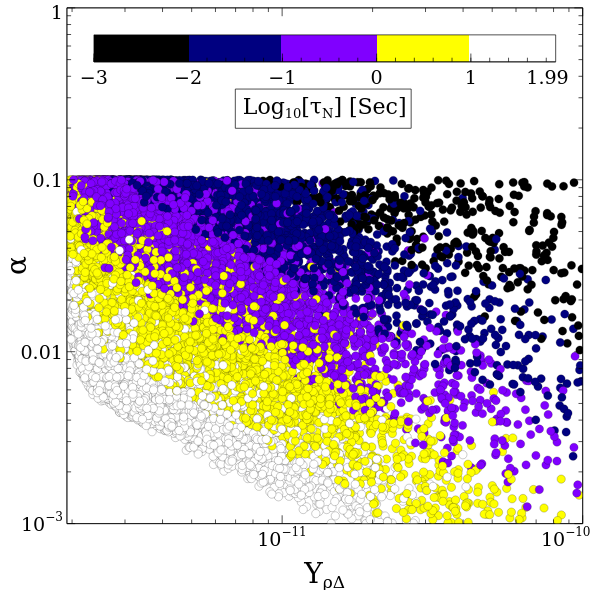}
	\caption{Scatter plot in 
		$Y_{\rho\Delta} - \alpha$ plane after
		demanding DM relic density in $3\sigma$ range. All the points satisfy BBN constraint   from $Y_{p}+ D/H + {}^4He/D$ and CMB spectral distortion\cite{Kawasaki:2017bqm}. }
	\label{vary-yrhodelta-alpha}
\end{figure}

In Fig.\,\ref{vary-yrhodelta-alpha}, we show scatter plots in the $Y_{\rho\Delta} - \alpha$ plane with the 
color variation  representing   the decay lifetime of $N$, $\tau_{N},$ where $N$ decays into $\rho H_1$.
The decay lifetime of $N$ is computed using the following
expression,
\begin{eqnarray}
\tau_{N} = \frac{16 \pi M_{N}}{Y^2_{\rho\Delta} \sin^{2}\alpha 
\left((M_{N} + M_{\rho})^{2} - M^2_{H_1} \right)}
\left[\left(1 - \left(\frac{M_{\rho} + M_{H_1}}{M_{N}} \right)^{2} \right)
\left(1 - \left(\frac{M_{\rho} - M_{H_1}}{M_{N}} \right)^{2} \right)\right]^{-\frac{1}{2}}\,.
\nonumber \\
\label{decay-lifetime-sec}
\end{eqnarray}
\begin{figure}[]
	\includegraphics[angle=0,height=7.5cm,width=7.5cm]{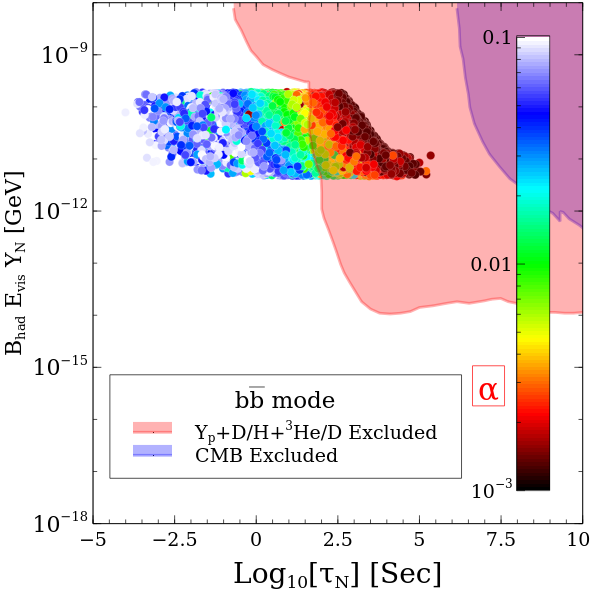}
	\includegraphics[angle=0,height=7.5cm,width=7.5cm]{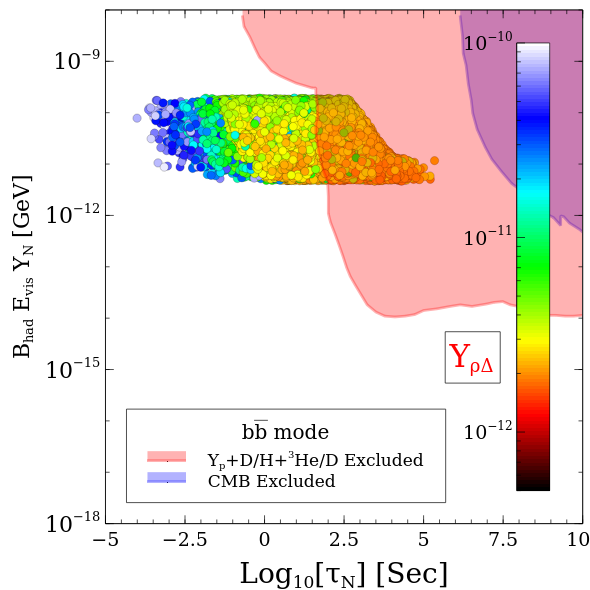}
	\caption{Scatter plots in life time of $N$ and injected hadronic energy through decay of $N$ after implementing the bound from BBN. The 
		colorbar in the LP corresponds to the variation of mixing angle whereas
		RP is for the Yukawa coupling. All points in the RP and LP satisfy the DM relic density constraints in $3\sigma$ range. The red shaded region denote the BBN constraints from $Y_{p}+ D/H + {}^4He/D$, and violet shaded region is excluded by CMB spectral distortion\cite{Kawasaki:2017bqm}. }
	\label{bbn_bound}
\end{figure}

As can be seen from the above equation the decay 
lifetime is inversely proportional to $Y^2_{\rho\Delta}$ and $\sin^2 \alpha$
{i.e.} lower values represent larger decay lifetime. The longer lifetime are constrained by BBN as discussed in section~\ref{Dm_BBN}, thus  the allowed points in Fig.\,\ref{vary-yrhodelta-alpha} correspond to 
 $\tau_N$  ranging from a very low value $10^{-3}$ upto  $ 97.7 $ seconds. Larger lifetimes are found in the  triangle shaped white region in the lower left corner which  is thus disallowed by BBN constraints .
Note that even if we were to impose the stronger constraint of $\alpha < 0.025$~corresponding to $v_{\Delta} < 3$ GeV~\cite{ParticleDataGroup:2020ssz} our allowed points would still cover the entire range of $Y_{\rho \Delta}$ values from $10^{-12}$ to $10^{-10}$.
Here we have chosen the model parameters  such that the decay mode $N \rightarrow \rho H_1$ is always kinematically allowed, see  Eq. \ref{parameter-range}. 
 For $M_{N} < M_{\rho} + M_{H_1}$, the three-body decay mode leads to a longer lifetime of the NLOP and most points are then in conflict with the BBN bound.

To illustrate explicitly  the impact of the BBN bounds, Fig.\,\ref{bbn_bound} shows  the points allowed by the relic density constraint in the  plane of the injected hadronic energy of $N$,  $\zeta_{had}=B_{had} E_{vis} Y_{N}$, and its lifetime  $\tau_N$. 
The injected energy to the visible sector per NLOP $N$ decay, $E_{vis}$, is given in Eq.~\ref{hadEvis}  where 
the visible energy release is due to the two-body decay of $N\rightarrow \rho H_1$ and $B_{had} \approx 0.57$
is calculated from Eq.~\ref{hadBr}. We find that for all points the injected hadronic energy falls in the range $\zeta_{had} \sim 4 \times 10^{-12} - 2.1 \times 10^{-10}$.  Those for which  $\tau_N > 100 $ sec are the ones that are disallowed by BBN bounds shown by the pink region in Fig.~\ref{bbn_bound}.
 In the LP we  show the color variation with respect to $\alpha$ and in the RP with $Y_{\rho\Delta}$. 
In the LP, we can see that lower values of $\alpha$ (black region) 
are ruled out from the BBN bounds which comes from the measurement of proton ($p$), deuterium (${}^{2}H$) and tritium (${}^{3}H$) abundance\cite{Kawasaki:2017bqm}.
As mentioned above, this is because lower values of $\alpha$
corresponds to the larger value of the decay lifetime. In the colorbar, $\alpha = 0.025$ corresponds to the $v_{\Delta} = 3$ GeV which is the upper bound obtained from the PDG data of electroweak precision measurements.
Note that a large fraction of the points allowed by BBN constraints have values of $\alpha$ within this limit. Thus using a more restrictive bound on $\alpha$ than $0.1$ used by us, will not change the main features of our analysis in any way.
 In the RP,
 we have shown the color variation in $Y_{\rho\Delta}$. Since we are taking $M_{\rho} < 1500$ GeV, the thermal contribution is small
 and most of the contribution comes from the freeze-in,
hence $Y_{\rho\Delta}$ lies in between $10^{-12}$ and $10^{-10}$ after 
 satisfying the DM relic density constraint. We infer from the  figure that although 
$N$ can have late decays, most of the parameter space is safe from BBN constraints.

\subsection{DM production in Scenario II : $M_\rho>M_N$} \label{S2}
We consider a scenario, where the RHN $N$ is lighter than  the triplet fermion $\rho$ hence serves as the DM.  
Contrary to the  previous scenario,  in this case DM  can only  be produced via   thermal and non-thermal freeze-in and there is no freeze-out contribution to the DM relic abundance.  Owing to its feeble coupling  $Y_{\rho \Delta}$ and being singlet under SM gauge group, $N$ never thermalizes with the  thermal bath. It is produced from the annihilations and decay of the bath particles, where the latter dominates the production of $N$. 
The annihilation contribution is tiny due  to  additional couplings, heavy propagators mediating $2\to2$ process and also additional numerical factors arising from phase space integral. In the considered  regime,  and for our chosen parameter space obeying $M_{H_2}> M_{\rho}+M_{N}$ and $M_{\rho}>M_{N}+M_{H_{1}}$,  $N$ is  produced  primarily from the  two body decay $H_2 \to \rho N$ and $\rho \to N H_1$.
 
\subsubsection{Solving for DM abundance}
Below, we qualitatively discuss thermal and non-thermal freeze-in contribution, before presenting an accurate evaluation of DM relic density by solving the Boltzmann equation. 

\begin{figure}[]
	\centering
	\includegraphics[angle=0,height=3.0cm,width=7.5cm]{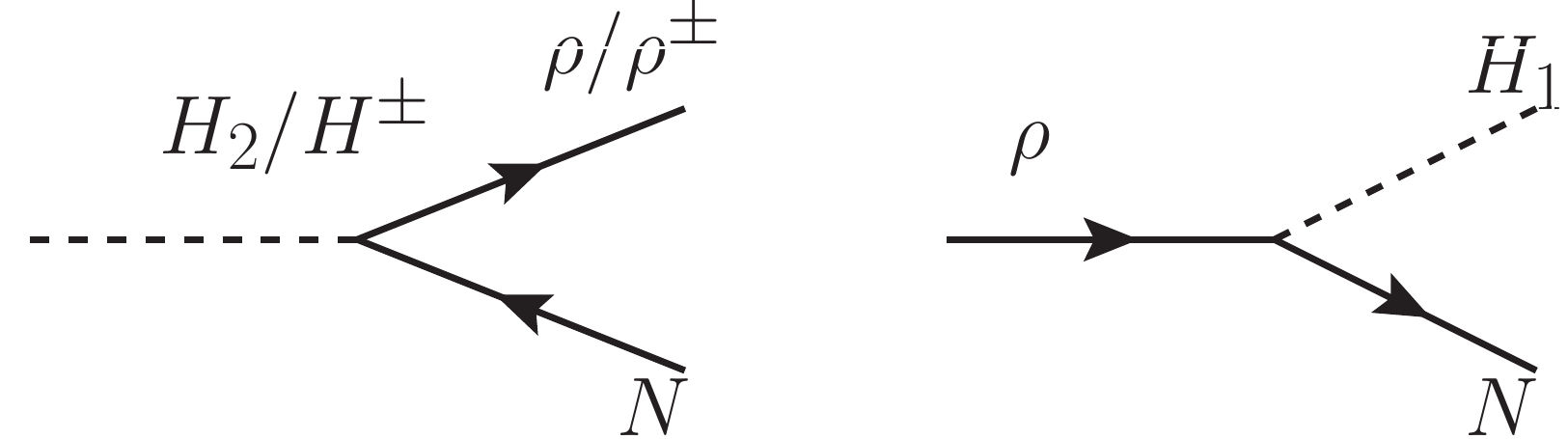}
	\caption{Feynmann Diagram for the dominant production of $N$ as well as its late production from the decay of $\rho$. \label{fdiaS2}}
\end{figure}
\begin{itemize}
\item
{\it \bf Thermal freeze-in of $N$}: At an early epoch of the universe, the production of $N$ is  governed by  thermal production from  $H_{2}$ and $\rho$ decay,  where the  mother particles  $H_2$ and $\rho$ are  in thermal equilibrium with the  thermal bath. The production of $N$ gradually increases, till the temperature of the thermal bath satisfies the condition $T> M_N$.  After this, the  production of DM through thermal freeze-in mechanism ceases.  The analytic expression for the thermal freeze-in $\Omega_{N}^{FI} h^2$ can be obtained by replacing $A$ to $H_2$ and $\rho$ respectively in Eq.~\ref{eq:omegadecay}. It is important to mention that the thermal freeze-in contribution from decay of $\rho$ is additionally suppressed by $\sin \alpha$ compared to the decay of $H_2$.
\item
{\it \bf Non-thermal freeze-in of $N$}: In addition to the  thermal freeze-in of $N$, production of $N$ also receives  additional contribution from the out of equilibrium decay of the NLOP  $\rho$ via $\rho \to N H_1$ decay mode. For our chosen DM mass $M_N$ and the mass of NLOP $M_{\rho}$, this decay is always open. There is also another decay mode $\rho \to  N f \bar{f}$ mediated via an off-shell $H_2$, which is subdominant in our case. The particle $\rho$ having gauge interaction was  in thermal equilibrium  with the rest of the plasma and at some  epoch, it decoupled from the thermal plasma after which the late decay of $\rho \to N H_1$ produced substantial DM abundance.   The non-thermal freeze-in contribution of the DM depends on the mass ratio  $M_N/M_{\rho}$, as well as the density of $\rho$ at the time of decoupling $T_d$ via 
\begin{equation}
\Omega_{N}^{NTFI} h^2= \frac{M_N}{M_{\rho}} \Omega_{\rho}^{FO} {h^2}
\label{s2superwimp}
\end{equation}   
where $\Omega_{\rho}^{FO} {h^2}$ is the abundance of $\rho$ at the decoupling temperature $T_d$, 
\begin{equation}
\Omega_{\rho}^{FO} {h^2}=\left[\frac{1.07\times 10^{9}}{g^{s}_{*}/\sqrt{g^{\rho}_{*}}}\right]\left[  \frac{x_f}{\langle \sigma_{eff} |v| \rangle} \right]\\
\end{equation}
with the decoupling temperature determine from  
\begin{eqnarray}
x_{f}&&= \frac{M_{DM}}{T_d}=\\ \nonumber &&\ln\left(0.038 g^{s}_{*}/\sqrt{g^{\rho}_{*}} M_{pl}M_{\rho}\langle \sigma_{eff} |v| \rangle \right)-\frac{1}{2}\ln\left(\ln\left(0.038 g^{s}_{*}/\sqrt{g^{\rho}_{*}} M_{pl}M_{\rho}\langle \sigma_{eff} |v| \rangle \right)\right) .
\end{eqnarray}. 
The analytical expressions for the thermal average cross-section $\langle \sigma_{eff} |v| \rangle$ are given in the Appendix \ref{appendix1}. Here we consider all possible annihilation channels involving $\rho,\rho^\pm$ namely, $\rho \rho \to W^+ W^-$,  $\rho^{\pm} \rho \to f^{\prime}\bar{f} $, $\rho^+ \rho^- \to f \bar{f}, W^+ W^-$.
As discussed earlier, the thermal average cross-section $\langle \sigma_{eff} |v| \rangle$ determining the $\rho$ annihilation into the bath particles decreases with the  increase in the mass of $\rho$. The abundance $\Omega_{\rho}^{FO} h^2$  being inversely proportional to the  thermal average cross-section therefore increases for a higher value of  $M_{\rho}$.  
\end{itemize}
Since both the thermal and non-thermal freeze-in productions can be sizeable in different regions of the   parameter space, therefore, in evaluating the  DM relic density, we  consider both of  these contributions together 
\begin{equation}
\Omega_N h^2 = \Omega^{FI}_N h^2+\Omega^{NTFI}_N h^2
\label{eq2b}
\end{equation}
We 
vary $M_N$ and $M_{\rho}$ in a wide range to demonstrate where  thermal and non-thermal freeze-in contributions dominate the DM production. 
 
In order to obtain the correct relic density of DM, we need to study the evolution of $\rho$ and $N$, which can be obtained by solving the coupled Boltzmann equations. In terms of co-moving number density of DM $N$ and NLOP  $\rho$, the two relevant  equations are: 
\begin{subequations}
	\begin{eqnarray}
	\frac{d Y_{\rho}}{d r}&&=- \sqrt{\frac{\pi}{45 G}}\,\frac{M_{Pl} 
		\sqrt{g_{*}(r)}}{r^{2}} \langle \sigma_{eff} |v| 
	\rangle \left( Y^2_{\rho} - (Y^{eq}_{\rho})^2 \right)\nn \\ && + \kappa(r) \theta (M_{H_{2}/H^{\pm}}-(M_{N}+M_{\rho/\rho^{\pm}})) \langle \Gamma_{H_{2}/ H^{\pm} \rightarrow N\,\rho/\rho^{\pm}}\rangle
	(Y_{H_2} - Y_{N} Y_{\rho})\,  \nn\\&& 
	 -\kappa(r) \theta (M_{\rho}-(M_{N}+M_{A})) \langle 
	\Gamma_{\rho \rightarrow N A} \rangle\,
	(Y_{\rho} - Y_{N} Y_{A})\,
	\label{s2_beqn1}
	\end{eqnarray}
	\begin{eqnarray}
	\frac{d Y_{N}}{d r} &&= \kappa(r) \theta (M_{H_{2}/H^{\pm}}-(M_{N}+M_{\rho/\rho^{\pm}}))\left[\langle \Gamma_{H_{2}/ H^{\pm} \rightarrow N\,\rho/\rho^{\pm}}\rangle
	(Y_{H_2} - Y_{N} Y_{\rho})\,\right]   
	+ \nn \\ && \kappa(r) \theta(M_{\rho/\rho^{\pm}}-(M_{N}+M_{H_{2}/H^{\pm}})) \left[\langle \Gamma_{\rho^{\pm}/\rho^{0} \rightarrow N H^{\pm}/ H_{2}} \rangle\,
	(Y_{\rho} - Y_{N} Y_{H^{\pm}/ H_{2}})\right]  + \nn \\ && \, \kappa(r) \theta (M_{\rho}-(M_{N}+M_{A}))\langle 
	\Gamma_{\rho \rightarrow N A} \rangle\,
	(Y_{\rho} - Y_{N} Y_{A})\,.\nonumber  \\
	\label{s2_beqn2}
	\end{eqnarray}
\end{subequations}
where $\kappa(r)$ is defined as $\frac{M_{Pl}\, r\, \sqrt{g_{\star} (r)}}{1.66\, M_{sc}^{2}\, g_{s}(r)}$. The expression for the thermally average effective cross-section
$\langle\sigma_{eff} |v|\rangle$ is given in Appendix \ref{appendix1}. The Boltzmann equation governing the evolution of $\rho$ is given by  Eq.(\ref{s2_beqn1}).
The first term in the Eq.(\ref{s2_beqn1}) represents the production
of $\rho$ through the freeze out mechanism, second term
is the production of $\rho$ from the decay of Higgs $H_2$ and the third term is the depletion of $\rho$ due to production of $N$ through the two body decay process $\rho \to N H_1$ or through three body decay process $\rho \to N f \bar{f}$. In the last term ``A" corresponds for two body or three body decay. The first term in Eq.(\ref{s2_beqn2})  governs the evolution of $N$ via the process $H_2 \to N \rho, H^{\pm} \to N \rho^{\pm}$,  the last term represents production of $N$ from the decay of $\rho \to N H_1$. After solving the Boltzmann equations, the DM relic density can be obtained from Eq. \ref{obsDM} replacing $Y_{\rho}$ by $Y_{N}$ and $M_{\rho}$ by $M_{N}$.
 
\begin{figure}[]
\includegraphics[angle=0,height=7.2cm,width=7.5cm]{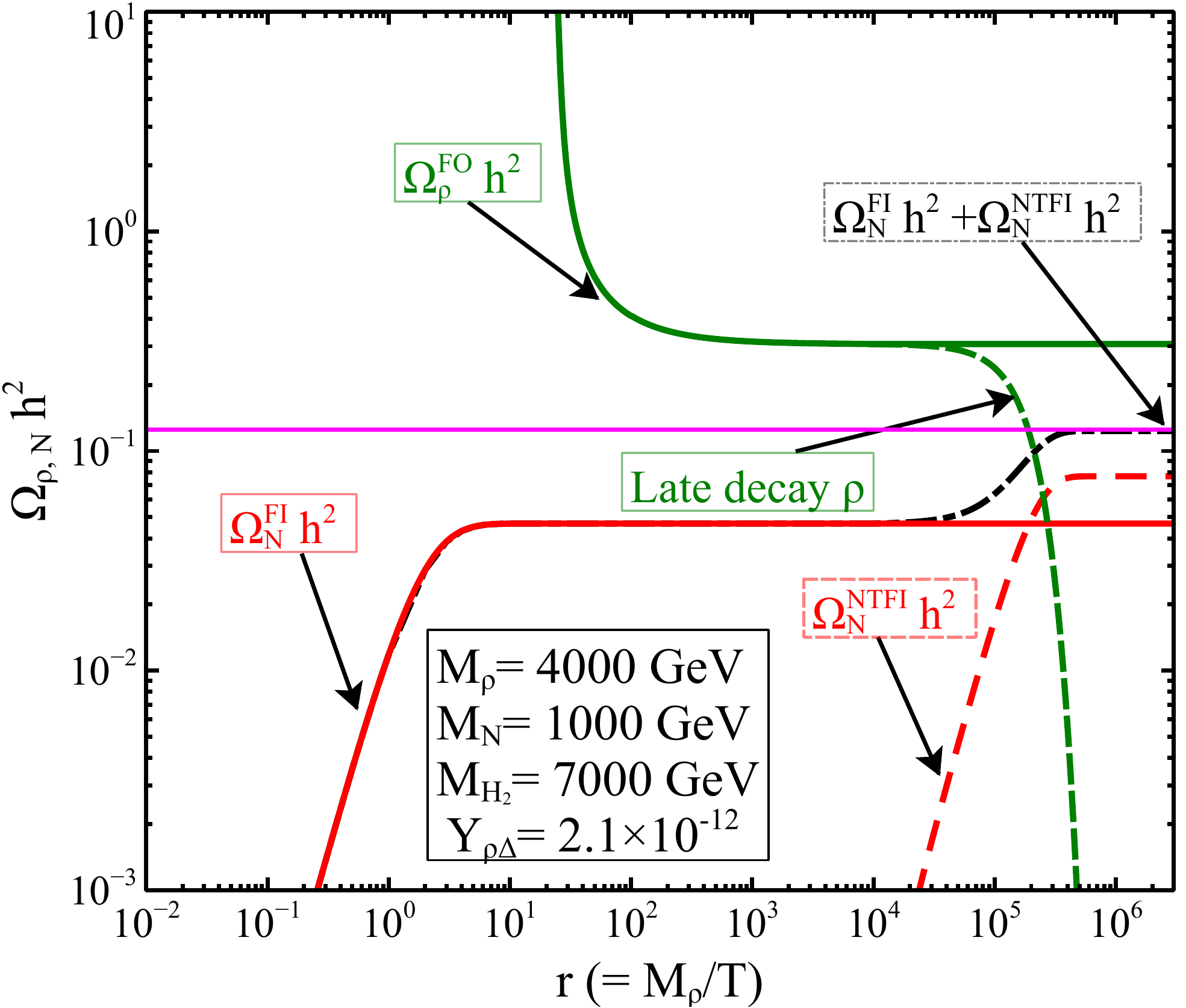}
\includegraphics[angle=0,height=7.2cm,width=7.5cm]{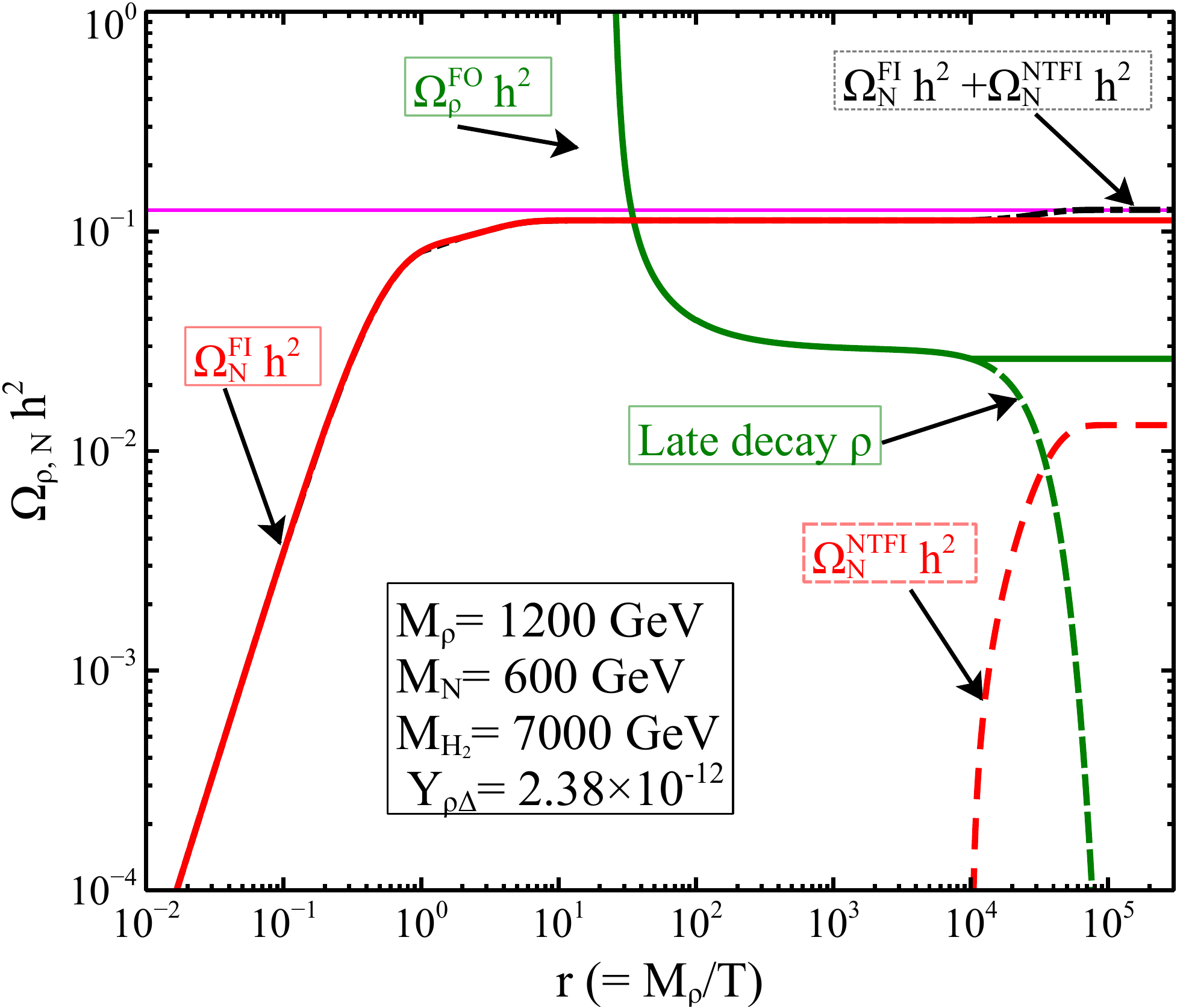}
	\caption{Evolution of $\rho$  and $N$ abundances including the decay of $\rho$. In the Right panel, the late decay of $\rho$ gives significantly large contribution. In the Left panel, thermal freeze-in is the most dominant.}
	\label{s2mechanism}
\end{figure}

\subsubsection{Results}
As we have discussed in Section \ref{S1}, the $\rho$ production through freeze-out increases with $M_\rho$,  and $\Omega_\rho^{FO}h^2$ exceeds the Planck value for the relic density  for $M_\rho>2.45$ TeV, see Fig.\ref{diff-production-mode}(upper left). Note that this statement applies in the limit $Y_{\rho\Delta}=0$ where $\rho$ is stable. When $Y_{\rho\Delta}\neq 0$, $\rho$ is unstable and the out-of equilibrium decay $\rho\rightarrow N H_1$  can contribute significantly to DM production. A larger contribution is expected for large $M_\rho$.
\begin{figure}[t!]
	\includegraphics[angle=0,height=7.0cm,width=7.5cm]{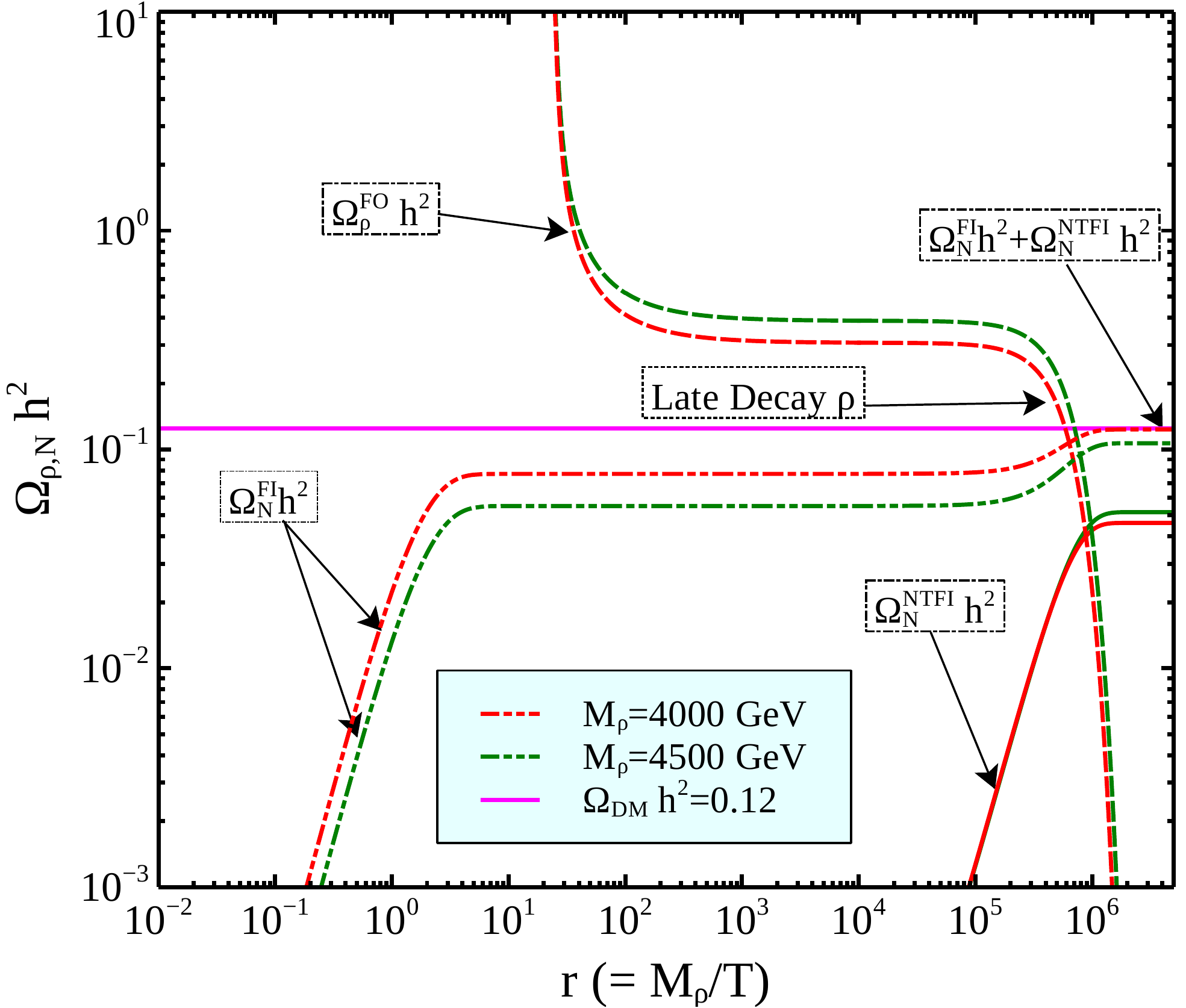}
	\includegraphics[angle=0,height=7.0cm,width=7.5cm]{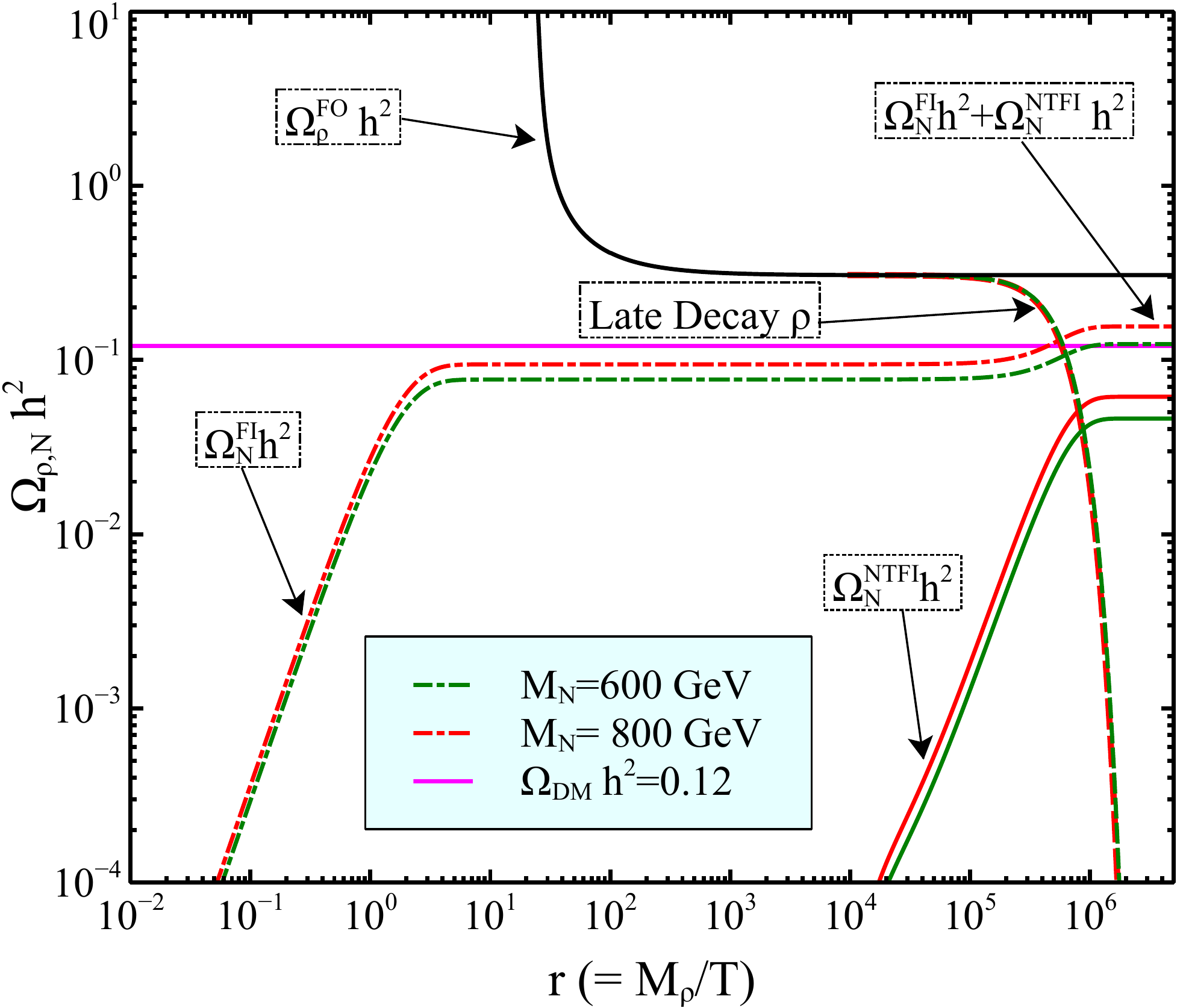}\\
	\includegraphics[angle=0,height=7.0cm,width=7.5cm]{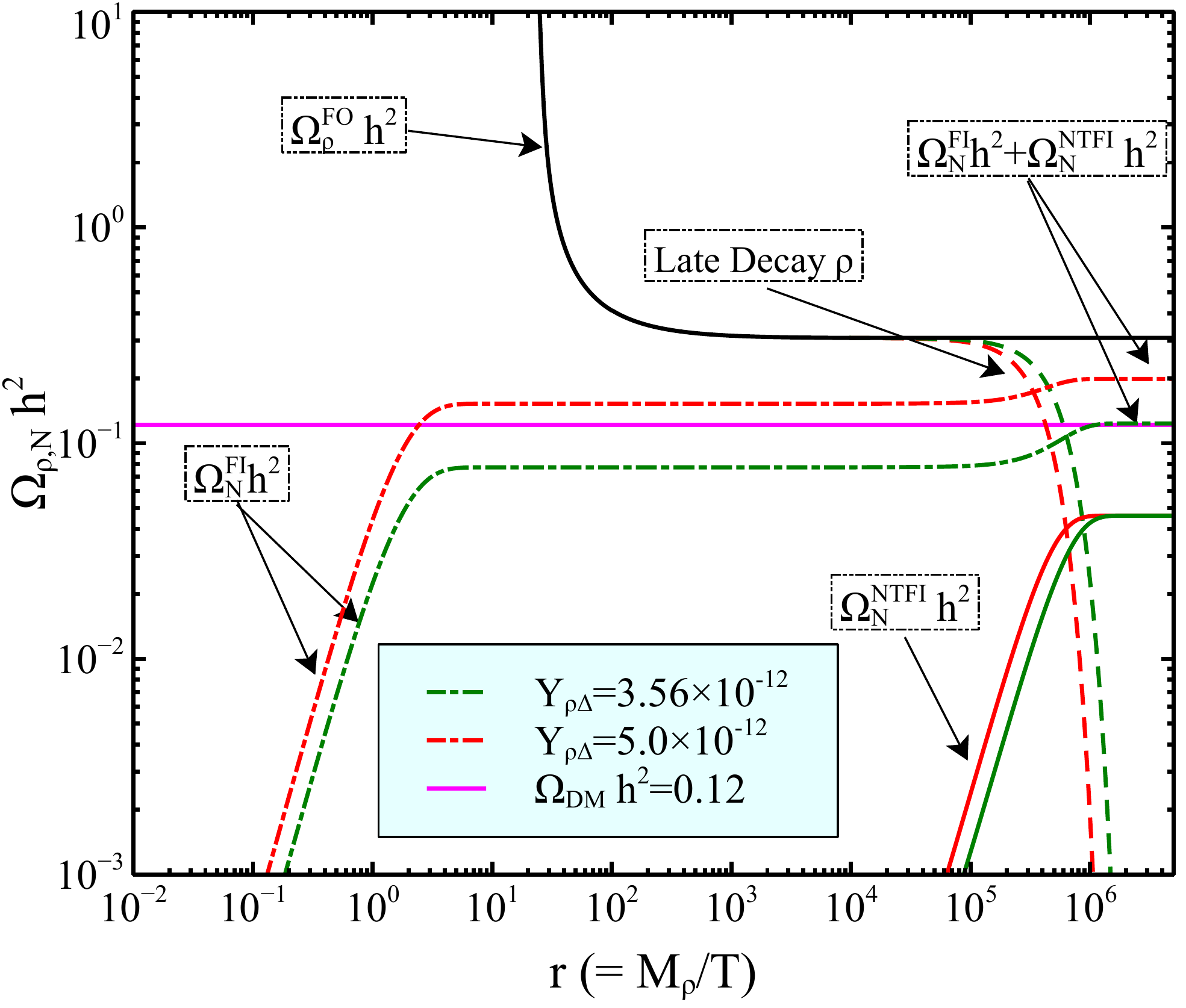}
	\includegraphics[angle=0,height=7.0cm,width=7.5cm]{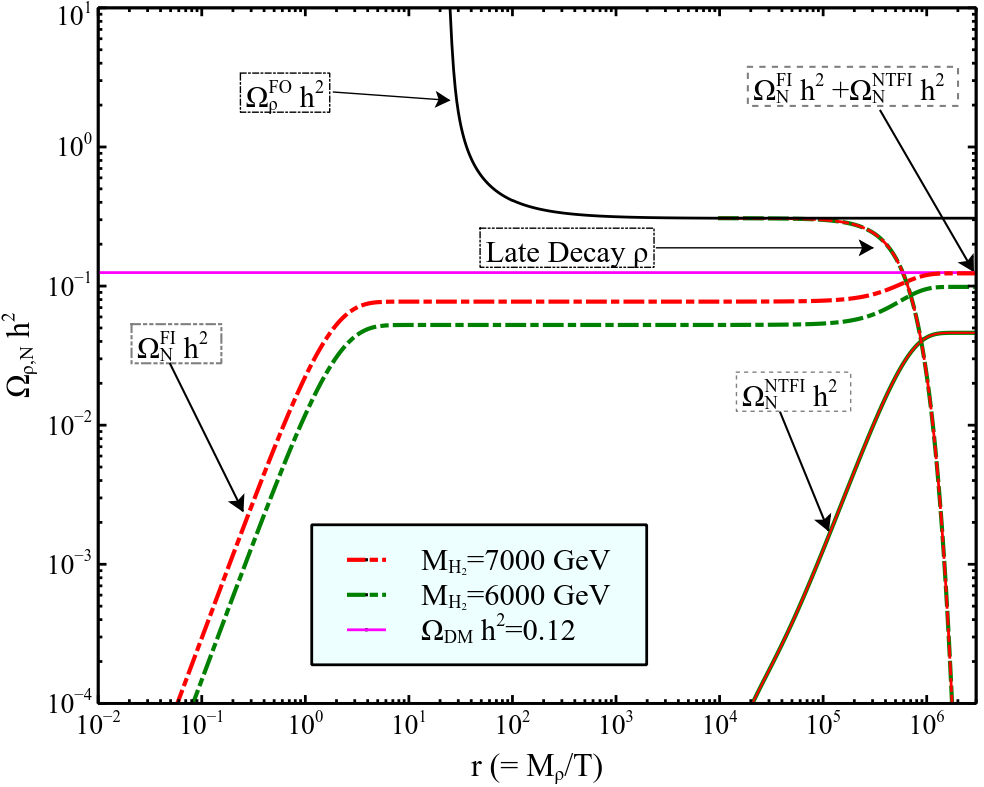}
	\caption{Upper panel shows the variation of DM and $\rho$ abundance  for different values of $M_{\rho,N}$. In upper RP, the topmost curve ($\Omega_{\rho}^{FO} h^2$) corresponds to $M_{\rho} = 4500$\, GeV and lowermost curve corresponds to $M_{\rho} = 4000$\, GeV. Whereas in upper LP, the lowermost and topmost curve of ($\Omega_{N}^{FI} h^2$) corresponds to dark matter mass $600$\, GeV and $800$\, GeV respectively. In the lower panel figures, we show the variation w.r.t different choices of Yukawa $Y_{\rho \Delta}$ and $M_{H_2}$. In lower LP, topmost and lowermost curve ($\Omega_{N}^{FI} h^2$) corresponds to $Y_{\rho\Delta}=\ 5.0\times10^{-12}$ and $Y_{\rho\Delta}=\ 3.56\times10^{-12}$ respectively. In lower LP,  topmost curve of $\Omega_{N}^{FI} h^2$ corresponds to $M_{H_{2}}=7000$\,  GeV. The following model parameter are kept fixed to, $M_{\rho}=4000$\, GeV, $M_N=600$\,  GeV, $M_{H_{2}}=7000$\,  GeV and $Y_{\rho\Delta}=\ 3.56\times10^{-12}$. }
	\label{s2rhoeff}
\end{figure}

The dominance of non-thermal freeze-in production of DM is illustrated in Fig.~\ref{s2mechanism} (left) for a benchmark point where $M_\rho=4000$ GeV, $M_{N}=1000$ GeV, $M_{H_2}=7000$ GeV and $Y_{\rho\Delta} =2.1\times10^{-12}$.
In evaluating the DM relic density, we assume that  the initial abundance of $N$ was zero. At high temperature DM $N$ is produced through processes $H_{2}\to \rho N$,$H^{\pm}\to \rho^{\pm} N,$ and $\rho \to N H_{1}$ via thermal freeze-in.
The decay $H_2 \rightarrow \rho N$ is the dominant one. The NLOP, $\rho$, first freezes-out and later undergoes out-of equilibrium decay $\rho\rightarrow  NH_1$ (green-dashed line) thus increasing significantly the relic density of $N$. The total production of $N$ taking into account both the thermal and non-thermal freeze-in is represented by the dot-dashed black line,
and shows comparable contributions from both processes.
When $M_\rho$ is much lighter than 2.4 TeV, its abundance is much suppressed and as expected the late decay contribution to $\Omega_{N} h^{2}$ is subdominant, this is illustrated in Fig.~\ref{s2mechanism}(right) where $M_\rho=1.2 {\rm TeV}$. 

Note that the non-thermal freeze-in contribution depends on the mass ratio $M_N/M_{\rho}$ and the abundance of $\rho$ particle, {\it i.e.} $\Omega^{FO}_{\rho} h^2$ at the time of decoupling. The mass of the heavy Higgs $H_2$,  of the charged Higgs $H^{\pm}$ and of  the  mixing angle $\sin \alpha$ have a negligible effect in determining $\Omega^{FO}_{\rho} h^2$, since the dominant annihilation channel of $\rho$ pairs is  $W^+ W^-$. The non-thermal freeze-in contribution does not depend on the Yukawa $Y_{\rho \Delta}$.  Contrary to that, the thermal freeze-in contribution  has a strong dependency on $M_{H_2}$, $M_{\rho}, M_N$ and  $Y_{\rho \Delta}$. 

In  Fig.~\ref{s2rhoeff}, the variation of $\rho$ and $N$ abundances on these four model parameters is displayed. 
In the upper two panels, we show $\Omega_{\rho}h^2$ and $\Omega_N h^2$ for different choices of $M_{\rho}$ and $M_N$, respectively. As discussed above, with an increase in $M_{\rho}$, $\Omega_{\rho}^{FO}h^2$ increases, thereby leading to a rise in the non-thermal freeze-in contribution $\Omega^{NTFI}_N h^2$ until the factor $M_N/M_{\rho}$ provides significant suppression factor. This  can be seen in the figure presented in the  LP. The thermal freeze-in production on the other hand decreases with the increase in $M_{\rho}$ due to phase space suppression in the decay processes $H_2 \to N \rho$. For small value of $r$, a linear increase of the DM abundance occurs due to  thermal freeze-in production  from $H_{2}$, $H^{\pm}$ and $\rho$ decay. Altogether, we show the total DM relic density taking into account both the thermal and non-thermal freeze-in contributions.  In the RP we show that both $\Omega^{FI}_N h^2$ and $\Omega^{NTFI}_N h^2$ increase with $M_N$, this occurs as both these contributions are proportional to $M_N$. In summary, the abundance $\Omega^{FI}_N h^2$ and $\Omega^{NTFI}_N h^2$  have different behaviour with respect to  variations of $M_{\rho}$ and $M_N$.

In the lower panel, we show the effect of different choices of $Y_{\rho \Delta}$ and $M_{H_2}$ on DM and $\rho$ abundance. The thermal freeze-in contribution is proportional to the dark sector Yukawa coupling, and hence increases with an increase of $Y_{\rho \Delta}$. This can be seen from the LP. On the other hand, the non-thermal contribution is independent of the $Y_{\rho \Delta}$. In the RP, we show the effect of different choices BSM Higgs mass on  DM and $\rho$ abundance. The abundance  $\Omega^{NTFI}_N h^2$ is independent of the BSM Higgs mass, as the dominant channel for $\Omega^{FO}_{\rho} h^2$ is $W^+W^-$.  $\Omega^{FI}_N h^2$  instead 
increases with the increase in $M_{H_2}$ since its thermal contribution is inversely proportional to  $M_{H_2}$. 

Among the different processes, the thermal production rate of DM is smaller for the process $\rho \to N H_1$ compared to process
$H_{2}\to\rho N$ and $H^{\pm}\to\rho^{\pm} N$ due to additional mixing angle between the  $H_1$ and $H_2$.
\begin{figure}[]
	\includegraphics[angle=0,height=7.0cm,width=7.5cm]{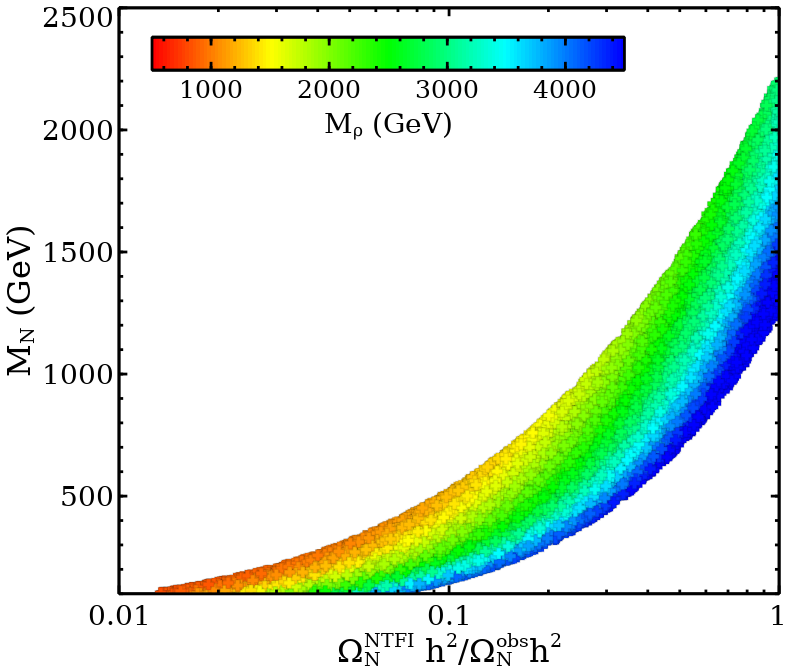}
	\includegraphics[angle=0,height=7.0cm,width=7.5cm]{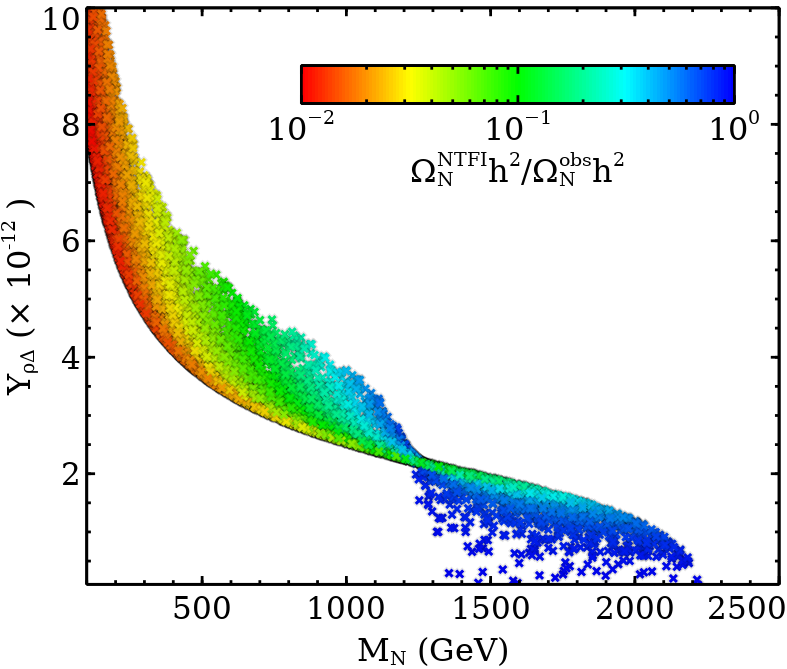}
	\caption{Left panel: Scatter plot in  DM mass $M_N$ and ratio of contribution from non-thermal freeze-in to the observed DM relic density $\Omega_{N}^{NTFI} h^{2}/\Omega_{N}^{obs} h^{2}$ plane which signifies the required variation of the NLOP mass $M_{\rho}$ in order to satisfy the correct DM relic density. Right panel: we show the variation of $\Omega^{NTFI}_N h^2/\Omega^{obs}_Nh^2$ as  scatter plot in $Y_{\rho\Delta}-M_{N}$ plane.} 
	\label{s2_scatter}
\end{figure}

\subsubsection{Scan on parameter space}
Three  important parameters in determining the DM abundance are $M_N$, $ M_{\rho}$ and $Y_{\rho\Delta}$. We therefore vary these parameters in the following range while keeping  $M_{H_2}=7000$ GeV,
\begin{equation}
10^{-11}<Y_{\rho \Delta}<10^{-15}, 100\,\textrm{GeV}\leq M_{N}\leq1800\ \textrm{GeV}\ {\rm and}\  600\,\textrm{GeV}\leq M_{\rho}\leq4500 \,\textrm{GeV}
\end{equation}

In the LP of Fig.~\ref{s2_scatter}, we show the points that reproduce the relic density  in the plane of  DM  mass and the ratio of non-thermal freeze-in to the observed DM relic density. We observe that for fixed mass of DM, the  ratio of non-thermal freeze-in to the observed relic density of DM increases as $M_{\rho}$ increases. As discussed above,  the abundance of  $\rho$ increases with $M_\rho$ and therefore leads to a large non-thermal freeze-in contribution to   the DM relic density. The RP of Fig.~\ref{s2_scatter} shows the scatter plot in the $Y_{\rho\Delta}-M_{N}$ plane. Clearly,  as the ratio of non-thermal freeze-in contribution  to the observed relic density of DM increases, a smaller value of $Y_{\rho\Delta}$ and a larger value of DM mass is required. The smaller value of the Yukawa coupling naturally leads to a suppressed thermal freeze-in contribution, which is essential to satisfy the DM relic density constraint. Contrary to that, a large Yukawa coupling is required to obtain a large thermal component of  $\Omega_{N}^{FI} h^2$, thereby leading to  a decrease in the ratio $\Omega_{N}^{NTFI} h^2/\Omega^{obs}_N h^2$.  The intersection point around $M_N=1250$ GeV corresponds to equal contribution of  $\Omega_{N}^{FI}h^2$ and $\Omega_{N}^{NTFI} h^2$.
\begin{figure}[]
	\includegraphics[angle=0,height=7.3cm,width=7.7cm]{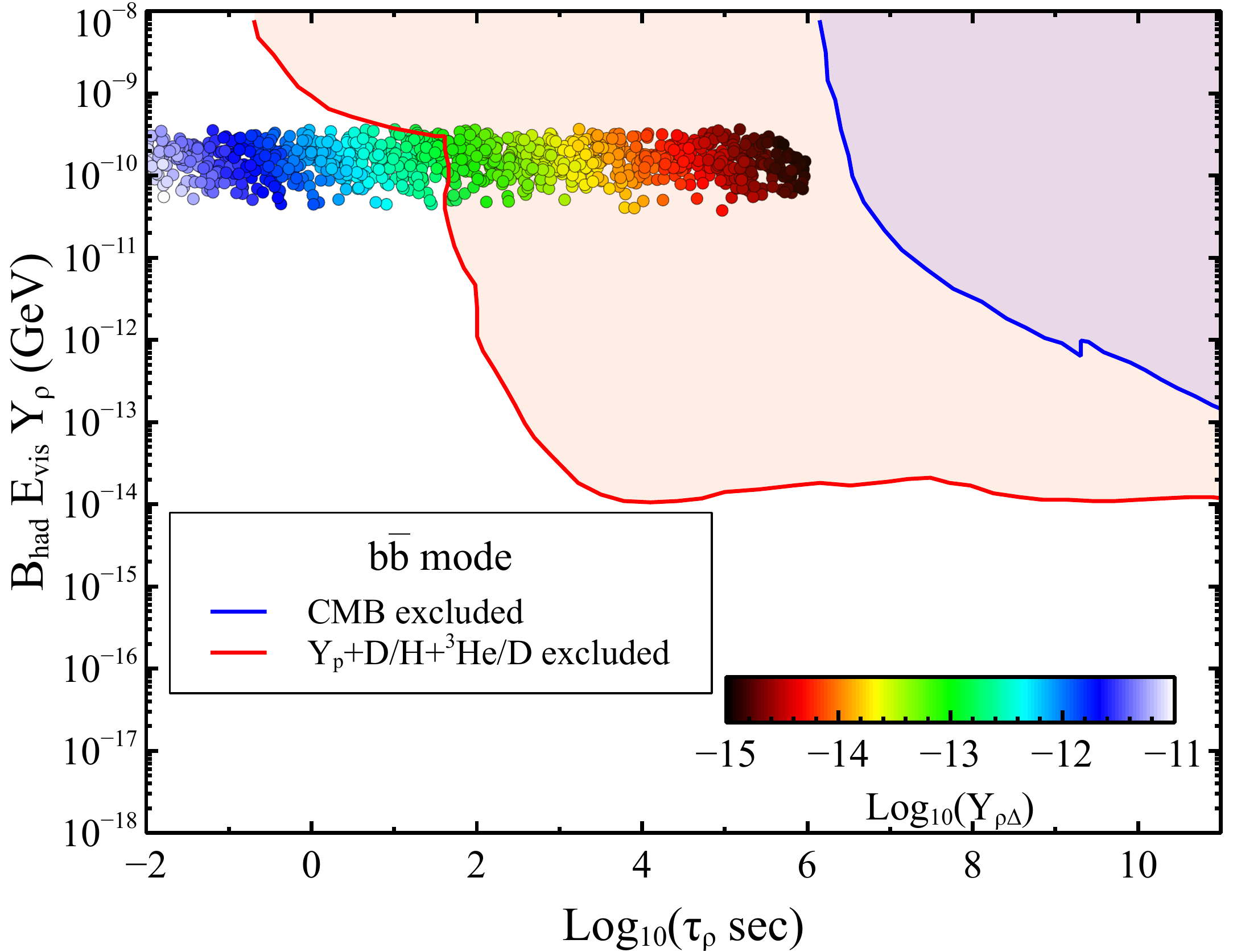}
	\includegraphics[angle=0,height=7.3cm,width=7.7cm]{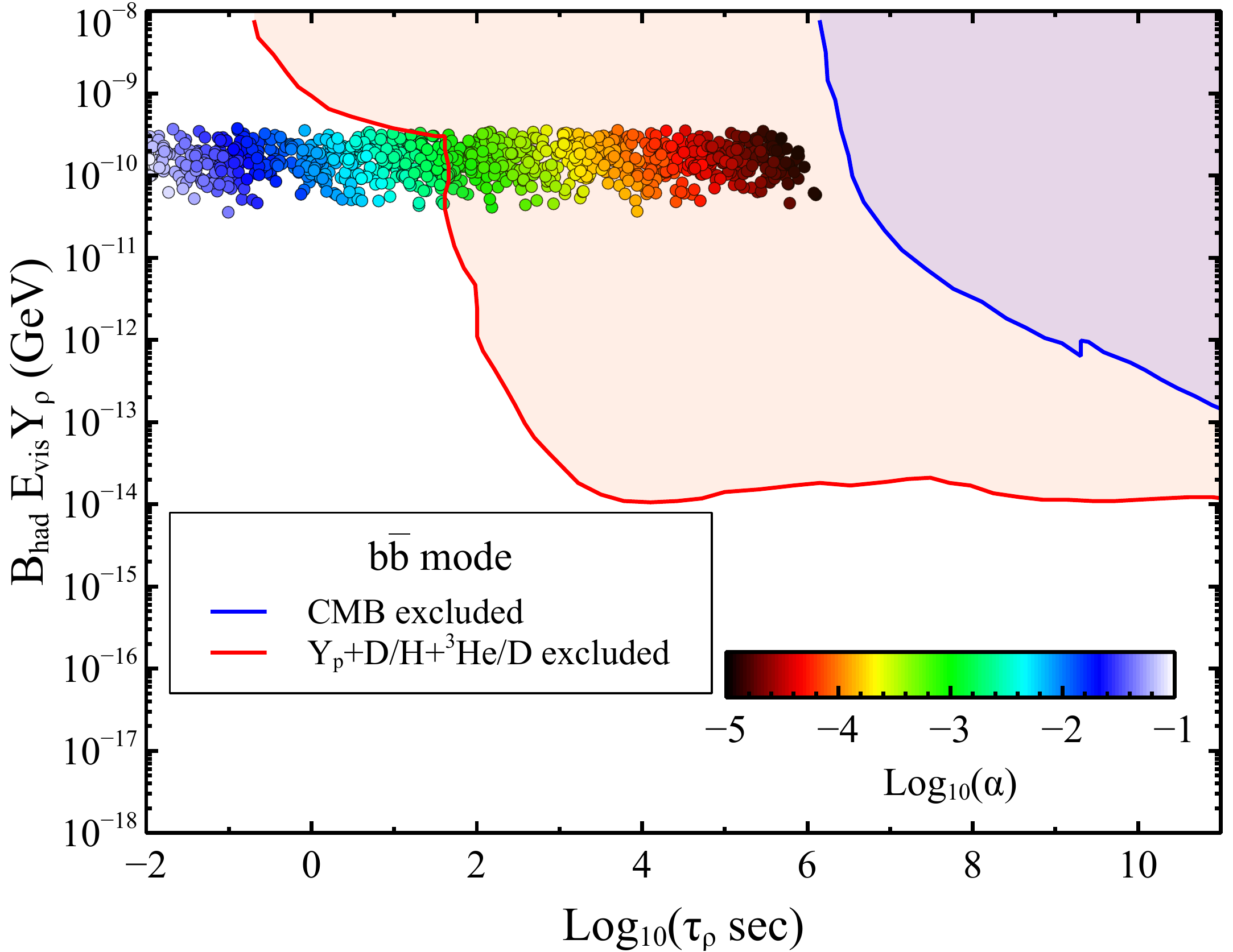}
	\caption{Scatter plot in the plane injected hadronic energy through out-of-equillibrium decay of $\rho$ plane and lifetime of $\rho$ for (LP), $10^{-15} < Y_{\rho \Delta}<10^{-11}$ and $\sin \alpha=0.1$ and (RP) $Y_{\rho \Delta} = 10^{-11}$ and  $10^{-5}<\sin \alpha < 10^{-1}$. All points in the RP and LP satisfy the DM relic density constraints in $3\sigma$ range. The solid red line denotes the BBN constraint which arises from observed light elements abundance \cite{Kawasaki:2017bqm}. The region above solid blue line is excluded by the CMB spectral distortion \cite{Kawasaki:2017bqm}.}
	\label{s2_scatter2}
\end{figure}
\newline
\noindent{\bf Impact of BBN constraints}

The NLOP $\rho$ decays only via $\rho \to N H_1$ mode,  producing significantly large non-thermal contribution to the DM relic density. The late decay of $\rho$ is therefore subject to cosmological constraints such as constraints arising from BBN discussed in section~\ref{Dm_BBN} or constraints on the amount of energy injected after $10^{10}$ sec that lead to spectral distortions of the CMB. The lifetime of $\rho$ has the following form 
\begin{eqnarray}
\frac{1}{\tau_{\rho}}=\frac{Y_{\rho\Delta}^{2} \sin^2 \alpha}{16\pi M_{\rho}}\left((M_{\rho}+M_{N})^{2}-M_{H_1}^2\right)\sqrt{1-\frac{2(M_{H_1}^2+ M_{N}^{2})}{M_{\rho}^{2}}+\frac{(M_{H_1}^2- M_{N}^{2})^{2}}{M_{\rho}^{4}}}
\label{fig:dcrho}
\end{eqnarray}

\begin{figure}[]
	\includegraphics[angle=0,height=7.5cm,width=7.6cm]{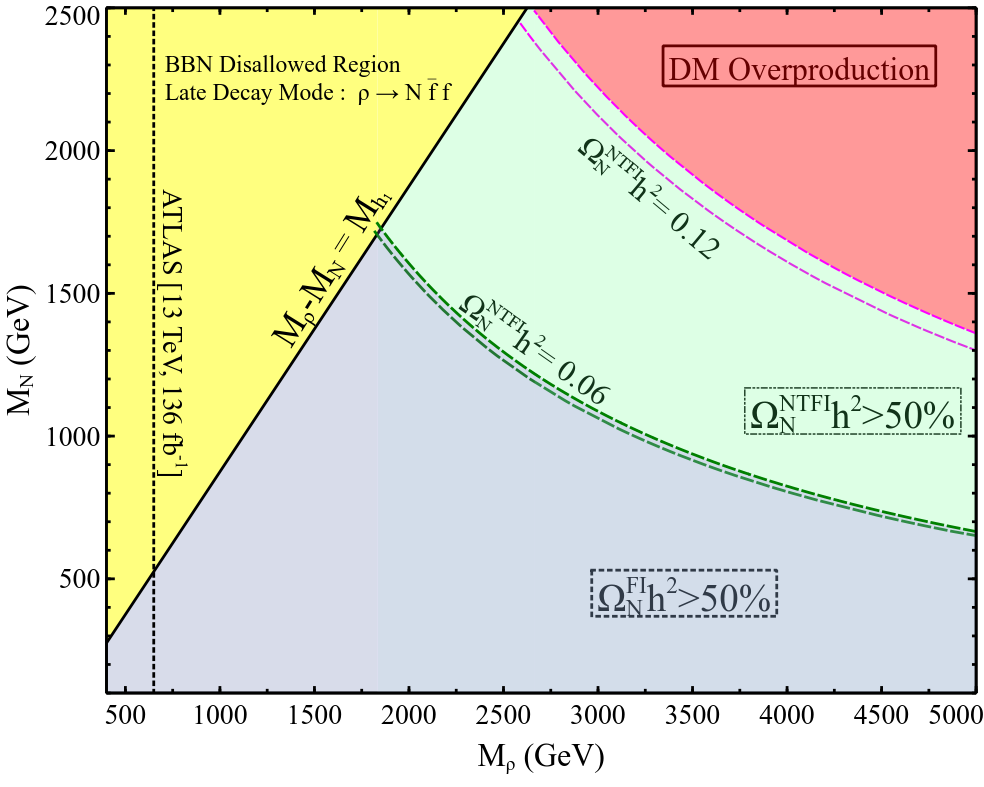}
	\includegraphics[angle=0,height=7.5cm,width=7.6cm]{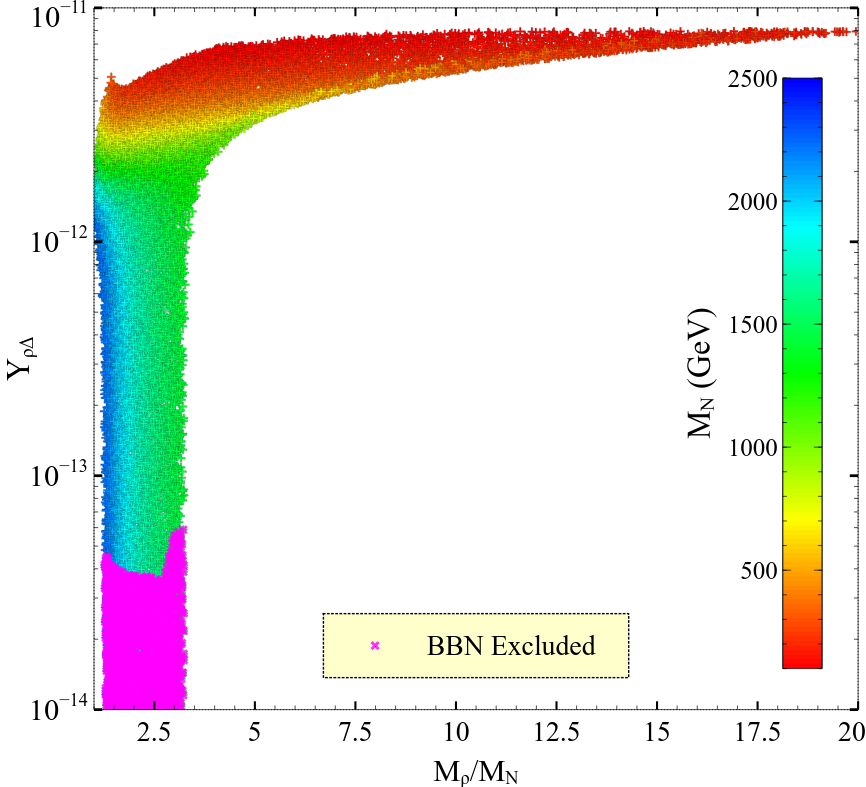}
	\caption{Left panel: Contour plot showing $\Omega_{N}^{FI}h^2>50\%$ and $\Omega_{N}^{NTFI}h^2>50\%$ in the  $M_{\rho}$-$M_{N}$ plane. The decay $\rho \to N H_1$ remains open in the region where $M_{\rho}-M_{N}>M_{H_1}$, as we consider that $\rho$ is not a stable particle. In the yellow region, $\rho\to N f\bar{f}$ contributes to $\Omega_{N}^{NTFI}h^2$ which takes place approximately around $10^{3}$-$10^{5}$ sec and is thereby disallowed by BBN. Right panel: Scatter plot in the $Y_{\rho\Delta}$ - $M_\rho/M_N$ plane where the magenta points are disallowed by the  BBN constraint on light elements \cite{Kawasaki:2017bqm,Jedamzik:2006xz}. }
	\label{s2_region}
\end{figure}
We consider two different constraints from a) CMB measurement and b) BBN constraints on light elements, namely the primordial mass fraction of $^4He$, $Y_p$, as well as the ratio of abundances of $D/H$ and $^3He/D$. These  exclusions are taken from reference \cite{Kawasaki:2017bqm}. In Fig.~\ref{s2_scatter2}, we show the constraint in terms of lifetime $\tau_{\rho}$ and  $B_{had}E_{vis}Y_{\rho}$. The color code correspond to  variations of Yukawa $Y_{\rho \Delta}$  (for the LP) and of the  Higgs mixing angle $\sin \alpha$ (for the RP).  The points excluded by BBN correspond to small values of 
$Y_{\rho \Delta}$  and/or $\sin \alpha$, as with smaller $Y_{\rho\Delta}$ and /or Higgs mixing angle, the lifetime of NLOP increases significantly.

In the LP of Fig.~\ref{s2_region}, we show  contour plots for  $\Omega_{N} h^{2}$  in the  $M_N- M_{\rho}$ plane, as well as the regions which satisfy $\Omega_{N}^{NTFI} h^{2}>50\% \Omega^{obs}h^{2}$ (green) and $\Omega_{N}^{FI} h^{2}>50\%  \Omega^{obs}h^{2}$ (grey), respectively.  $\Omega_{N}^{NTFI} h^{2}$ and $\Omega_{N}^{FI} h^{2}$ are  evaluated from Eq.~\ref{eq:omegadecay}  and Eq.~\ref{s2superwimp} respectively  
where in the latter the partial decay width of $H_{2}$  is given by,
\begin{eqnarray}
\Gamma_{H_{2} \rightarrow N\,\rho}=\frac{Y_{\rho\Delta}^{2} M_{H_{2}}}{8\pi}\left(1-\frac{(M_{\rho}+M_{N})^2}{M_{H_{2}}^{2}}\right)\sqrt{1-\frac{2(M_{\rho}^2+ M_{N}^{2})}{M_{H_{2}}^{2}}+\frac{(M_{\rho}^2- M_{N}^{2})^{2}}{M_{H_{2}}^{4}}}
\label{fig:scatter1}
\end{eqnarray}

In the RP of Fig.~\ref{s2_region}, we show the allowed points after the relic density constraint  in the $Y_{\rho\Delta}-(M_{\rho}/M_{N})$ plane for different masses of N. We can see that $Y_{\rho\Delta}$  increases, as $M_{\rho}/M_{N}$ increases and it increases due to decrease in $M_N$. This occurs due to the increase in thermal freeze-in contribution to the total relic density as  $\Delta M (= M_{\rho}-M_{N})$ increases. For a small value of $\Delta M$, the non thermal freeze-in contributes significantly to the total relic density, thus negligible contribution from thermal freeze-in mechanism is required. Therefore, small value of $Y_{\rho\Delta}$ is required when compared to the region with a large ratio of $M_{\rho}/M_{N}$. This in turn leads to an increase in  the lifetime of $\rho$ and gets severely constrained from BBN. The magenta points which corresponds to $Y_{\rho\Delta}\lessapprox 5\times10^{-14}$  are disallowed by BBN. 



\subsection{DM production in Scenario III: a lighter scalar sector }
\label{sec:lightH2}

\begin{figure}[h]
 	\includegraphics[angle=0,height=7.2cm,width=7.5cm]{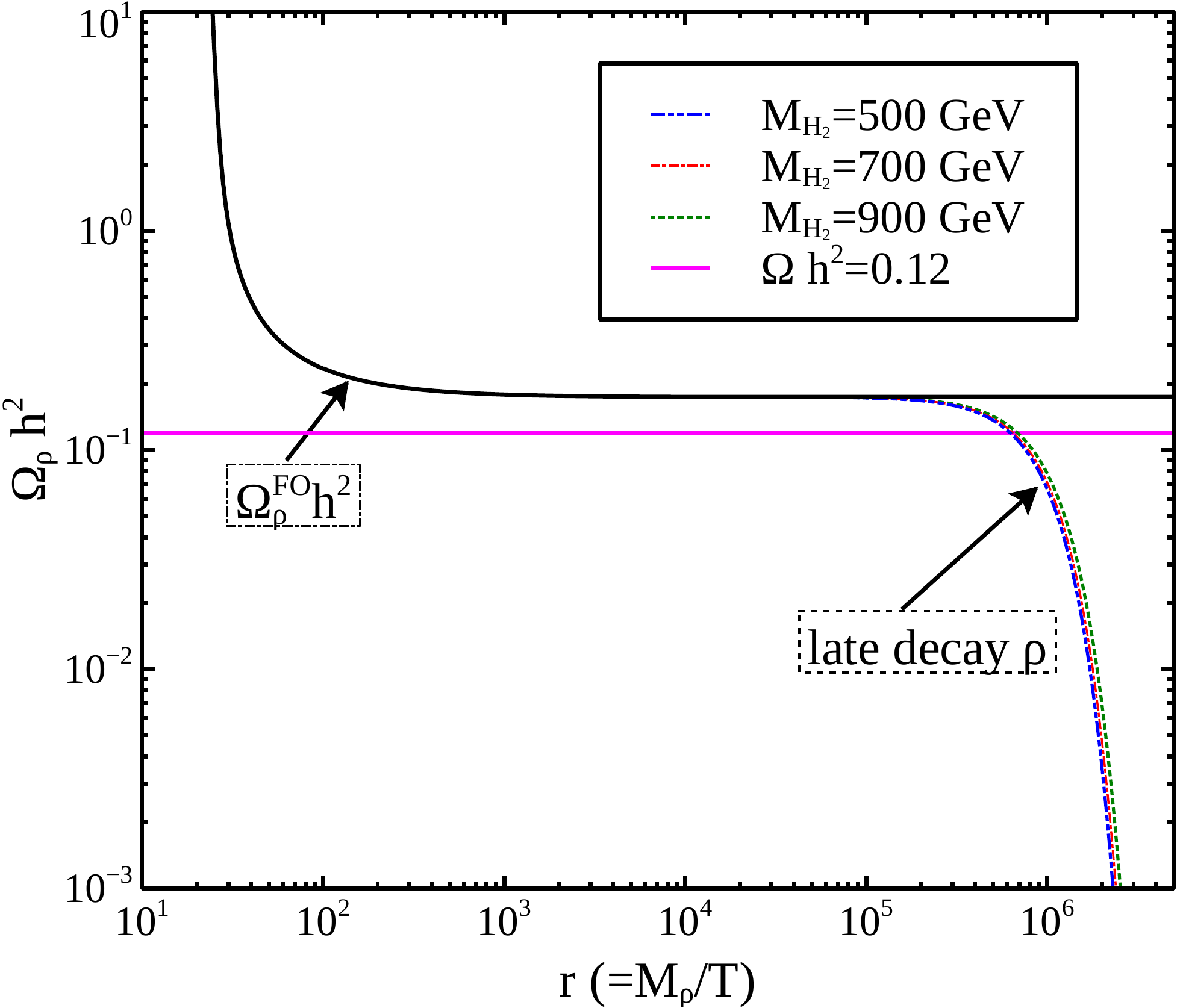}
 	\includegraphics[angle=0,height=7.2cm,width=7.5cm]{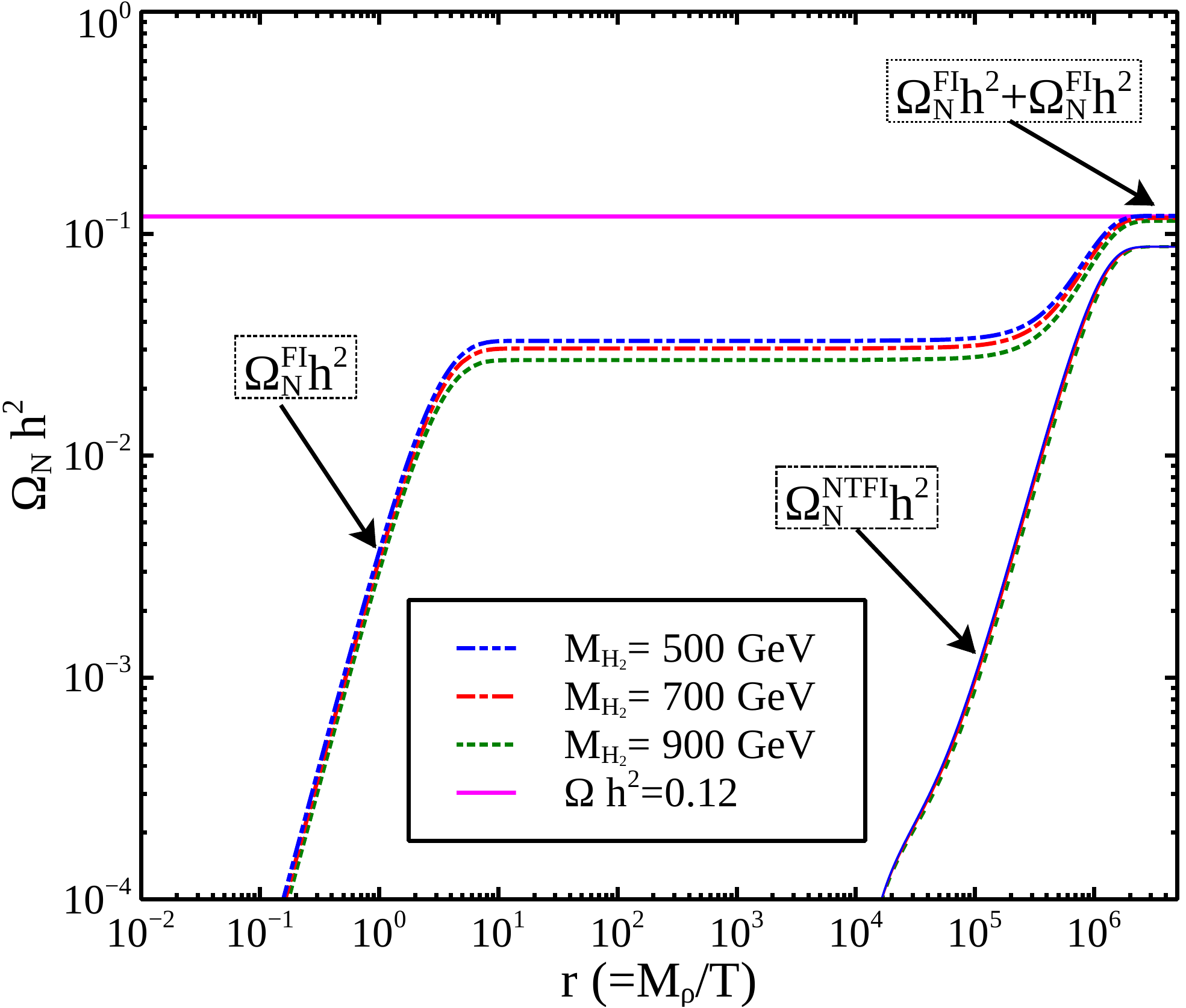}
 	\caption{Variation of the relic density of $\rho$(LP) and DM(RP) for the three different values of $M_{H_2}$. In the RP, the contributions from thermal and non-thermal freeze-in  are shown explicitly. The model parameter are kept fixed to, $M_{\rho}=3000\  GeV,$ $M_N=1500\  GeV,$ and $Y_{\rho\Delta}= 5\times10^{-13} $.    }
 	\label{DM_low_mrho}
 \end{figure}

In the previous sections, we have primarily focussed on    DM production from the decay of BSM particles. Considering the chosen values for the  DM mass, the BSM Higgs 
masses need to be at least a  few TeV's for  decay processes to be kinematically allowed.    One major drawback of having a heavy 
BSM Higgs is that they are difficult  to be observed at the LHC. To include the possibility of lighter BSM Higgs masses,  in this section we deviate from the assumption of decay dominance in DM production and also consider DM production from annihilation of SM/BSM particles. 
First we illustrate the impact of a lighter $H_2$ on the DM relic density including only decay processes. For this 
 we consider a scenario where $N$ is the DM and is produced   from the processes $\rho \to N H_2$ and $\rho \to N H_1$, while $H_2 \to \rho N$  is kinematically forbidden. We choose $M_\rho=3000\,\, {\rm GeV}, M_N=1500\,\, {\rm GeV}$ 
 and different values of  $M_{H_2}$ such that $M_{H_2}<M_{\rho}<M_{N}$.  
 In Fig.\ref{DM_low_mrho}, we show both the evolution of $\Omega_\rho h^{2}$ from freeze-out and its out-of equilibrium decays and of $\Omega_N h^{2}$ from thermal as well as non-thermal freeze-in mechanism. 
 At low temperatures, the number density of $\rho$ is decreased by two processes, $i.e,$ $\rho\to N H_{2}$ and $\rho\to N H_{1}$ contrary to the previous scenario where only the latter process was kinematically open.
 The corresponding production of $N$ from these decays,  $\Omega_{N}^{NTFI}h^{2}$  as computed from Eq.\ref{s2superwimp} shows no dependence on  $M_{H_{2}}$. The thermal freeze-in contribution on the other hand increases with a decrease in the mass of $H_{2}$. This occurs due  to a decrease  in the phase space suppression factor for the dominant process $\rho\to N H_{2}$.

Although decays will typically contribute to DM formation, we found that for a lighter BSM Higgs the annihilation contribution can be as large as $60\%$ and   that the DM relic density can be satisfied even for a few hundred GeVs  BSM neutral Higgs $H_2$ and charged Higgs $H^{\pm}$. 
To take into account the annihilation contribution,  we implement  the  model in 
Feynrules \cite{Alloul:2013bka}
and generated the CalcHEP files \cite{Belyaev:2012qa} that are  fed to micrOMEGAs
\cite{Belanger:2018ccd, Belanger:2022qxt}. In studying the annihilation contribution,
we consider  two scenarios, a) $\rho$
is DM  and $M_{\rho} < M_{N}$ and b) $N$ is DM and $M_{N} < M_{\rho}$. For both of these scenarios,  we consider 
the parameter space where  the DM relic density varies  in the following range,
\begin{eqnarray}
0.05 \leq \Omega_{DM} h^{2} < 0.123\,.
\label{eqdmrange}
\end{eqnarray}
In choosing this range we require the DM in this model to constitute at least  40\% of the total DM content.

\subsubsection{Fusion dominated scenario: $M_\rho< M_N$ and $M_{H_2}<M_\rho+M_N$}

In this regime, the triplet  $\rho$ is a WIMP DM which is weakly coupled with the  bath particles.  As we have pointed out before, the relic density constraint for the freeze-out production of $\rho$ is satisfied only for $M_{\rho}>2.4$ TeV. For lighter $M_{\rho}$, the thermal contribution to relic density is sub-dominant, while  other production mechanisms can give  large contributions.

We impose the  DM relic density constraint in Eq.~\ref{eqdmrange} and  vary the model parameters
in the following range,
	\begin{eqnarray}
	&& 10^{-12} \leq Y_{\rho\Delta} \leq 10^{-9}\,\,,
	10^{-3} \leq \sin \alpha \leq 10^{-1}\,,\,
	200\,\,{\rm GeV}\,\leq \Delta M  \leq 2000\,\,{\rm GeV}\,,
	\nonumber \\ 
	&&~~~~~~~~~~~~~\,\,700\,\,{\rm GeV} \leq M_{\rho} \leq 1600\,\,{\rm GeV}\,,\,\, 
	125\,\,{\rm GeV}\,\, \leq M_{H_2} \leq 1000\,\,{\rm GeV}\,.
	\end{eqnarray}
	where $\Delta M= M_{N} - M_{\rho}$. 
	 Note that, since $\rho$ is a thermal DM,  it is subject to constraints from direct and indirect detection experiments, which we discuss in Section.~\ref{rho_lhc}. Moreover, the lower value  chosen for $M_{\rho}$ is in  agreement with the 
disappearing track search limit from the  LHC \cite{ATLAS:2022rme}. Additionally,  the chosen $\sin \alpha$ and $M_{H_2}$ range are also in agreement with the LHC constraints which we discuss in Section.~\ref{boundscal}.  Below, we discuss the impact of our chosen  range  for the  model parameters on DM observables.  
\begin{figure}[]
	\centering
	\includegraphics[angle=0,height=8.0cm,width=10.5cm]{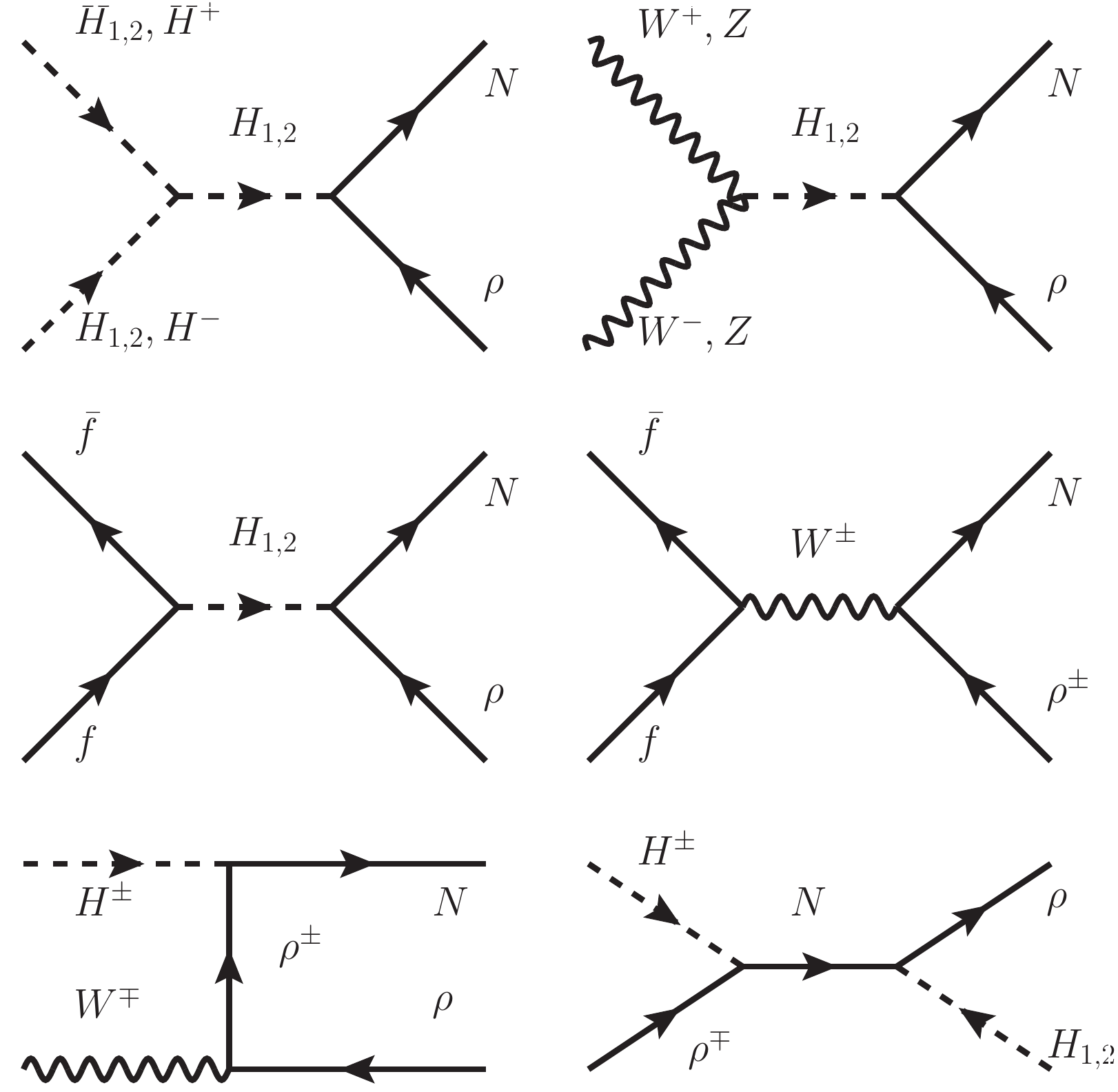}
	\caption{ Production channels for N.  }
	\label{feyndiag}
\end{figure}

The processes contributing to annihilation are shown in Fig.~\ref{feyndiag} and include $N$ and $\rho$ production processes such as  $W^+ W^-, ZZ \to H_{1,2} \to N \rho, H^+ W^- \to H_{1,2} \to N \rho$,  $N\rho^\pm$ production as well as annihilation channels of the type
		$A B \rightarrow N \rightarrow \rho H_i$, where $A, B = \rho, \rho^{\pm}, H_{i}$ (i = 1, 2).
For the masses we consider,    $H_2$ is produced off-shell and hence annihilation processes mediated by $H_2$ are suppressed. Annihilation processes mediated by $H_1$ are further suppressed by the small  mixing angle. Moreover  the decay mode, $H_{2} \rightarrow N \rho$  is  not kinematically allowed. Thus, the  major annihilation contribution arises from  processes involving $\rho H_i$ in the final state where the mediator $N$ is produced on-shell. The production of $N$  can be considered as $2 \to 1$ fusion process $A, B \to N$, and then the produced $N$ undergoes two body decay. The corresponding Feynman diagram is the last diagram of Fig.~\ref{feyndiag}. 

	The on-shell production of $N$ through $2 \rightarrow 1$ production mode
	can be approximately expressed as \cite{Hall:2009bx},
	\begin{eqnarray}
	\Omega_{N} h^{2} \simeq \frac{1.09 \times 10^{24}
		\lambda^2_{N\rho h_i}}
	{16\pi M^2_{N}} \left( (M_{N} + M_{\rho})^{2} - M^2_{H_i} \right)
	\sqrt{1 - \left(\frac{M_{\rho} + M_{H_i}}{M_{N}} \right)^{2} + 
		\left(\frac{M^2_{\rho} - M^2_{H_i}}{M^2_{N}} \right)^{2} } \nonumber \\
	\end{eqnarray}
	where $\lambda_{N \rho H_2 (H_1)} = Y_{\rho\Delta} \cos \alpha\,\, (\sin \alpha)$.
	Once  $N$ is produced on-shell,  it will eventually decay to DM 
	 contributing to its relic density, $\Omega_{\rho} h^{2} = 
	\frac{M_{\rho}}{M_{N}}\, \Omega_{N} h^{2}$. For a very tiny $Y_{\rho\Delta}$, the decay of $N$ can however alter the BBN prediction.
	
\begin{figure}[]
	\centering
	\includegraphics[angle=0,height=7.5cm,width=7.5cm]{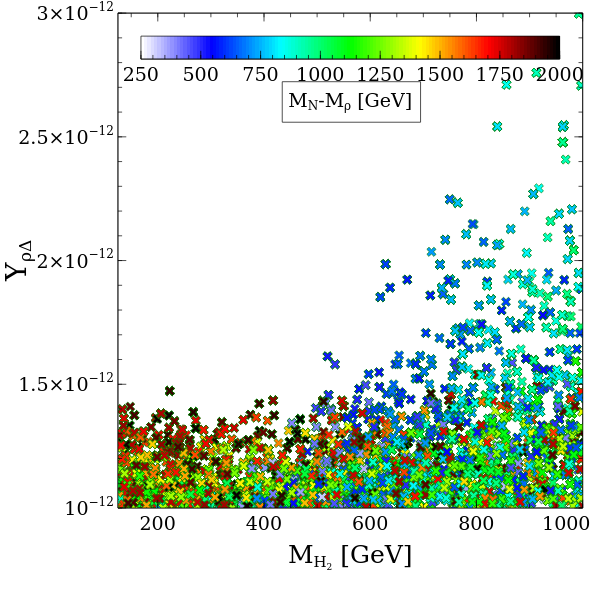}
	\includegraphics[angle=0,height=7.5cm,width=7.5cm]{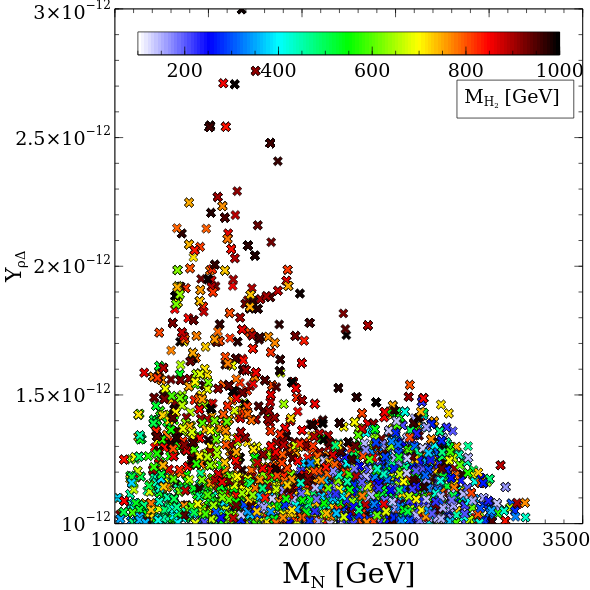}
	\caption{Allowed points in the  $M_{H_2} - Y_{\rho\Delta}$ plane  (LP) and in 
		the $M_{N} - Y_{\rho\Delta}$ plane (RP).
		The color bar in the LP and RP correspond to $M_{N}-M_{\rho}$
		and $M_{H_2}$, respectively. All the points in LP and RP satisfy BBN constraints. }
	\label{vary-mh2-yrho-mn}
\end{figure}

 The masses of  the BSM Higgs, of  $N$ and DM as well as  the Yukawa coupling  are the most important model parameters to determine the relic density. In Fig.\,\ref{vary-mh2-yrho-mn},  we show scatter plots in the 
	$M_{H_2} - Y_{\rho\Delta}$ and $M_{N}-Y_{\rho\Delta}$ planes after
	demanding the  DM relic density to be within the range mentioned in Eq.~\ref{eqdmrange}. The color bar 
	in the LP and RP are for the mass difference $(M_{N} - M_{\rho})$
	and the BSM Higgs mass $M_{H_2}$, respectively.
 On the LP we can see that for  $H_{2}$ mass  below 500 GeV, there are 
	mostly green, red, yellow and black points which correspond to $M_{N} - M_{\rho} > 750$ GeV. Since for this regime, there is no  phase space suppression
	for the production of $N$, hence, $N$ is produced on-shell and decay abundantly to DM. Therefore,  there is no noticeable  variation 
	in the required $Y_{\rho\Delta}$ coupling.  Once we go beyond 
	$M_{H_2} > 500$ GeV,  there are points that correspond to any mass splitting.  The points with the larger mass splitting (red,
	yellow and green points)  are confined to values   $Y_{\rho
		\Delta}\approx (1-4 ) \times 10^{-12}$ for the same reason as above. However for the smaller mass splitting (blue and cyan points),  higher values of $Y_{\rho\Delta}$ are required, as in this region production of on-shell $N$  faces kinematic  suppression. Therefore, to obtain the DM relic density for $M_{H_2} > 500$ GeV, 
	 higher values of $Y_{\rho\Delta}$  are required to  enhance the production of $N$. In the entire region, the primary decay mode is $N \to \rho H_2$ since
the	other allowed decay mode $N \rightarrow \rho H_1$ is  suppressed due to  small 
	neutral Higgs mixing angle. 
	In the RP, we show  the allowed points  in the $M_{N} - Y_{\rho\Delta}$
	plane where the color bar represents variation in  $M_{H_2}$. For  $M_{N} < 2000$ GeV, the production of  $N$ can encounter phase-space  suppression hence requiring much higher values for  $Y_{\rho\Delta}$. 
	 For  $M_{N} > 2000$ GeV,
	 phase space factor do not have any effect.  The variation in Yukawa is solely guided by the chosen  range of DM relic density. This is evident, that  for both the LP and RP, the increase in $Y_{\rho \Delta}$ occurs for similar values of $M_{H_2} $ between  $800 - 1000$ GeV, as for the chosen range of $M_{\rho}$ and $M_{N}-M_{\rho}$, only in this range production of $N$ encounters phase-space suppression.  
\begin{figure}[]
		\centering
		\includegraphics[angle=0,height=7.3cm,width=7.5cm]{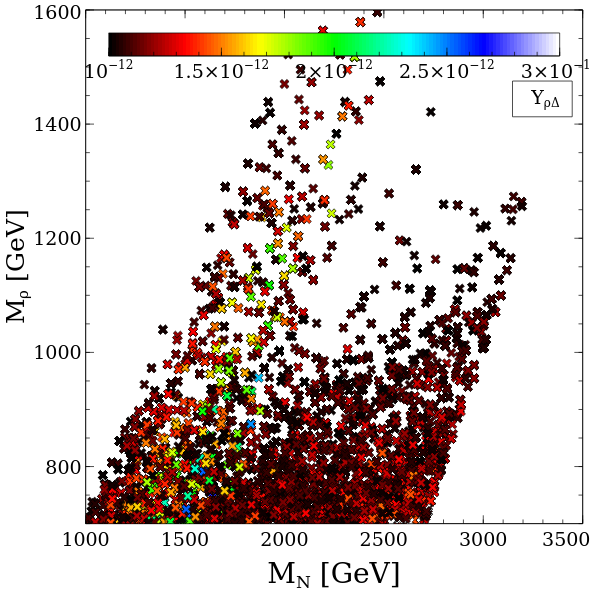}
		\includegraphics[angle=0,height=7.3cm,width=7.5cm]{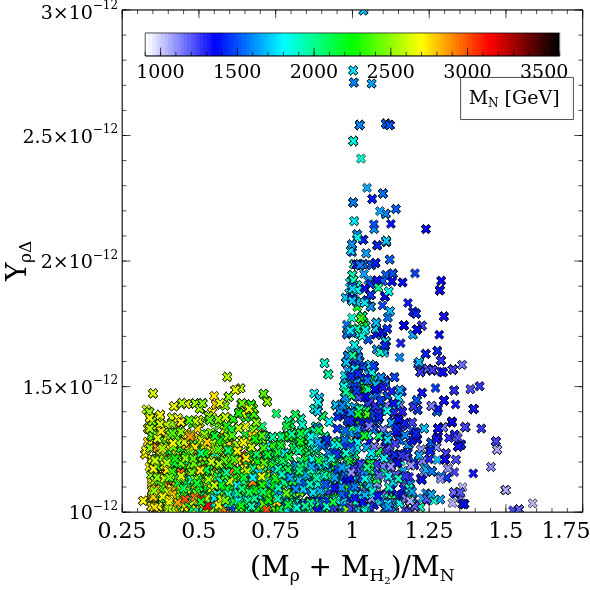}
		\caption{LP (RP): Variation in the $M_{N} - M_{\rho}$ 
			$\left(\frac{M_{\rho}+M_{H_2}}{M_{N}} - Y_{\rho\Delta}\right)$ plane where the color variation is for the Yukawa coupling $Y_{\rho\Delta}$ 
			($M_{N}$). All the points in LP and RP satisfy BBN constraints.}
		\label{vary-mrho-mn-phase-space}
	\end{figure}
		In the LP of Fig.~\ref{vary-mrho-mn-phase-space}, we show a  scatter 
	plot in the $M_{N} - M_{\rho}$ plane where the color points represent
	the variation in  the Yukawa coupling $Y_{\rho\Delta}$. There is a sharp correlation between 
	$M_{N}$ and $M_{\rho}$ as the DM relic density follows the relation
	$\Omega_{\rho} h^{2} = \frac{M_{\rho}}{M_{N}} \Omega_{N} h^{2}$. 
	At $M_{N} \sim 1500$ GeV, a  higher value of $Y_{\rho\Delta}$ is required to satisfy the relic density, 
	as is evident from the few green and yellow points. This occurs due to  phase space suppression in the production of $N$. For higher values of
	$M_{N} \sim 3000$ GeV, the maximum value of  DM mass reaches upto  1350 GeV.  For  a heavier  $\rho$, the relic density from freeze-out contribution also becomes large, and hence after taking into account the late decay contribution from $N$,   the points which lie in the $M_{\rho} > 1350$ GeV range produce overabundant $\Omega_{DM} h^2 > 0.12 $ and hence are disallowed. 
Similarly the region corresponding to  a high $M_N/M_{\rho}$ ratio is disallowed, since it leads to underabundant DM. On the other hand, the white region towards the left is where $N$ production encounters phase-space suppression and hence DM relic density is very low. In the RP, we  
	show  scatter plot in the $\frac{M_{\rho} + M_{H_2}}{M_{N}} - 
	Y_{\rho\Delta}$ plane and the color variation
	represents  different values of $M_{N}$. The points which satisfy 
	$\frac{M_{\rho} + M_{H_2}}{M_{N}} > 1$ correspond to  the region where 
	$N \rightarrow \rho H_2 $ is kinematically not allowed.  hence, in this region   $N \rightarrow \rho
	H_1$ decay is dominant. Moreover in this region there is no  phase-space suppression in the production of $N$ and a moderate value of $Y_{\rho\Delta}$ is required  when $\frac{M_{\rho} + M_{H_2}}{M_{N}} \gg 1$  to maintain the DM relic density in the desired range.
	On the other hand, for the region $ \frac{M_{\rho} + M_{H_2}}{M_{N}} \sim 1$, 
	$N$ production is kinematically suppressed, which leads to  a sudden rise in the  required $Y_{\rho\Delta}$
	 in order to obtain the DM relic density. For  $ \frac{M_{\rho} + M_{H_2}}{M_{N}} \ll 1$, the mass of $N$ is very large. In this region, $N \to \rho H_2$ decay is open and in addition, production of $N$ is also not limited by kinematics. Therefore, a moderate Yukawa is required.  
	
	\subsubsection{Substantial Annhilation Contribution: $M_N < M_\rho$  and $M_{H_2}<M_\rho+M_N$}
	
	In this regime,  $N$ is  a FIMP DM.	This scenario  is challenging  to probe via  direct and indirect DM detection experiments. However, as we will see, the decay of $\rho$ to visible particles along with DM can be probed at colliders. To obtain the DM relic density
	in the aforementioned range, Eq.~\ref{eqdmrange}, we have varied the model parameters 
	as follows,  
	\begin{eqnarray}
	&& 10^{-12} \leq Y_{\rho\Delta} \leq 10^{-8}\,,\,\,\,
	10^{-3} \leq \sin \alpha \leq 10^{-1}\,,\,
	300\,\,{\rm GeV}\,\leq M_{\rho} \leq 1200\,\,{\rm GeV}\,,
	\nonumber \\ 
	&&\,\,10^{-4}\,\,{\rm GeV} \leq M_{N} \leq 100\,\,{\rm GeV}\,,\,\, 
	125\,\,{\rm GeV}\,\, \leq M_{H_2} \leq 1000\,\,{\rm GeV}\,.
	\end{eqnarray}
	For  most of the above mentioned parameters, the chosen range   is similar to the previous scenario, except for $M_N$, which in this case can also be in the sub-GeV range. As we will see, this has implication for the detection of this scenario at future experiments that search for long-lived particles (LLPs), such as MATHUSLA~\cite{Curtin:2018mvb}. 
		
	\begin{figure}[]
		\centering
		\includegraphics[angle=0,height=7.5cm,width=7.5cm]{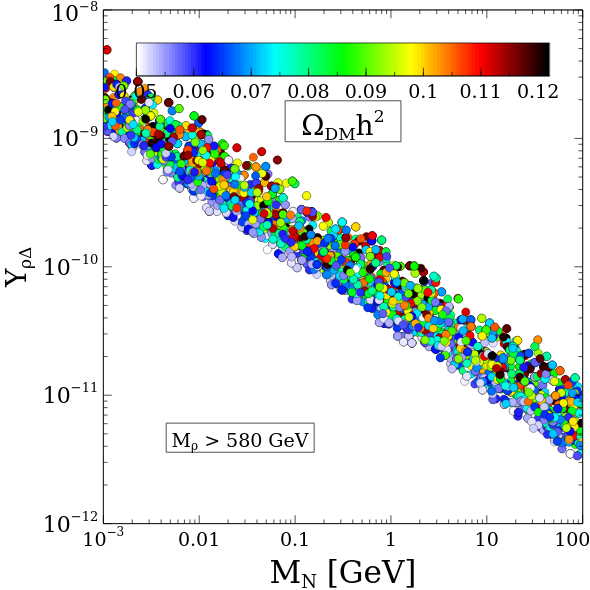}
		\includegraphics[angle=0,height=7.5cm,width=7.5cm]{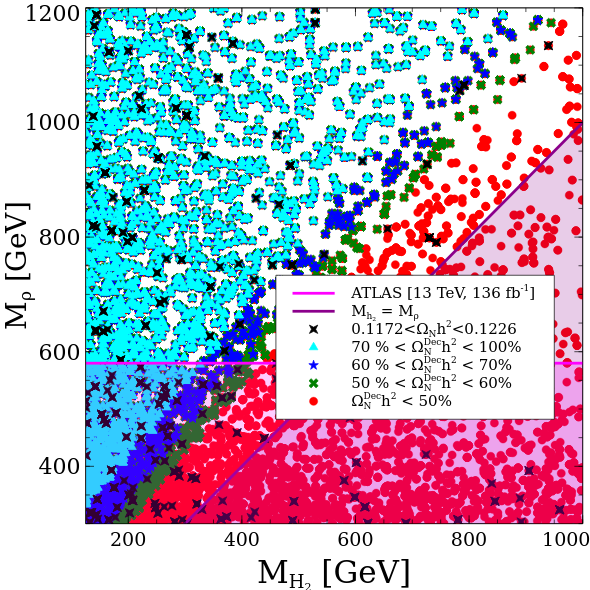}
		\caption{LP and RP show the scatter plots in $M_{N} - Y_{\rho\Delta}$
			and $M_{H_{2}} - M_{\rho}$ planes. In the LP the color variation are 
			due to the variation in DM relic density. All the points in LP and RP satisfy BBN constraints.}
		\label{vary-mn-yrhodelta-mh2rho}
	\end{figure}
	
	 Note that, similar to the previous scenario, $H_{1,2} \to \rho N$ decay is kinematically forbidden, although standard freeze-in contribution and late decay contribution from  $\rho \to N H_{2}$ can be large in part of the parameter space. The other significant contribution arises from the annihilation channels $A B \rightarrow \rho N$ mediated via  
	$H_2$, where $A, B=W^{\pm},Z,\rho^{\pm}, H_{1,2}, H^{\pm}$ etc., see Fig.~\ref{feyndiag}.
	Contrary to the previous scenario, these contributions can be large due to the choice of  a lower DM mass.  In the LP of Fig.~\ref{vary-mn-yrhodelta-mh2rho}, we show scatter plot in the $M_{N} - Y_{\rho\Delta}$ plane where the color bar represents 
	variation in the DM relic density.
	In this scenario, DM is primarily produced from the decay of $\rho$, so the DM
	relic density can be expressed as, $\Omega_{N}h^{2} \propto 
	\frac{M_{N} Y^2_{\rho\Delta}}{M_{\rho}}$. To obtain  the DM relic
	density in a specific range, $M_{N}$ and $Y_{\rho\Delta}$ must follow 
	$\sqrt{M_N} \sim 1/Y_{\rho\Delta}$. This dependency is clearly visible  in  the 
	figure. For a fixed value of $M_{N}$ increase in 
	$Y_{\rho\Delta}$ leads to larger DM relic density, which is  depicted by the red and black points. In the RP, we show the scatter plot in the $M_{\rho} - M_{H_2} $ plane. The different colors points represent the different contributions from the
	decay mode. The region below the dark magenta line corresponds to 
	$M_{\rho} < M_{H_2}$ where mostly  $A B \rightarrow \rho N$ (A, B are the 
	other bath particles) mediated via 
	$H_2$ dominates.
	The region below the red line is ruled out from the 13 TeV  disappearing track searches of the LHC with 136
	$\rm{fb}^{-1}$ data \cite{ATLAS:2022rme}. For a fixed value of $M_{H_2}$ when  $M_{\rho}$ increases the 
	decay contribution dominates because of less phase space suppression 
	for $\rho \rightarrow N H_{1,2}$ decay.

	\begin{figure}[]
		\centering
				\includegraphics[angle=0,height=7.5cm,width=7.5cm]{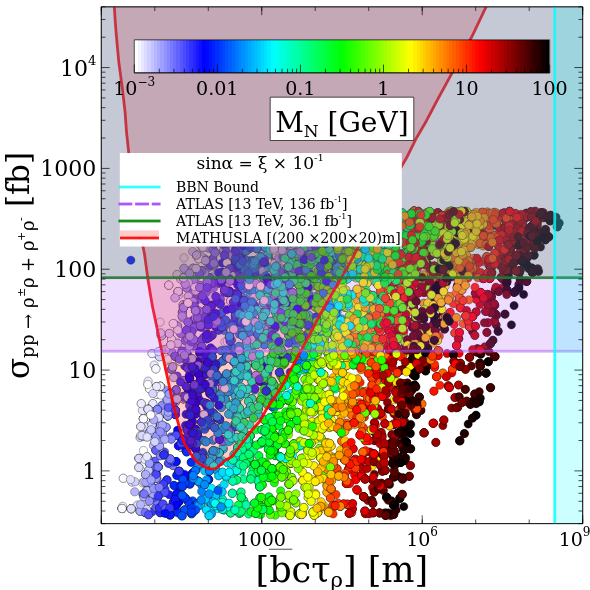}
				\includegraphics[angle=0,height=7.5cm,width=7.5cm]{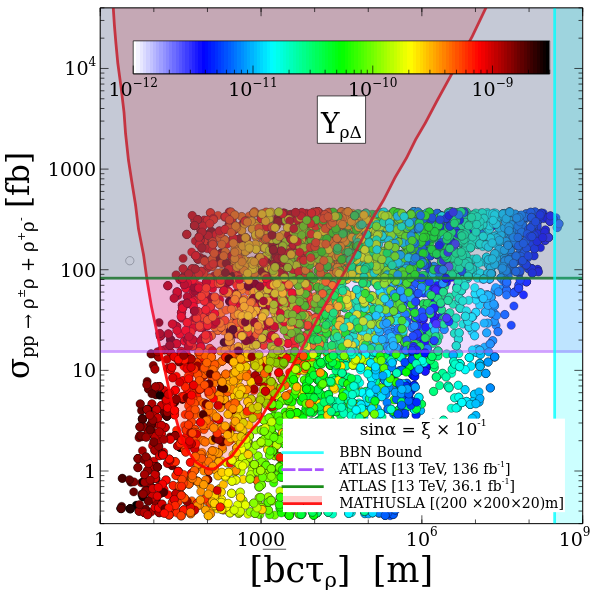}
	\caption{LP and RP show the scatter plot in the decay length 
			($[\bar{b} c \tau_{\rho}] \times \xi^2$) and production cross section
			of $\rho$ ($\sigma_{pp \rightarrow \rho^{+} \rho}$) plane. In the 
			LP color variation is for   $M_{N}$  and $Y_{\rho\Delta}$ for the RP.}
		\label{vary-mathusla}
	\end{figure}
	Although a FIMP DM is challenging to detect at DM direct and indirect detection experiment, and the NLOP in such scenario leaves no signature at the detector unless it is charged, detectors such as MATHUSLA can probe the long lived NLOP.  In Fig.\,\ref{vary-mathusla}, we show  the  detection prospects 
	for our model parameter space at the future MATHUSLA detector with a volume 
	[$200\,m \times 200\, m \times 20\,m$]. We display  the production cross section of the 
	LLP, $\sigma(p p \to \rho^+ \rho^-+\rho^+\rho^0)$, at the HL-LHC as function of the lifetime in the lab-frame
	 $\bar{b}c\tau_{\rho}$ where $\bar{b}$ is the average boost factor \cite{Curtin:2018mvb} and $c\tau_{\rho}$ is the decay length of 
	the long lived particle $\rho$. The color bar represents the variation in $M_N$  ($Y_{\rho \Delta}$)  LP (RP).  
	The region above the green line (which corresponds to 
	$M_{\rho^{\pm}} = 400$ GeV)  is ruled out by the 
	LHC 36.1 $\rm{fb}^{-1}$ data from disappearing track searches \cite{ATLAS:2017oal} and  the region above the  magenta dashed
	line (which corresponds to $M_{\rho^{\pm}} = 580$ GeV) is ruled out by an updated search with a higher luminosity  136 $\rm{fb}^{-1}$ 
	data \cite{ATLAS:2022rme}.
	The red solid line represents the projected  sensitivity of  MATHUSLA taken from \cite{Curtin:2018mvb}. 
	While  the region with a  cross-section of $\rho$ production above  12 fb  is already ruled out by current LHC search, part of the remaining parameter space  can be probed by MATHUSLA. In particular the region with a very light DM,  $M_N \sim 1\ \rm{MeV}-1\ \rm{GeV}$(represented by white, blue and green points)  as can be seen in  LP of Fig.~\ref{vary-mathusla} falls within the sensitivity reach of MATHUSLA.  Note that, large
	 decay lengths correspond to the higher values of DM mass  and to the small values of $Y_{\rho\Delta}$, as can be seen from  RP of Fig.~\ref{vary-mathusla}.
	  As we have mentioned before  the decay width of $\rho$ is proportional to 
	$Y^2_{\rho\Delta}$ and hence the decay length is  inversely proportional to it. Moreover since $\Omega_{N} h^{2} \propto M_{N} Y^2_{\rho\Delta}$, therefore 
	$M_{N}$ and $Y_{\rho\Delta}$
	are anti-correlated.
	 The main important message  from Fig.\,\ref{vary-mathusla} is that  the MATHUSLA detector can be sensitive to  DM mass in the MeV
	range while for heavier  DM, the NLOP  $\rho$ will decay outside the detector.  
	Finally we have checked for both mass ranges {\it i.e.} $M_{\rho} < M_{N}$ and $M_{\rho} > M_{N}$ that our allowed parameter spaces are safe from the Lyman-$\alpha$ constraints, in particular following Ref. \cite{Cembranos:2005us}, we  obtain a free streaming length  less than $0.5$ Mpc.
	
 \section{Collider Prospects of triplet fermions and BSM Higgs}\label{collider1}
 Before concluding, we present a brief discussion on fermion triplet ($\rho_{3}$) and BSM Higgs signature with the ATLAS and CMS detectors at the LHC.
 In our model, the dark sector comprises of SM singlet $N$ and  $\rho_3$ which is a part of electroweak multiplet. The mass of $\rho^{\pm}$  and $\rho$ of $\rho_3$ are degenrate at tree level to preserve the gauge invariance. However, the degeneracy is broken due to quantum correction from electroweak gauge boson and tends to make $\rho^{\pm}$ slightly heavier than $\rho$. The mass splitting $\Delta M(= M_{\rho^{+}}-M_{\rho})$ is proportional to $\alpha_{2} M_{W} \sin^{2}(\frac{\theta_{w}}{2}) $ which is numerically in order of MeV value. The $\Delta M$ increases as mass of $\rho$ increases and after which it becomes approximately constant for large mass of $\rho$. Contrary to fermion triplet, the charged and neutral component of scalar triplet can have mass splitting at tree level due to extra renormalizable interaction with SM Higgs field. However, the mass splitting can't be too large at tree level due to perturbative constraint (see Fig.\ref{unitarity-plot}).\\
 Since $\rho^{\pm}$ and $\rho$ form a compressed mass spectrum, $\rho^{+}$ can decay to $\rho$ along with charged lepton and neutrino when $\Delta M$ is less than the pion mass. For $\Delta M > m_{\pi}$, $\rho^{\pm}$ decays to $\rho$ and $\pi^{\pm}$ and the branching fraction is approximately $98 \%$ for this decay mode\cite{Cirelli:2005uq}. For our considered mass range of $\rho^{\pm},\ \rho $, the above mentioned decay primarly occurs. The decay width of $\rho_{\pm}$ has the following form \cite{Cirelli:2005uq},
 \begin{equation}
 \Gamma (\rho^{\pm} \to \rho \pi^{\pm})= \frac{8 G_{F}^{2}V_{ud}^{2} \Delta M^{3} f_{\pi}^{3}}{4\pi}\sqrt{1-\frac{m_{\pi}^{2}}{\Delta M^{2}}}
 \end{equation}
 The mass splitting between $\rho^{\pm}$ and $\rho$  ensures $\rho$ to be the DM component and makes $\rho^{\pm}$  to be  a long lived particle. The decay product of $\rho$ ,i.e pions, are very soft and is stopped by the magnetic field of the detector. Thus, the pions leaves short track in the detector after which it disappears and hence can be considered as a MET. This disappearing track signature can be used to probe the DM multiplet ($\rho$) having compressed mass spectrum. 
 \subsection{Disappearing track searches}
 There are dedicated LHC analyses of the disappearing track signature for the super symmetric particles such as charginos which are applicable to our model. The production as well as decay process of $\rho$ is similar to charginos. The ATLAS and CMS collaborations set bounds on the  lifetime of charginos as a function of its mass. The maximum excluded mass of $\rho$ from the ATLAS disappearing track search with $136 \ \rm{fb}^{-1}$ data is approximately around $860$ GeV \cite{ATLAS:2022rme}. For our model, this puts a bound on the charged fermions($\rho^{+}$) which is $m_{\rho}>580$ GeV (see Fig \ref{ATLAS-CMS}). The HL-LHC with $3 \ \rm{ab}^{-1}$ of data will also be able to probe upto 870 GeV \cite{Dainese:2019rgk}.
   \begin{figure}[]
 	\centering
 	\includegraphics[angle=0,height=7.0cm,width=7.5cm]{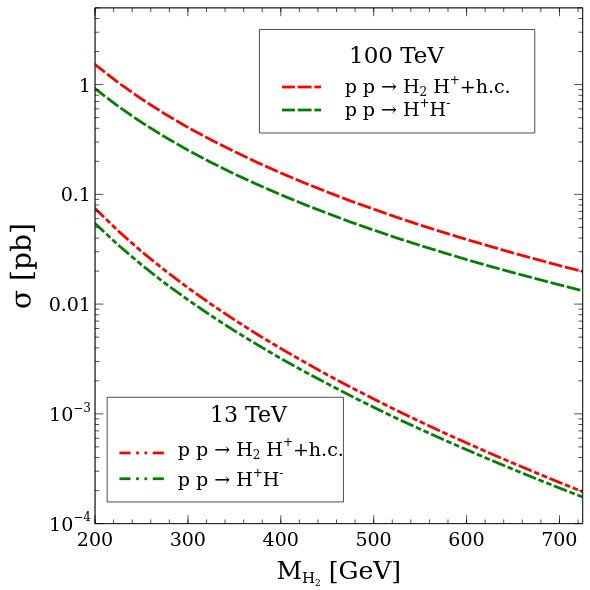}
 	\includegraphics[angle=0,height=7.0cm,width=7.5cm]{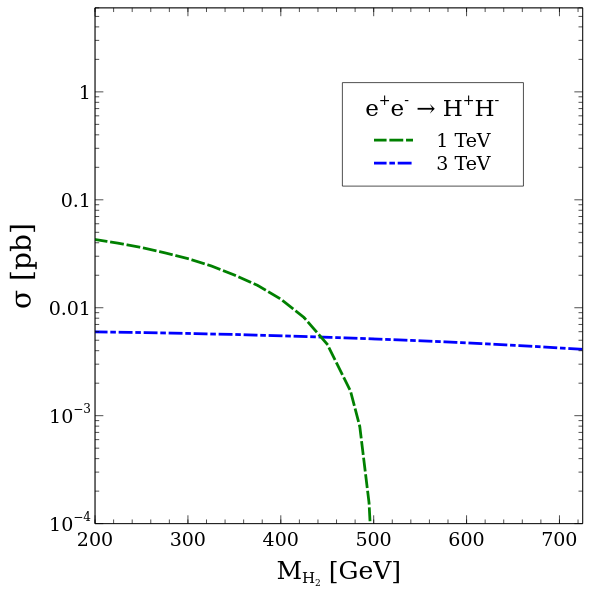}
 	\caption{LP: LHC Production rate of $H_2 H^{+}$ and $H^{+} H^{-}$ for $\sqrt{s}=13$ TeV LHC. RP: Production of $H^{+}H^{-}$ at the $e^{+}e^{-}$
 	lepton collider for $\sqrt{s}=1$ TeV and $\sqrt{s}=3$ TeV.
 	The maximum luminosity goal for 100 TeV FCC-hh collider is 20 $ab^{-1}$ \cite{Hinchliffe:2015qma}
 	whereas the luminosity goal for 3 TEV $e^{+}e^{-}$ CLIC collider 
 	is 5 $ab^{-1}$ \cite{Robson:2018zje}.}
 	\label{ppxsec}
 \end{figure} 
 
 \subsection{Scalar Triplet}
 The BSM Higgs plays an important role in the production of DM for both the scenarios where $\rho$ and $N$ are the DM candidates. Therefore, it becomes crucial to determine whether our choice of paramter spaces are allowed by LHC searches.\\
The primary production process for the neutral BSM Higgs in our scenarios at LHC is through Drell Yan production. The gluon-gluon fusion (ggF) and vector boson fusion (VBF) production mode for $H_{2}$ is suppressed compared to Drell Yan Production due to the $\sin \alpha$ suppression in the coupling of $H_{2}$ to heavy quarks and vector bosons ($W^{\pm}, Z$). Additionally, we also have pair production of the neutral BSM Higgs at LHC via off-shell SM Higgs. This production rate of $H_{2}$ depends on $\lambda_{4}+ 2\lambda_{1} $. In our analysis, we have assumed $\sin\alpha = \sin\delta$ which ensures that $\sin\alpha<0.1$ for $v_\Delta<12\ $GeV and also guarantees production of $H_2$ via off-shell SM Higgs is suppressed. As a result, the primary production of BSM Higgs is via neutral or charged current Drell Yan production. In the LP of Fig. \ref{ppxsec}, we show the cross-sections for $p p \to H^{+} H^{-} $ and $p p \to H_{2} H^{+} $ for $\sqrt{s}=13 $ TeV LHC and $\sqrt{s}=100 $ TeV FCC-hh \cite{Golling:2016gvc} colliders whereas in the RP we show the  $e^{+}e^{-}\to H^{+}H^{-}$ 
production cross-section for c.o.m energy $\sqrt{s} = $ 1 TeV at ILC \cite{Baer:2013cma} and $\sqrt{s} = $ 3 TeV
at CLIC \cite{CLICDetector:2013tfe}. 
  
 \begin{itemize}
 	\item  Production Process
 	\begin{itemize}
 		\item Drell-Yan pair production:- $q q^{'} \to Z,\gamma \to H^{\pm}H^{\mp},H_{2}H_{2}$,\  $q q^{'} \to W^{\pm} \to H^{\pm}H_{2}$
 		\item gg Fusion:- $gg \to H_{1} \to H^{\pm}H^{\mp},H_{2}H_{2}$, $gg \to H_{2} $
 		\item VB Fusion:- $ q q^{'}  \to H_{2}jj$
 	\end{itemize}
 \end{itemize}

 The different LHC contraints applicable for the BSM Higgs and its mixing are as follows,
 
 \begin{itemize}
 	\item The signal strength of the SM Higgs  is given by,
 	\begin{equation}
 	\mu_{H_{1}\to xx}=\frac{\sigma_{H_1}}{\sigma_{H_1}^{SM}}\frac{Br(H_{1}\to xx)}{Br^{SM}(H_{1}\to xx)},
 	\end{equation}
 	where $H_{1}\to x x$ represents decay mode of the SM Higgs. The branching ratio of $H_1$ is almost identical to SM Higgs, $i.e, Br(H_{1}\to xx)\sim Br^{SM}(H_{1}\to xx)$\footnote{As the decay mode of $H_{1}\to \rho N$ is kinematically forbidden in our analysis.}. The primary production mode of SM Higgs is through ggF and VBF. The prodution cross-section of $H_1$ can be written as $\sigma_{H_1}=\cos^{2}{\alpha}\ \sigma^{SM}_{H_1}$. Therefore, the above signal strength can be approximated as,
 	\begin{equation}
 	\mu_{H_{1}\to xx}\sim \cos^{2}{\alpha}.
 	\end{equation} 
 	
 	\begin{figure}[]
 		\centering
 		\includegraphics[angle=0,height=6.5cm,width=7.5cm]{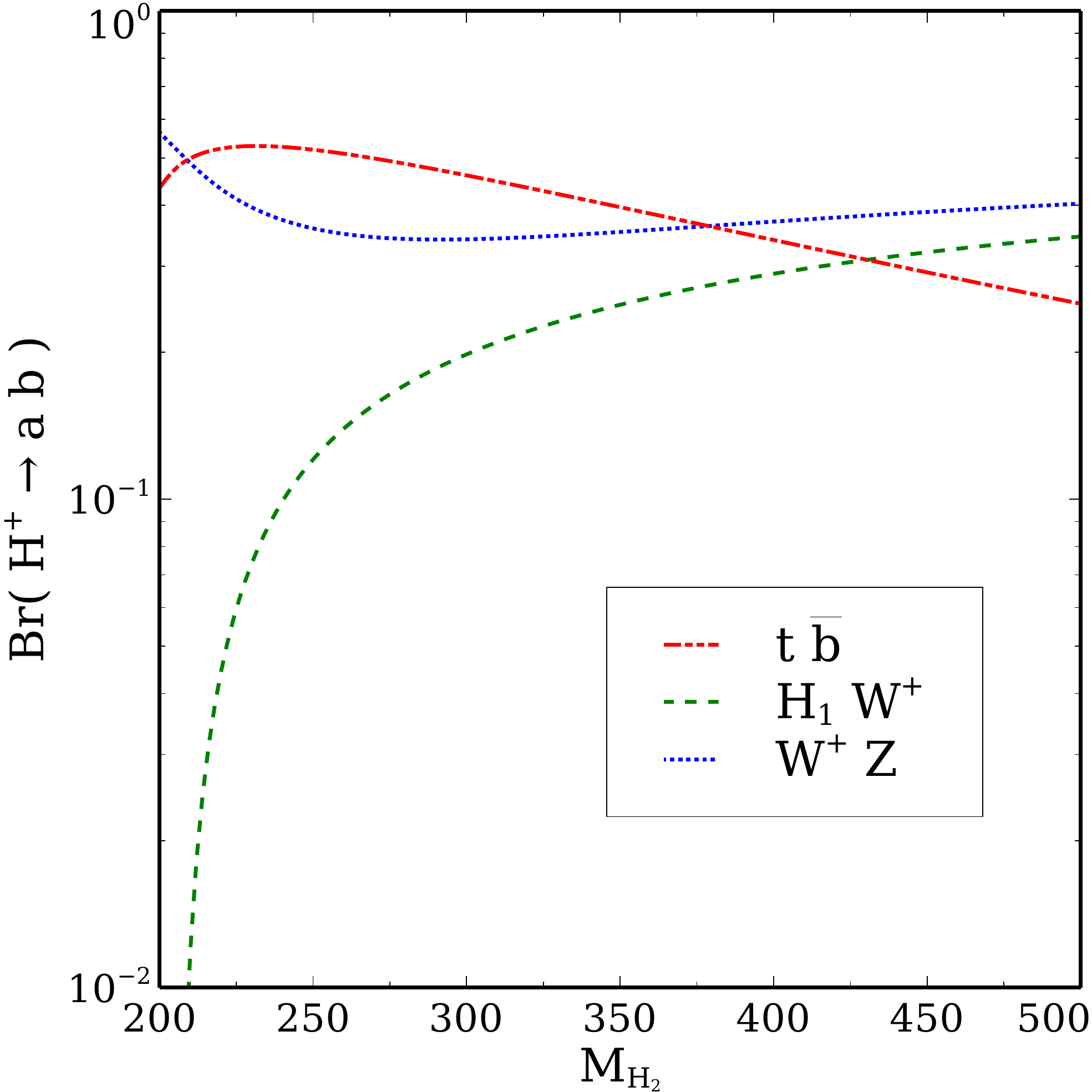}
 		\includegraphics[angle=0,height=6.5cm,width=7.5cm]{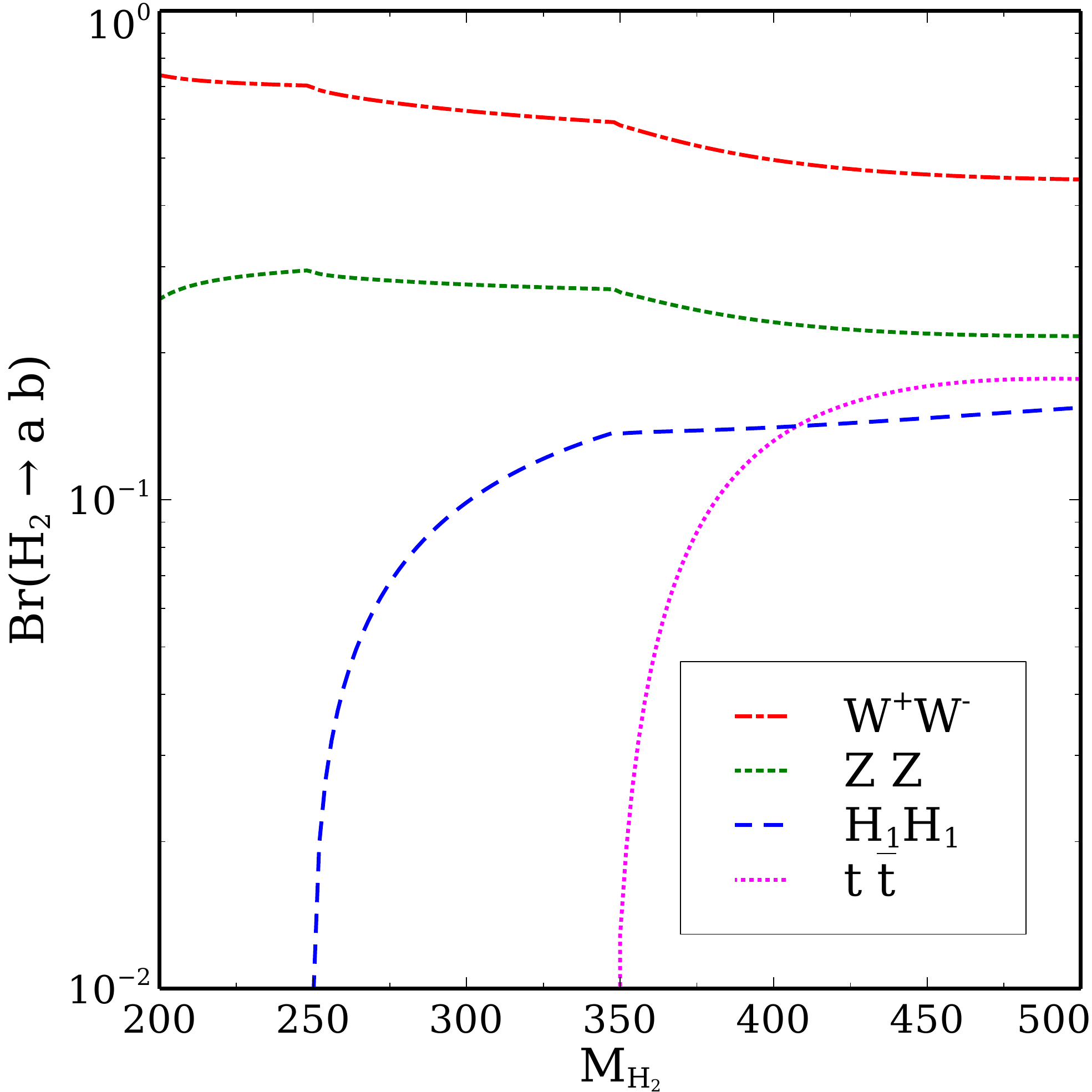}
 		\caption{Branching fractions for $H_2$ and $H^{+}$ for different masses of $H_{2}$.}.
 		\label{BrFrac}
 	\end{figure} 
 	
 	The global signal strength of $H_1$ is $\mu =1.06\pm0.07$ as measured by $\sqrt{s}$=13 TeV LHC \cite{ATLAS:2020qdt}. This measurement puts an upper bound on the BSM-SM Higgs mixing angle and demands that $\sin\alpha$ should be smaller than 0.36. In our analysis, we have considered $\sin\alpha\leq0.1$ and our choosen parameters are in agreement with LHC constraints.
 	\item The LHC searches for resonance BSM Higgs production through ggF and VBF and its decay to SM states. The main decay channels of BSM Higgs $H_{2}$ are $W^{+}W^{-}, ZZ, t\bar{t}$ and $H_1H_1$. The branching fraction for both $H^{+}$ and $H_{2}$ is shown in Fig.\ref{BrFrac}. The ATLAS and CMS searches puts a limit on production cross-section of $H_2$ produced through ggF and VBF times branching fraction of $H_2$. We have checked into following searches, $pp\to H_2 \to ZZ$ \cite{ATLAS:2020tlo}, $pp\to H_2 \to W^{+}W^{-}+ ZZ$\cite{ATLAS:2020fry} and $pp\to H_2 \to H_1H_1$\cite{ATLAS:2019qdc} and found that searches doesn't constraint our parmater space at all.

 \item Higgs Diphoton rate:- The introduction of the real triplet scalar will lead to a correction to the SM Higgs diphoton rate via the addition of a newly charged scalar in the loop. For the light charged scalar, the SM Higgs diphoton rate is enhanced because of constructive interference between Higgs triplet and SM contribution. The enhancement in the diphoton rate will be  proportional to coupling strength $\lambda_{4}+ 2\lambda_{1}$ of the scalar potential. The SM Higgs partial width to diphoton is given by \cite{Wang:2013jba},
 \begin{equation}
 \Gamma_{H_{1}\to \gamma\gamma}^{SM}=\frac{\alpha^{2} M_{H_1}^2}{256 \pi^{2}v^2}\left|\sum_{f} N_{f}^{c}Q_{f}^{2}y_{f}A_{1}^{\gamma\gamma}(r_{f})+y_{W}A_{2}^{\gamma\gamma}(r_{W}) \right|^{2}.
 \end{equation}
 In our scenario, the partial width of $H_1$ to diphoton is given by,
 \begin{equation}
 \Gamma_{H_{1}\to \gamma\gamma}^{HTM}=\frac{\alpha^{2} M_{H_1}^2}{256 \pi^{2}v^2}\left|\sum_{f} N_{f}^{c}Q_{f}^{2}y_{f}A_{1}^{\gamma\gamma}(r_{f})+y_{W}A_{2}^{\gamma\gamma}(r_{W})+Q_{H}^{2}\frac{v g_{hH^{+}H^{-}}}{m_{H_2}^2}A_{3}^{\gamma\gamma}(r_{H_2}) \right|^{2}.
 \end{equation}
  
 where $r_{i}=m_{H_{1}}^{2}/4M_{i}^{2}$, $Q_{F,H_{2}}$ are the electric charges of fermion and scalar. The vertex 
 $g_{hH^{+}H^{-}}$ takes the  form,
 \begin{eqnarray}
 g_{hH^{+}H^{-}} = - (\lambda_{3} + 2 \lambda_{2}) v_{\Delta} \sin\alpha 
 - \frac{v \cos\alpha }{2} \left[ 
 \left( \lambda_{4} + 2 \lambda_{1} \right) \cos^{2}\delta + 
 \lambda_{h} \sin^2\delta \right]\,.
 \end{eqnarray} 
From the expression of $g_{hH^{+}H^{-}}$, it is clear that considering
$\delta\neq \alpha $  would only have a mild impact on 
$g_{hH^{+}H^{-}}$ and hence on $R_{\gamma\gamma}$. 

  The loop functions are given by,
 \begin{align*}
 A_{1}^{\gamma\gamma}(x)&=2\left(x+(x-1)f(x)\right)x^{-2},\\
 A_{2}^{\gamma\gamma}(x)&=-\left(2x^{2}+3x+3(2x-1)f(x)\right)x^{-2},\\
 A_{3}^{\gamma\gamma}(x)&=-\left(x-f(x)\right)x^{-2},\\
 f(x)&=\begin{cases}
 (\sin^{-1}\sqrt{x})^{2} \, \, \, x\leq 1,\\
 -\frac{1}{4}\left[log \frac{1+\sqrt{1-x^{-1}}}{1-\sqrt{1-x^{-1}}}-i\pi\right] \, \, \, \, x>1.
 \end{cases}\\
 \end{align*}
 \begin{figure}[]
 	\centering
 	\includegraphics[angle=0,height=7.5cm,width=9.0cm]{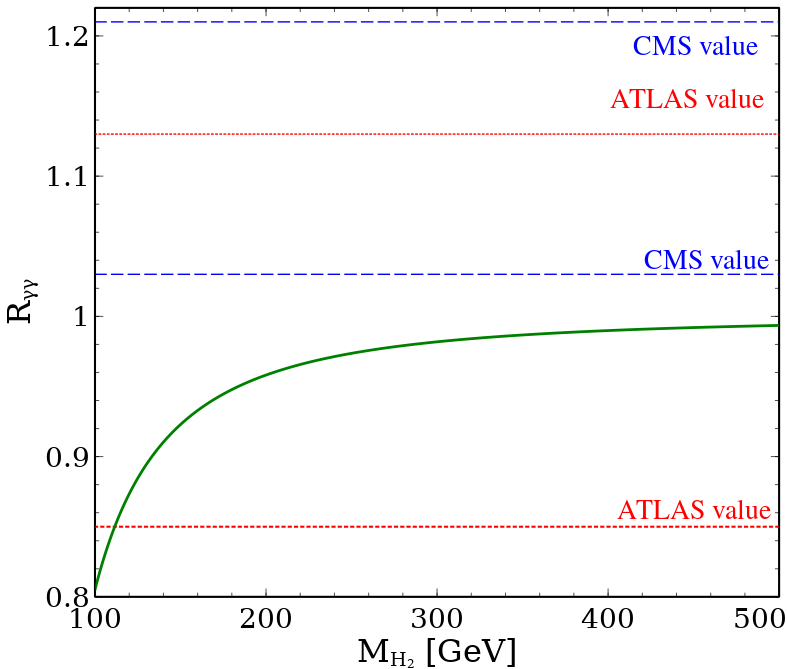}
 	\caption{Higgs diphoton Rate for different masses of $H_{2}$ assuming $\sin \alpha =\sin \delta =0.1$.}
 	\label{diphoton}
 \end{figure}
 The signal strength of  SM Higgs to diphoton process is  given by,
 \begin{equation}
 R_{\gamma\gamma}=\frac{\Gamma^{HTM}_{H_{1}\to \gamma\gamma}}{\Gamma^{SM}_{H_{1}\to \gamma\gamma}}
 \end{equation}
 Recent $R_{\gamma\gamma}$ measurements by the ATLAS \cite{ATLAS:2018hxb}  and CMS \cite{CMS:2021kom} are,
 \begin{equation}
 R^{ATLAS}_{\gamma\gamma}=0.99 \pm 0.14\,  \, \, \, \, \,  R^{CMS}_{\gamma\gamma}=1.12^{+0.09}_{-0.09}
 \end{equation}
 
 For our analysis, we have assumed $\sin\alpha=\sin\delta$ which ensures that the quartic coupling $\lambda_{4}+2\lambda_{1}$ is positive and independent of the mass of $H_{2}$. Due to this, $H_{1}\to \gamma\gamma$ is suppressed as the destructive interference occurs between the SM and Higgs triplet contribution. In Fig.\ref{diphoton}, we have shown that $H_{1}\to \gamma\gamma$ which increases with the increase in mass of $H_2$. 
\end{itemize}

In the LP and RP of Fig. \ref{BrFrac} we have 
shown the branchings of
$H^{+}$ and $H_2$ to different SM final states. In generating
the plots, we have assumed that $M_{H^{\pm}} = M_{H_2}$.   
As can be seen from the LP, in the mass range 
$200 \,\,{\rm GeV} < M_{H_2} < 375\,\,\,{\rm GeV}$, the dominant 
decay mode  is $H^{+} \rightarrow t \bar{b}$  while in the 
mass range $375 \,\,{\rm GeV} < M_{H_2} < 500\,\,\,{\rm GeV}$, the
dominant mode is $H^{+} \rightarrow W^{+} Z$. Therefore, for the lower mass range  
 the signature for $H^{+}H^{-}$ production at either $pp$ or $e^{+}e^{-}$ 
colliders would be either $4j+4b$ or $2l+4b+\cancel{E}_{T}$.
On the other hand for the mass range $375 \,\,{\rm GeV} < M_{H_2} < 500\,\,\,{\rm GeV}$, 
the final states are either $6l + \cancel{E}_{T}, 2l + \cancel{E}_{T},
4j + 4l \,\,\,{\rm and}\,\,\,4j + \cancel{E}_{T}$. In RP we show that for $H_2$  the dominant decay
mode is $W^{+}W^{-}$, for the whole mass range. Therefore, one can search for 
$pp \rightarrow H^{+}H_2$ in  the following final states $6j+2b$, $3l+2b+\cancel{E}_{T}$, $5l+\cancel{E}_T$
and $6j+\cancel{E}_T$.
Note that, as seen in  Fig. \ref{ppxsec},
the production cross-section for  $H^{+}H^{-}$ at $pp$ collider falls sharply at high masses for $\sqrt{s} = 13$ TeV and
is lower than in $e^{+}e^{-}$ collisions. Thus it could be advantageous to study this process at a high energy $e^+e^-$ collider than the 
$\sqrt{s} = 13$ TeV pp collider. Moreover, we need a full fledged collider 
study in order to compare the signal superiority between the 
proposed $e^{+}e^{-}$ for $\sqrt{s} = 1, 3$ TeV and
100 TeV FCC-hh colliders. The detailed study of the collider 
prospects for different
final states including the associated backgrounds is left for future work.

\section{Conclusion}\label{conclusion}

In this work, we have extended the SM by adding one singlet fermion, three
$SU(2)_{L}$ triplet fermions and one triplet scalar with hypercharge zero
for all the additional particles. The new particles can solve 
two well-accepted SM problems namely a
dark matter candidate and the origin of the neutrino mass. When the triplet
scalar acquires a vev, it generates a mixing between the neutral component  of 
the triplet fermion and the singlet fermion which are odd under 
$\mathbb{Z}_2$ symmetry. The lightest of these particles becomes a suitable 
dark matter candidate while  the NLOP eventually decays to the DM candidate. The NLOP is typically long-lived.
The remaining two triplet fermions which are even under the  
$\mathbb{Z}_2$ symmetry take part in the generation of neutrino masses and  the bounds on the 
neutrino oscillation parameters can be very easily satisfied. 
In exploring the viable parameter space of the model we impose the strict range on the DM relic density
obtained from PLANCK observations of the CMB. We have also 
taken into account  bounds on the mass of the fermion triplet 
 from the disappearing track search at the ATLAS and CMS detectors as well as current bounds from   
direct and indirect searches of DM which apply only to the WIMP DM. Morever we found that the neutral and charged components of the scalar triplet are nearly degenerate when imposing a perturbative bound
on the quartic couplings.

To investigate the possible DM formation mechanisms, we have considered two 
mass regimes corresponding to the lightest stable DM particle being the triplet dominated
fermion $\rho$ or the singlet dominated fermion $N$.  In both regimes, the NLOP can decay to the
DM candidate and a scalar particle thus injecting extra hadronic energy in the Universe.
Depending on the coupling strength,  the late decay might happen during the BBN and thus alter the light elements (H, D, He, Li) abundance. We have imposed constraints from BBN   and found the model to be viable in large areas of parameter space.
 
For the mass range $M_{\rho} < M_{N}$, DM can be produced  by thermal freeze-out and by the  late decay of the NLOP $N$. The presence of the late decay production mode means that  the correct value for the DM relic density for the triplet can be reached  even for masses below $1$ TeV while the freeze-out mechanism by itself requires a mass above 2.4 TeV.
The NLOP $N$ can be produced  by the freeze-in mechanism from the decay of the BSM Higgs $H_{2}$,
the annihilation of bath particles and fusion processes. The dominant production mode depends on the masses of the 
NLOP and of other particles involved in its production.
Since the NLOP never reaches thermal equilibrium, in order
to properly take into account its decay contribution to DM, 
we have determined  the NLOP distribution function and used it in the subsequent DM production.
   
In the  mass regime $M_{N} < M_{\rho}$, DM never reaches 
thermal equilibrium with the cosmic soup, so it is produced
by the freeze-in  mechanism through the decay, annihilation and fusion 
of the bath particles.  Depending on the mass  of DM and other new particles
either/all of them actively contribute to DM production. Moreover,
DM is also produced from the late decay of the NLOP $\rho$ by the superWIMP 
mechanism. In this case we showed that DM could be detected indirectly at the MATHUSLA
detector from the late decay of NLOP or by reconstructing the 
displaced vertex and missing energy searches at the ATLAS and CMS detectors.

In conclusion in both regimes there remains a  possibility for detection of DM
even if it is produced mainly  by the freeze-in mechanism. Moreover we have  briefly
discussed the possibility of triplet fermion and Higgs detection prospects
at the LHC, a more extensive study of collider prospects is
left for future work.
\section*{Acknowledgements}
This work was funded in part by the Indo-French Centre for the Promotion of
Advanced Research (Project title: Beyond Standard Model Physics with Neutrino and Dark Matter at Energy, Intensity and Cosmic Frontiers, Grant no: 6304-2).
This work used the Scientific Compute Cluster at GWDG, 
the joint data center of Max Planck
Society for the Advancement of Science (MPG) and University of 
G\"{o}ttingen. AR acknowledges SAMKHYA: High-Performance Computing Facility provided by the Institute of Physics (IoP), Bhubaneswar.

\appendix
\section{Appendix}\label{appendix}
\subsection{ Analytical expressions for thermal average cross-section} \label{appendix1}
The analytical expression for DM annihilation and co-annihilation 
channels are as follows \cite{Ma:2008cu},
\begin{eqnarray}
\sigma (\rho \rho) |v| \simeq \frac{2 \pi \alpha^2_L}{M^2_{\rho}}\,,
\,\,\,
\sigma (\rho \rho^{\pm}) |v| \simeq \frac{29 \pi \alpha^2_L}{8 M^2_{\rho}}
\,,\,\,\,\sigma (\rho^{+} \rho^{-}) |v| \simeq \frac{37 \pi \alpha^2_L}{8 M^2_{\rho}}\,,\,\,\,\sigma (\rho^{\pm} \rho^{\pm}) |v| \simeq \frac{\pi \alpha^2_L}{M^2_{\rho}}\,. 
\end{eqnarray}    
Thermal average of the above cross sections takes the following form,
\begin{eqnarray}\label{Appendix}
\langle \sigma_{eff} |v| \rangle = \frac{g^2_{0}}{g_{eff}} \sigma(\rho \rho)
+ 4 \frac{g_{0} g_{\pm}}{g^2_{eff}} \sigma(\rho \rho) (1 + \epsilon)^{3/2} 
e^{- \epsilon x} +  \frac{g^2_{\pm}}{g^2_{eff}}
\left[ 2 \sigma(\rho^{+}_{3} \rho^{-}_{3}) + 2 \sigma(\rho^{\pm}_{3} \rho^{\pm}_{3}) \right] (1 + \epsilon)^{2} e^{- 2 \epsilon x} \nonumber \\
\end{eqnarray}
where $g_{0} = g_{\pm} = 2$, $\epsilon = \frac{\Delta}{M_{\rho}}$
and $g_{eff} = g_{0} + 2 g_{\pm} (1 + \epsilon)^{3/2} e^{-\epsilon x}$ 
($\Delta = M_{\rho^{\pm}_{3}} - M_{\rho}$)\,.

\subsection{ Collision function} \label{appendix2} The collision function for the production of
the next to lighest neutral fermion from the decay of
the BSM Higgs has the following form,
\begin{eqnarray}
\mathcal{C}^{h_i \rightarrow N \rho} &=&
\dfrac{r}{16\pi M_{sc}}\dfrac{\mathcal{B}^{-1}(r)\,\,|M|^2}
{\xi_p \sqrt{\xi_p^2\mathcal{B}(r)^2+
\left(\dfrac{M_{N}\,r}{M_{sc}}\right)^2}}
\times
 \left(e^{-\sqrt{\left(\xi_{k}^{\rm min}\right)^2
\mathcal{B}(r)^2+\left(\frac{ M_{H_2}\,r}{M_{sc}}\right)^2}}
\,-\,e^{-\sqrt{\left(\xi_{k}^{\rm max}\right)^2
\mathcal{B}(r)^2+\left(\frac{ M_{H_2}\,r}{M_{sc}}\right)^2}}
\right) \,.\nonumber \\
\label{ch2zblzbl-final}
\end{eqnarray}
where $\mathcal{B}(r) = \displaystyle \left( \displaystyle 
\frac{g_{s}\left(\displaystyle \frac{M_{sc}}{r_0}\right)}
{g_{s}\left(\displaystyle \frac{M_{sc}}{r}\right)} \right)^{\frac{1}{3}}$ and $r_0$
is the initial value of $r$.
The amplitude for the process $h_{2} \rightarrow N \rho$ can be expressed as,
\begin{eqnarray}
|M|^{2} = 2 \lambda^2_{N\rho h_i} M^2_{h_i} \left(1 - x^2 \right) \theta(1 - x)
\end{eqnarray}
where $x = \frac{M_{\rho} + M_{N}}{ M_{H_2}}$, $\lambda_{N\rho h_{2}} = 
Y_{\rho \Delta} \cos \theta$ and 
$\lambda_{N\rho h_{1}} = Y_{\rho \Delta} \sin \theta$\,. The parameters
$\xi^{min}_{k}$ and $\xi^{max}_{k}$ can be expressed as\footnote{$k$ is
the momentum used in the phase space integration},
\begin{eqnarray}
\xi_k^{\rm min} (\xi_p,r)&=&\dfrac{M_{sc}}{2\,\mathcal{B}(r)\,r\,M_{N}}
\left| \,\eta (\xi_p,r)-\dfrac{\mathcal{B}(r)
\times M_{H_2}^2}{M_{N} \times M_{sc}}\,\xi_p\,r
\right| \,,\nn \\
\xi_k^{\rm max} (\xi_p,r)&=&\dfrac{M_{sc}}{2\,\mathcal{B}(r)\,r\,M_{N}}
\bigg( \,\eta (\xi_p,r)+\dfrac{\mathcal{B}(r)
\times M_{H_2}^2}{M_{N}\times M_{sc}}
\,\xi_p\,r \,\bigg)\,,
\end{eqnarray}
where $\eta (\xi_p,r)$ is given by
$
\eta(\xi_p,r)= \left(\frac{ M_{H_2}\,r}{M_{sc}}\right)
\,\sqrt{\dfrac{ M_{H_2}^2}{M_{N}^2}-4}\,\,
\sqrt{\xi_p^2\,\mathcal{B}(r)^2+
\left(\frac{M_{N}\,r}{M_{sc}}\right)^2}\,.
$
\subsection{ Decay Widths} \label{appendix3} The decay width for
the process $N\rightarrow \rho h_1$ can be expressed as,
\begin{eqnarray}
\Gamma_{N \rightarrow \rho h_{1}} = \frac{\lambda^2_{N\rho h_1} 
\left((M_{N} + M_{\rho})^{2} - M^2_{H_1} \right)}{16 \pi M_{N}}
\sqrt{\left(1 - \left(\frac{M_{\rho} + M_{H_1}}{M_{N}} \right)^{2} \right)
\left(1 - \left(\frac{M_{\rho} - M_{H_1}}{M_{N}} \right)^{2} \right)}
\nonumber \\
\end{eqnarray}

\bibliographystyle{JHEP}
\bibliography{bibitem.bib}
\end{document}